\documentclass[english,aps,prd,amsfonts,preprintnumbers,superscriptaddress,showpacs,floatfix,nofootinbib]{revtex4}
\usepackage[T1]{fontenc}
\usepackage[latin9]{inputenc}
\usepackage{amsmath}
\usepackage{graphicx}
\usepackage{amssymb}

\makeatletter
\@ifundefined{textcolor}{}
{%
 \definecolor{BLACK}{gray}{0}
 \definecolor{WHITE}{gray}{1}
 \definecolor{RED}{rgb}{1,0,0}
 \definecolor{GREEN}{rgb}{0,1,0}
 \definecolor{BLUE}{rgb}{0,0,1}
 \definecolor{CYAN}{cmyk}{1,0,0,0}
 \definecolor{MAGENTA}{cmyk}{0,1,0,0}
 \definecolor{YELLOW}{cmyk}{0,0,1,0}
 }

\usepackage{epsfig}% Include figure files
\usepackage{dcolumn}% Align table columns on decimal point
\usepackage{bm}% bold math
\@ifundefined{definecolor}
 {\usepackage{color}}{}
\usepackage{verbatim}

\newcommand{\ud}{\,\text{d}}

\def\thebiblio#1{
\begin{center}\bf \large References
\end{center}
\list
{[\arabic{enumi}]}{\settowidth\labelwidth{#1.}\leftmargin\labelwidth
 \advance\leftmargin\labelsep
 \usecounter{enumi}}
 \def\newblock{\hskip .11em plus .33em minus -.07em}
 \sloppy
 \sfcode`\.=1000\relax}%\nofiles

\makeatother

\usepackage{babel}

\begin{document}

\title{Particle Production of Vector Fields: Scale Invariance is Attractive}

\author{Konstantinos Dimopoulos}
\email{k.dimopoulos1@lancaster.ac.uk}
\affiliation{Physics Department, Lancaster University, Lancaster LA1 4YB, U.K.\\
}

\author{Jacques M. Wagstaff}
\email{j.wagstaff@lancaster.ac.uk}
\affiliation{Physics Department, Lancaster University, Lancaster LA1 4YB, U.K.\\
}

\date{\today}
%---------------------------------------------------------------------------------------------------------------------
%---------------------------------------------------------------------------------------------------------------------

\begin{abstract}
In a model of an Abelian vector boson with a Maxwell kinetic term and
non-negative mass-squared it is demonstrated that, under fairly general
conditions during inflation, a scale-invariant spectrum of perturbations for the components of
a vector field, massive or not, whose kinetic function (and mass) is modulated
by the inflaton field is an attractor solution. If the field is massless, or
if it remains light until the end of inflation, this attractor solution
also generates anisotropic stress, which can render inflation weakly
anisotropic. The above two characteristics of the attractor solution can source
(independently or combined together) significant statistical anisotropy in the
curvature perturbation, which may well be observable in the near future.
\end{abstract}
\maketitle
\begin{widetext}
%---------------------------------------------------------------------------------------------------------------------
%---------------------------------------------------------------------------------------------------------------------
\section{Introduction}

Cosmic inflation is arguably the most compelling way to overcome
(or at least ameliorate)
the so-called horizon and flatness fine-tuning problems for the initial
conditions of the Hot Big Bang cosmology \cite{inf}.
However, this success of inflation is
guaranteed if inflation lasts long enough regardless of the specifics which
cause the accelerated expansion in the first place. As a result one cannot
use the horizon and flatness problems as a discriminator between different
models of inflation. To this end, much attention has been paid to another
byproduct of an inflationary period; namely the generation
of the curvature perturbation $\zeta$ in the Universe, which is the source of
structure formation and the observed CMB primordial temperature perturbation.
The characteristics of $\zeta$ open an observational window on the dynamics of
inflation, which is no more that ten e-folds and corresponds to the so-called
cosmological scales \cite{book}.
Using this, inflation has become a testable theory with
each model realisation offering specific predictions that can be
observationally falsified.

Contrast with observations has strengthened the case for inflation. Firstly, in
the last decade, the rival theory (that of cosmic strings) for the generation
of $\zeta$ was observationally rejected \cite{leandros}.
Secondly, the basic predictions of
inflation appear to be supported by observational evidence. Indeed, inflation
generically seems to produce a predominantly scale-invariant spectrum of
predominantly Gaussian curvature perturbations which are also predominantly
statistically homogeneous and isotropic \cite{book}.
Simple, single-field inflationary models are able
to produce such a $\zeta$. However, in recent years the precision of
cosmological observations has increased to the point that allows us to explore
deviations from the above ``vanilla'' predictions of inflation, which has put
increasing tension on inflationary models and resulted in a shift in the
thinking of model-builders, away from the simple single-field toy-models and
more towards more complex model realisations of inflation which make
better use of the rich content of realistic theories beyond the standard model.
A prominent example is the so-called curvaton hypothesis \cite{curv}, which
removes
the responsibility for the generation of $\zeta$ from the inflaton field and
assigns it to another field (the curvaton) which is unrelated to the physics
of inflation (it does not affect the inflationary dynamics) and can be
associated with physics at much lower energy scales, e.g. TeV physics explored
by collider experiments such as the LHC.

Observations now indicate that there is a significant deviation from
scale-invariance in $\zeta$. Indeed, the spectral index of the curvature
perturbation is \mbox{$n_s-1=-0.037\pm 0.014$} according to the latest WMAP
data \cite{wmap}.
This observation reveals the existence of inflationary dynamics, which
agrees with expectations. It also means that models of inflation that generate
a blue spectrum of perturbations are most probably rejected. In
conjunction with the upper bounds on tensor perturbations, the above result
has also falsified some red-tilted models too, such as all monomial models
of chaotic inflation higher than quadratic. Similarly, non-Gaussian signatures
appear to offer a second test for inflationary models. Indeed, the latest WMAP
results produce a tentative observation of non-Gaussianity in the squeezed
configuration, which gives for the non-linearity parameter
\mbox{$f_{\rm NL}=32\pm 21$} \cite{wmap}.
If this result is confirmed it will falsify all
single-field models of inflation, which predict \mbox{$|f_{\rm NL}|\ll 1$}.

Along these lines deviations from the remaining ``vanilla'' predictions of
inflation should be investigated, namely statistical homogeneity and isotropy.
Observations already offer some tentative evidence that $\zeta$ may not be
exactly homogeneous and isotropic. For example, a 10\% difference in power
between hemispheres has been reported which would amount to a deviation from
homogeneity \cite{inhom}.
Furthermore, tantalising evidence of a preferred direction
(the so-called Axis of Evil \cite{AoE}) on the microwave sky, amounting to an
alignment of the low multipoles of the CMB, has been much discussed
\cite{acker}. This
alignment persists beyond foreground removal \cite{forg} and is statistically
extremely unlikely \cite{hansen}.
If such a preferred direction does exist then this could amount to
statistical anisotropy in the curvature perturbation, which can become a
powerful discriminator between inflationary models and mechanisms for the
generation of $\zeta$.

Observations currently allow up to 30\% statistical anisotropy in the
power-spectrum of $\zeta$ \cite{GE}.\footnote{In Ref.~\cite{GE} statistical
anisotropy of magnitude
\mbox{$(29\pm3)\%$} is actually found in the spectrum of the WMAP data at
9-$\sigma$. However, the preferred direction is suspiciously close to the
ecliptic plane so the authors acknowledge that this observation is probably
influenced by some unknown systematic (see also Ref.~\cite{ref}). Hence, this
finding can only be taken as an upper bound on the statistical anisotropy of
the primordial signal.}
Similarly, statistical
anisotropy in the bispectrum can be predominantly anisotropic even if
statistical anisotropy in the spectrum is negligible. If this is
the case, then non-Gaussianity will feature an angular dependence on the
microwave sky \cite{fnlanis,cesar}.
The forthcoming observations of the Planck satellite may well
provide conclusive evidence of non-zero statistical anisotropy in the spectrum
and/or bispectrum of the CMB temperature perturbations. In particular, the
bound on statistical anisotropy in the spectrum will be reduced to 2\% if
such anisotropy is not observed \cite{planck}.
Similarly, the observational bounds on
$f_{\rm NL}$ are expected to tighten by at least an order of magnitude,
so statistical anisotropy in the bispectrum may well be found if
non-Gaussianity is observed. Thus, it is imperative to investigate whether and
how inflation can generate a statistically anisotropic signal.

Traditionally, the dynamics of inflation and the generation of $\zeta$ have
been
studied using fundamental scalar fields only. Such a setup, however, cannot
produce a preferred direction on the microwave sky. The simplest way to do so
is to involve at least one vector field, which naturally selects a preferred
direction. Using a vector field to contribute to the curvature perturbation was
pioneered by Ref.~\cite{vecurv}, where  it was shown that a massive Abelian
vector field can successfully play the role of the curvaton. Since then,
substantial interest was ignited for the involvement of vector fields in
inflation and the generation of the curvature perturbation. The first
comprehensive study of how vector fields can contribute to $\zeta$ and generate
statistical anisotropy can be found in Ref.~\cite{stanis}.

Involving a vector field can give rise to statistically anisotropic $\zeta$ in
two ways. One possibility is that the contribution of the vector field to the
density during
inflation may result in non-negligible anisotropic stress. This can produce
statistically anisotropic perturbations of the inflaton scalar field which
can eventually give rise to statistical anisotropy in $\zeta$
\cite{anisinf0,anisinf,anisinf+}. In this case, the
perturbations of the vector field itself are unimportant (and can be ignored)
and the only requirement is that its contribution to the energy density remains
non-negligible throughout inflation, or at least during the period when the
cosmological scales exit the horizon.
The second possibility is orthogonal to the previous one, in that it assumes
a negligible anisotropic stress during inflation and it focuses instead on the
contribution to $\zeta$ of the vector field perturbations, which can give rise
to statistical anisotropy if the particle production process for the vector
field is itself anisotropic \cite{stanis,varkin}.
This is the generic expectation for a vector field
that undergoes particle production and obtains a superhorizon spectrum of
perturbations for some or all of its components \cite{stanis}.

To undergo particle production and obtain superhorizon spectra of perturbations
for its components, the vector field needs to have its conformal invariance
appropriately broken during inflation. Mechanisms for such a breaking of
conformality were originally developed in the effort to generate a coherent
primordial magnetic field during inflation (e.g. see Ref.~\cite{pmfrev,mine}
and references therein). One proposal suggested coupling
non-minimally an Abelian vector field to gravity through a term of the form
$\frac16RA^2$, where $R$ is the scalar curvature \cite{TW}. The model was shown
to generate scale-invariant spectra for all vector field components
(introducing
a small non-zero mass gives rise to a longitudinal component) and particle
production was anisotropic at the level of 100\%
\cite{stanis,nonmin}. Thus, if such a vector field generates
a subdominant contribution to $\zeta$, it can generate observable statistical
anisotropy. This model, however, may suffer from
instabilities such as ghosts \cite{peloso}
(see however Ref.~\cite{RA2save}).

To overcome the instability issue another model was put forward, which employs
a vector field with a varying kinetic function \cite{varkin}.
This possibility has been
massively explored for the generation of a primordial magnetic field
\cite{gaugekin}. In this
model, particle production of the transverse components of the vector field can
produce a scale-invariant superhorizon spectrum of perturbations if the kinetic
function of the vector field scales as \mbox{$f\propto a^{-1\pm 3}$} during
inflation (or at least when the cosmological scales exit the horizon), where
$a$ is the scale factor of the Universe \cite{varkin,sugravec}. If the
vector field has a non-zero mass $m$ it makes sense to talk about the
longitudinal component as well. To obtain a scale invariant spectrum for this
one, the additional requirement is \mbox{$m\propto a$} \cite{varkin}.
We also need the mass of the physical (in contrast to comoving) vector field
\mbox{$M\equiv m/\sqrt f$} to be small (i.e. \mbox{$M\ll H$}, with $H$ being
the Hubble parameter) when the cosmological scales exit the horizon. Under the
above conditions, if the vector field remains light until the end of inflation
then the power spectra of the
transverse and longitudinal components respectively are \cite{varkin}
\begin{equation}
{\cal P}_\perp=\left(\frac{H}{2\pi}\right)^2\quad{\rm and}\quad
{\cal P}_\|=\left(\frac{H}{2\pi}\right)^2\left(\frac{H}{3M}\right)^2,
\end{equation}
so that \mbox{${\cal P}_\perp\ll{\cal P}_\|$} and particle production is
strongly anisotropic. In contrast, if the vector field becomes heavy by the
end of inflation (i.e. \mbox{$M\gtrsim H$}) then the vector field begins
coherent oscillations around zero and the power spectra of the
transverse and longitudinal components obtain the average value \cite{varkin}
\begin{equation}
\overline{\cal P}_\perp=\overline{\cal P}_\|=
\frac12\left(\frac{H}{2\pi}\right)^2\left(\frac{H}{3M}\right)^2,
\end{equation}
which means that particle production becomes isotropic. This possibility
arises only if $M$ is growing during inflation, which corresponds to
\mbox{$f\propto a^{-4}$} (the other possibility, namely \mbox{$f\propto a^2$},
results in \mbox{$M=\,$constant}). One can translate the
above spectra of vector field perturbations into a contribution to $\zeta$
through a variety of mechanisms. In Ref.~\cite{varkin} the vector curvaton
mechanism of Ref.~\cite{vecurv} was used. In Ref.~\cite{soda} the end of
inflation
mechanism was used instead for the same vector field model with \mbox{$m=0$}.

One expects the modulation of $f$ and $m$ to be due to a degree of freedom
which varies during inflation. The most natural candidate for this is the
inflaton field itself. However, in this case, there is a coupling between the
vector field and the inflaton, which can backreact onto the inflaton's
dynamics. Such backreaction was first considered in Ref.~\cite{anisinf}
where it was shown that,
for a specific functional form of the kinetic function, a model of quadratic
chaotic inflation renders the scaling \mbox{$f\propto a^{-4}$} an attractor
solution. The role of the backreaction to vector field particle production in
quadratic chaotic inflation with a massless, Abelian gauge field was also investigated in Ref.~\cite{Kanno:2009ei} with the aim to explore the generation 
of a primordial magnetic field. The same attractor behaviour was found.
Furthermore, the backreaction on inflation from a massless, Abelian vector field whose kinetic function is modulated by the inflaton was studied in Ref.~\cite{Emami:2010rm}, where an attractor behaviour which results in weakly anisotropic inflation was also observed.

In this paper, we perform a stability analysis to obtain the
criteria under which the conditions \mbox{$f\propto a^{-4}$} and
\mbox{$m\propto a$}, which render the vector field perturbation spectra scale
invariant (and which at first glance may seem somewhat artificial),
become attractor solutions for a generic functional form of the dependence of
the kinetic function $f(\phi)$ and the scalar potential $V(\phi)$ on the
inflaton field $\phi$. We find that the conditions imposed on the model
parameters are such that the above scalings for $f$ and $m$ become attractor
solutions for 50\% of the parameter space (in the sense that for a given
model parameter the condition is one upper or lower bound only) and can be
attained for natural values. The criterion boils down to the requirement that
the backreaction onto the inflaton field dynamics should not be decreasing
during inflation, because were it so, the vector field could never affect the
inflationary dynamics and no attractor behaviour would arise. Thus, we show
that
the conditions which guarantee the generation of scale invariant spectra of
vector field perturbations (with amplitudes as shown above) can be naturally
attained if the kinetic function and mass of the vector field are modulated by
the inflaton during inflation.

Our findings also show that the attractor solution can generate a non-zero
anisotropic stress when the vector field is still light. This anisotropic
stress can be kept small (negligible) by choosing the model parameters
appropriately. In this case, the assumption of an approximately isotropic
inflationary expansion is justified and we can rely on the vector field
perturbations to produce statistical anisotropy as discussed in
Ref.~\cite{stanis,varkin}.
However, we can choose the model parameters such that the anisotropic stress
of the attractor solution is non-negligible in which case the inflationary
expansion itself would be anisotropic.
In this case, the vector field perturbations
can play a negligible role (and can be ignored) and we can produce the
statistical anisotropy from the anisotropic expansion itself as in
Refs.~\cite{anisinf,anisinf+}.
In our paper we do not make a choice between these two options, or indeed the
third possibility in which both the above two
sources of statistical anisotropy can be important. Instead we quantify the
anisotropic stress at the attractor solutions as well as demonstrate that the
scaling of $f$ and $m$ results to scale invariance.

An interesting byproduct of the vector field backreaction on
the inflationary dynamics is that it provides extra resistance in the roll of
the inflaton down its potential. This can be used to maintain slow-roll even if
the curvature of the scalar potential along the inflaton direction is
substantial. Thereby we can overcome the infamous $\eta$-problem of inflation
which arises by K\"{a}hler corrections to the scalar potential in supergravity
theories.

The structure of our paper is as follows.
In Sec.~\ref{Sec-EqofMotion}
we outline our model and obtain the equations of motion for all the fields
involved.
In Sec.~\ref{dyna}
we define a set of dimensionless quantities which we use to perform
a dynamical analysis of our model, without specifying the functional
dependence of $f(\phi)$, $m(\phi)$ and $V(\phi)$.
In Sec.~\ref{Sec-masslessVFcase}
we study the case of a massless vector field, in which only the transverse
components are physical. We obtain the standard slow-roll critical point,
which is an attractor for half of the parameter space where the
vector field backreaction does not grow during inflation. We also find another
critical point which is our vector scaling solution, which exists for the
remaining half of the parameter space, when the vector field backreaction
to the inflaton does affect the inflaton's roll, while also rendering the
standard slow-roll critical point unstable.
In Sec.~\ref{Sec-massiveVFcase} we consider a non-zero mass for the vector
field and investigate whether our vector scaling solution is spoiled or not.
Firstly, we study the case of a light vector field (\mbox{$M\ll H$}) where
we find that our vector scaling solution trifurcates into three critical points
with weak flows between them. The inflationary expansion at these points
can be anisotropic. Afterwards, we study the case of a heavy vector field
(\mbox{$M\gtrsim H$}) which is undergoing harmonic oscillations. Again we
find a vector scaling solution, which now corresponds to zero anisotropy.
In Sec.~\ref{summary} we display a summary of our results.
In Sec.~\ref{examples} we apply our findings to four different concrete
examples and we numerically verify our analytic results. The first example
assumes an exponential dependence of $f(\phi)$, $m(\phi)$ and $V(\phi)$
on the inflaton field, which could be motivated by string theory
considerations. %This case corresponds to constant model parameters.
The second example considers chaotic inflation and reproduces the findings of
Ref.~\cite{anisinf} in the massless field case. The third and fourth examples
are supergravity inspired and assume that the inflaton is a flat direction of
supersymmetry, lifted only by K\"{a}hler corrections to the scalar potential.
The vector field is then a Higgsed gauge field with $f$ being the gauge
kinetic function associated with the gauge coupling as \mbox{$f\sim 1/g^2$}.
In these examples we clearly demonstrate how the backreaction of the vector
field can overcome the $\eta$-problem of inflation in supergravity.
Finally, in Sec.~\ref{conclusions} we discuss our results and present our
conclusions.

While this paper was being written up Ref.~\cite{sodanew} appeared, which
performs a
similar stability analysis for the massless version of our model assuming an
exponential functional dependence for $f(\phi)$ and $V(\phi)$. As such it
corresponds to part of the first of our examples. However, the authors of
Ref.~\cite{sodanew}
have explored deeper the inflationary dynamics and were mostly
interested in the possibility of anisotropic inflationary expansion,
completely ignoring the perturbations of the vector field and what the vector
scaling solution implies for them. Our results agree with their findings
where there is overlap.

In our paper we consider natural units, where \mbox{$c=\hbar=k_B=1$} and
Newton's gravitational constant is \mbox{$8\pi G=m_P^{-2}$}, with $m_P$ being
the reduced Planck mass.

%---------------------------------------------------------------------------------------------------------------------
%---------------------------------------------------------------------------------------------------------------------
\section{Equations of Motion}\label{Sec-EqofMotion}

Consider the Lagrangian density for a massive Abelian vector field plus a scalar field together with Einstein gravity
\begin{equation}\label{Lagrangian}
{\cal L}=-\frac{m_{P}^{2}}{2}R
+\frac{1}{2}(\partial_{\mu}\phi)(\partial^{\mu}\phi)-V(\phi)
-\frac{1}{4}fF_{\mu\nu}F^{\mu\nu}+\frac{1}{2}m^{2}A_{\mu}A^{\mu},
\end{equation}
where $f$ is the vector field kinetic function, $m$ is the mass of the vector field, the field strength tensor is \mbox{$F_{\mu\nu}=\partial_{\mu}A_{\nu}-\partial_{\nu}A_{\mu}$}
and $V(\phi)$ is the scalar potential.
The above can be
the Lagrangian density of a massive Abelian gauge field, in which case $f$ is
the gauge kinetic function. However, we need not restrict ourselves to gauge
fields only. If no gauge symmetry is considered the argument in support of the
above Maxwell type kinetic term is that it is one of the few (three) choices
\cite{carroll} which avoids introducing instabilities, such as ghosts
\cite{peloso}. Also, we note here that a massive vector field which is
not a gauge field is renormalizable only if it is Abelian \cite{tikto}.

It was shown in Ref. \cite{vecurv} that, as inflation homogenises the vector field $\partial_{i}A_{\mu}=0$, the temporal component $A_{0}=0$.\footnote{If \mbox{$m=0$} then we can set \mbox{$A_{0}=0$} by a gauge choice.}
Therefore, without loss of generality, the spacial components of the vector field can be lined-up
in the $\textit{z}-\textrm{axis}$ $A_{\mu}=(0,0,0,A_{z}(t))$.
We can then assume a Bianchi-I background with a residual isotropy in the plane perpendicular to the vector field expectation value
\begin{equation}\label{metric}
\ud s^{2}=\ud t^{2}-e^{2\alpha(t)}\left[e^{2\sigma(t)}(\ud x^{2}+\ud y^{2})+e^{-4\sigma(t)}\ud z^{2}\right],
\end{equation}
where $a\equiv e^{\alpha}$ is the isotropic scale factor and $\sigma$ parameterises the amount of anisotropy. A constant $\sigma$ corresponds to a flat FRW Universe with the Hubble rate given by $H=\dot{\alpha}$. One can shift the values of $\alpha$ and $\sigma$ by a constant factor without changing the physics of the system.

In Ref. \cite{varkin} we showed that the scaling $f\propto a^{-1\pm 3}$ and
$m\propto a$ leads to scale invariant spectra of
superhorizon perturbations for the vector field components, assuming
approximately isotropic quasi-de Sitter inflation.
Such perturbations can then contribute to the curvature perturbation of the Universe $\zeta$, e.g. through the vector curvaton mechanism \cite{vecurv}. The modulation of the kinetic function $f$ and mass $m$ of the vector field comes from some degree of freedom which varies during inflation. The most natural choice for such a degree of freedom is of course the inflaton field itself, but other choices are also possible. In this paper, we consider the possibility that the kinetic function and mass of the vector field
are some function of the inflaton field $\phi$, i.e. $f=f\left(\phi\right)$ and $m=m\left(\phi\right)$. If this is so, then, under certain conditions, we demonstrate that the scaling solutions for the kinetic function and mass $f\propto a^{-4}$ and $m\propto a$, which give rise to scale invariant perturbation spectra, are in fact dynamical attractor solutions.

The Einstein equations derived from Eq. (\ref{Lagrangian}) are
\begin{eqnarray}
\dot{\alpha}^{2}-\dot{\sigma}^{2} & = & \frac{1}{3m_{P}^{2}}\left[\frac{1}{2}\dot{\phi}^{2}+V(\phi)-\frac{1}{2}g^{zz}f\dot{A}_{z}^{2}
-\frac{1}{2}g^{zz}m^{2}A_{z}^{2}\right],\label{fried}\\
\ddot{\alpha}+3\dot{\alpha}^{2} & = & \frac{1}{3m_{P}^{2}}\left[3V(\phi)-\frac{1}{2}g^{zz}f\dot{A}_{z}^{2}-g^{zz}m^{2}A_{z}^{2}\right],\\
\ddot{\sigma}+3\dot{\alpha}\dot{\sigma} & = & \frac{1}{3m_{P}^{2}}\left[-g^{zz}f\dot{A}_{z}^{2}+g^{zz}m^{2}A_{z}^{2}\right],\label{Einstein03}
\end{eqnarray}
where $g^{zz}=-e^{-2\alpha(t)+4\sigma(t)}$. The scalar field equation of motion is
\begin{equation}\label{phieq2}
\ddot{\phi}+3\dot{\alpha}\dot{\phi}+V'(\phi)+\mathcal{B}_{A}=0,
\end{equation}
where the prime denotes derivative with respect to $\phi$ and $\mathcal{B}_{A}$
is a source term due to the modulation of the kinetic function and mass of the
vector field from this scalar field.
We define this backreaction of the vector field to the scalar field dynamics as
 $\mathcal{B}_{A}=\mathcal{B}_{A-f}+\mathcal{B}_{A-m}$,
where
\begin{equation}
\mathcal{B}_{A-f}\equiv\frac{1}{2}g^{zz}f'\dot{A}_{z}^{2}\qquad\text{and}\qquad
\mathcal{B}_{A-m}\equiv-g^{zz}mm'A_{z}^{2},\label{baf-decomp}
\end{equation}
which we call the $f$-term and $m$-term backreaction respectively.

The equation of motion for the vector field is
\begin{equation}\label{veceq}
\ddot{A}_{z}+\left(\dot{\alpha}+4\dot{\sigma}+\frac{\dot{f}}{f}\right)\dot{A}_{z}
+\frac{m^{2}}{f}A_{z}=0,
\end{equation}
and the vector field energy density is given by $\rho_{A}\equiv\rho_{{\rm {kin}}}+V_{A}$, where
\begin{equation}
\rho_{{\rm {kin}}}\equiv-\frac{1}{2}g^{zz}f\dot{A}_{z}^{2}\qquad\text{and}\qquad
 V_{A}\equiv-\frac{1}{2}g^{zz}m^{2}A_{z}^{2}.\label{vecerg01}
\end{equation}
 The equation describing the acceleration of the Universe is
 \begin{equation}
\frac{\ddot{a}}{a}=\ddot{\alpha}+\dot{\alpha}^{2}=-2\dot{\sigma}^{2}+\frac{1}{3m_{P}^{2}}\left[-\dot{\phi}^{2}+V(\phi)-\rho_{{\rm {kin}}}\right].
\end{equation}
Notice how the vector field potential term $V_{A}$, has no influence
on the acceleration of the Universe, therefore only the growth of
the vector kinetic term can spoil an inflating Universe.

It will also be useful to define the following energy density and
backreaction ratios
\begin{equation}\label{Ratios}
\mathcal{R}\equiv\frac{\rho_{A}}{\rho_{\phi}}\qquad\text{and}\qquad\mathcal{R_{B}}\equiv\frac{\mathcal{B}_{A}}{V'(\phi)},
\end{equation}
where the energy density of the homogeneous scalar field
is given by the usual expression $\rho_{\phi}\equiv\frac{1}{2}\dot{\phi}^{2}+V(\phi)$.

The backreaction, Eq. (\ref{baf-decomp}) is dynamic with respect to the vector field, as it is
a function of $\dot{A}_{z}$, but not with respect to the scalar field, since
it is not a function of $\dot{\phi}$. Thus, the backreaction may be
interpreted as to only modify the effective slope of the potential,
$V'_{\textrm{eff}}\equiv V'+\mathcal{B}_A$ seen by the inflaton.

If we consider gauge symmetry then $f$ is the gauge
kinetic function where $f\sim1/g^{2}$ and $g$ is the gauge coupling.
Consequently, because we assume canonical normalisation
$f\rightarrow1$ at the end of inflation,
we require that the kinetic function is
always decreasing in time, $\dot{f}(t)<0$ so that the vector field
remains weakly coupled. Note, however, that the vector field does not need to be a gauge field necessarily.

Successful particle production of vector fields during inflation requires that the physical field is effectively massless at horizon exit, $m/\sqrt{f}\ll H$. However,
%for a vector field contribution to the curvature perturbation, employing the curvaton mechanism, we require that
the vector field can become heavy at some point after horizon exit. In this case, particle production can be isotropic \cite{varkin}. A heavy vector field oscillates and acts like a pressureless isotropic fluid which
does not cause anisotropic stress \cite{vecurv}. Therefore, to consider the effects of a non-zero mass for the vector field, we also require the mass to be increasing in time, $\dot{m}(t)>0$, otherwise it will always be negligible.

Due to the above, we notice that the backreaction always has an opposite sign
to the potential slope. This is because the scalar field rolls down its
potential, so if $V'(\phi)>0$ then $\dot{\phi}<0$ therefore $f'(\phi)>0$
and $m'(\phi)<0$, however if $V'(\phi)<0$ then $\dot{\phi}>0$ therefore
$f'(\phi)<0$ and $m'(\phi)>0$. This ensures that $\mathcal{B}_{A}$
always has an opposite sign to the potential slope, $V'(\phi)$. We conclude that the backreaction will always reduce the effective
potential slope experienced by the inflaton and slow down the scalar
field as it rolls down its potential. The effective potential slope as seen by the scalar field is then given by
\begin{equation}\label{Veff01}
V'_{\textrm{eff}}=V'+\mathcal{B}_A=V'(1+\mathcal{R_{B}}).
\end{equation}

%---------------------------------------------------------------------------------------------------------------------
%---------------------------------------------------------------------------------------------------------------------

\section{Dynamical Analysis}\label{dyna}

The Einstein equations Eqs. (\ref{fried})-(\ref{Einstein03}) together with the scalar and vector field equations Eqs. (\ref{phieq2}) and (\ref{veceq}) form a system of coupled non-linear differential equations. Analytical solutions to this system are extremely difficult if not impossible to obtain. However we may understand the
qualitative behaviour of the solutions by analysing the phase-space of the system. Our aim is to identify attractor solutions which
may result from the effect of the vector field backreaction on the scalar field dynamics.

Let us first define a new set of dimensionless, expansion normalised variables
\begin{equation}
\Sigma\equiv\frac{\dot{\sigma}}{\dot{\alpha}}\:\text{,}\quad x\equiv\frac{\dot{\phi}}{\sqrt{6}m_{P}\dot{\alpha}}\:\text{,}\quad y\equiv\frac{\sqrt{V(\phi)}}{\sqrt{3}m_{P}\dot{\alpha}}\:\text{,}\quad z\equiv\frac{\sqrt{\rho_{\textrm{kin}}}}{\sqrt{3}m_{P}\dot{\alpha}}\:\text{,}\quad s\equiv\frac{\sqrt{V_{A}}}{\sqrt{3}m_{P}\dot{\alpha}}\quad\text{and}\quad\mu\equiv\frac{m}{\dot{\alpha}\sqrt{f}}\:.
\end{equation}
The Friedmann constraint, Eq. (\ref{fried}) then becomes
\begin{equation}
\Sigma^{2}+x^{2}+y^{2}+z^{2}+s^{2}=1.
\end{equation}
We can write the Einstein equations and the scalar and vector
field equations as a system of first order coupled non-linear differential
equations
\begin{eqnarray}
\frac{\ud\Sigma}{\ud\alpha} & = & -3\Sigma+2z^{2}-2s^{2}+\Sigma\left(3\Sigma^{2}+3x^{2}+2z^{2}+s^{2}\right),\label{fullsystem} \\
\frac{\ud x}{\ud\alpha} & = & -3x-\lambda y^{2}+\Gamma z^{2}-\Xi s^{2}+x\left(3\Sigma^{2}+3x^{2}+2z^{2}+s^{2}\right), \\
\frac{\ud y}{\ud\alpha} & = & \lambda xy+y\left(3\Sigma^{2}+3x^{2}+2z^{2}+s^{2}\right),\\
\frac{\ud z}{\ud\alpha} & = & -2z-2z\Sigma-\Gamma zx-\mu s+z\left(3\Sigma^{2}+3x^{2}+2z^{2}+s^{2}\right), \\
\frac{\ud s}{\ud\alpha} & = & -s+2s\Sigma+\Xi sx+\mu z+s\left(3\Sigma^{2}+3x^{2}+2z^{2}+s^{2}\right), \\
\frac{\ud\mu}{\ud\alpha} & = & \left[\left(\Xi-\Gamma\right)x+\left(3\Sigma^{2}+3x^{2}+2z^{2}
+s^{2}\right)\right]\mu.\label{fullsystem-m} \end{eqnarray}
Where we have defined the dimensionless model parameters, which are the only model-dependent terms in the system above, as
\begin{equation}\label{model_param}
\lambda(\alpha)\equiv\sqrt{\frac{3}{2}}m_{P}\left(\frac{V'}{V}\right)\:\text{,}\qquad
\Gamma(\alpha)\equiv\sqrt{\frac{3}{2}}m_{P}\left(\frac{f'}{f}\right)\qquad\text{and}\qquad
\Xi(\alpha)\equiv\sqrt{6}m_{P}\left(\frac{m'}{m}\right).
\end{equation}
Following the arguments at the end of Sec. \ref{Sec-EqofMotion}, we see that
$\lambda$ and $\Gamma$ have the same sign but opposite
to $x$ and $\Xi$. For definiteness we can choose $\lambda$
and $\Gamma$ to be positive and $x$ and $\Xi$ to be
negative without loss of generality.

It will also be useful to use the slow-roll parameters defined by
\begin{equation}\label{slow-roll-param}
\epsilon(\phi)\equiv\frac{m_{P}^{2}}{2}\left(\frac{V'}{V}\right)^{2}=\frac{1}{3}\lambda^{2}\qquad\text{and}\qquad\eta(\phi)\equiv m_{P}^{2}\left(\frac{V''}{V}\right)=\sqrt{\frac23}m_P\lambda'+2\epsilon.
\end{equation}
However because of the backreaction affecting the scalar potential
slope $V'(\phi)$, it will also be useful to define the slow-roll
parameters in the Hamilton-Jacobi formalism
\begin{equation}\label{slow-roll-paraHJ}
\epsilon_{\textrm{H}}\equiv-\frac{\dot{H}}{H^{2}}
=-\frac{\ddot{\alpha}}{\dot{\alpha}^{2}}
=3\Sigma^{2}+3x^{2}+2z^{2}+s^{2}\qquad\text{and}\qquad
\eta_{\textrm{H}}\equiv-\frac{\ddot{H}}{2H\dot{H}}
=\epsilon_{\textrm{H}}-\frac{1}{2\dot{\alpha}}\frac{\dot{\epsilon_{\textrm{H}}}}{\epsilon_{\textrm{H}}}.
\end{equation}

In the framework of our expansion normalised variables, we find that
\begin{equation}\label{scalingSol1}
\frac{1}{f}\frac{\ud f}{\ud \alpha}=2x\Gamma\qquad\text{and}\qquad
\frac{1}{m}\frac{\ud m}{\ud \alpha}=x\Xi.
\end{equation}
The scaling solutions required to generate a scale-invariant vector field spectrum of perturbations, $f\propto a^{-4}$ and $m\propto a$ are therefore respectively given by the solutions
\begin{equation}\label{scalingSol}
x=-\frac{2}{\Gamma}\qquad\text{and}\qquad x=\frac{1}{\Xi}.
\end{equation}

The full system above, Eqs. (\ref{fullsystem})-(\ref{fullsystem-m}), is 5-dimensional and the phase-space analysis becomes highly
complicated. In 2 dimensions, from an analytic point of view one can
say a lot more about a system of differential equations. For systems
of higher dimension one cannot do much more than analyse the stationary
points and their stability, see Ref.~\cite{WainwrightEllis}. The stationary points (or critical points)
of the system are defined by $\frac{\ud x}{\ud\alpha}=\frac{\ud y}{\ud\alpha}=...=0$,
and their stability is established by analysing the behaviour of the
solutions close to the critical points.

%---------------------------------------------------------------------------------------------------------------------
%---------------------------------------------------------------------------------------------------------------------
\subsection{Stability}

To investigate the stability of the critical point solutions we will
consider perturbations about the critical points
\begin{equation}
x=x_{c}+u\,,\qquad y=y_{c}+v\,,\qquad\Sigma=\Sigma_{c}+w\,,\quad\cdots
\end{equation}
 and substituting into the system of equations we obtain the linearized
system
\begin{equation}
\frac{\ud}{\ud\alpha}\left(\begin{array}{c}
u\\
v\\
w\\
\vdots\end{array}\right)=\mathcal{M}\left(\begin{array}{c}
u\\
v\\
w\\
\vdots\end{array}\right),
\end{equation}
where the matrix $\mathcal{M}=\mathcal{M}(x_{c},y_{c},\Sigma_{c},\cdots)$.
The general solution for the evolution of linear perturbations can
be written as
\begin{eqnarray}
u(\alpha) & = & u_{1}e^{m_{1}\alpha}+u_{2}e^{m_{2}\alpha}+u_{3}e^{m_{3}\alpha}+\cdots\nonumber\\
v(\alpha) & = & v_{1}e^{m_{1}\alpha}+v_{2}e^{m_{2}\alpha}+v_{3}e^{m_{3}\alpha}+\cdots\label{linearSol}\\
w(\alpha) & = & w_{1}e^{m_{1}\alpha}+w_{2}e^{m_{2}\alpha}+w_{3}e^{m_{3}\alpha}+\cdots\nonumber\\
\vdots\quad & = & \quad\vdots\nonumber
\end{eqnarray}
where $u_{i},\, v_{i},\, w_{i}$ are constants of integration and
the index $i$ gives the dimension of the sub-system under consideration.

The critical point is considered to be perturbatively stable when
the real parts of all the eigenvalues $m_{i}$ of the matrix $\mathcal{M}$
are negative. If $\text{Im}(m_{i})\neq0$ then the point is a stable
spiral, otherwise it is a node. Generally the eigenvalues will be
a mixture of positive and negative numbers, in this case the critical
point is an unstable saddle point. If however the real parts of one
or more of the eigenvalues is zero then the point is non-hyperbolic
whose stability cannot be established using the linearisation procedure
above, more sophisticated methods such as the center manifold theorem
has to be used. In this paper we only consider hyperbolic critical
points.

Now it is important to note that we may write down these solutions to the linearised system Eq. (\ref{linearSol}), if and only if the eigenvalues $m_i$, which will generally depend on the model parameters Eqs. (\ref{model_param}), are {\em constant or slowly varying} compared to the Hubble scale $H$. This may be achieved if the model parameters are constant or slowly varying themselves, or if the eigenvalues depend weakly on the model parameters. The method to establish the stability of the critical points relies on this assumption, and therefore the validity of this assumption has to be verified once we have obtained the eigenvalues of the matrix $\mathcal{M}$.

In general, we find that the local dynamics of the critical points depend on the dimensionless model parameters, $\lambda$, $\Gamma$ and $\Xi$. When small continuous changes in these model parameters result in dramatic changes in the dynamics, the critical point is said to undergo a bifurcation. The values of the parameters which result in a bifurcation at the critical point can often be located by examining the linearised system. The bifurcations are located at the parameter values for which the real part of an eigenvalue is zero. We will identify the bifurcations in our system as these may correspond to interesting changes in the cosmological scenarios described by the system.

%---------------------------------------------------------------------------------------------------------------------
%---------------------------------------------------------------------------------------------------------------------
\section{The Massless Vector Field Case}\label{Sec-masslessVFcase}

There has been recent wide interest in the cosmological implications of a massless $U(1)$ field whose kinetic function is modulated by the inflaton
\cite{anisinf,anisinf+}. Authors in Ref. \cite{anisinf} have shown that for a certain kinetic functional form and potential an attractor solution exist which corresponds to a slow-roll phase of inflation that maintains prolonged anisotropy. We will show however using dynamical analysis that these scaling solutions can be obtained for a wide class of scalar potentials and kinetic functions.

Consider a massless vector field, $s,\mu,\Xi=0$. The full system of equations Eqs. (\ref{fullsystem})-(\ref{fullsystem-m}), is then reduced to the 3-dimensional sub-system
\begin{eqnarray}
\frac{\ud\Sigma}{\ud\alpha} & = & -3\Sigma+2\left(1-\Sigma^{2}-x^{2}-y^{2}\right)+\Sigma\left[3\Sigma^{2}+3x^{2}
+2\left(1-\Sigma^{2}-x^{2}-y^{2}\right)\right],\label{System-m0} \\
\frac{\ud x}{\ud\alpha} & = & -3x-\lambda y^{2}+\Gamma\left(1-\Sigma^{2}-x^{2}-y^{2}\right)+x\left[3\Sigma^{2}+3x^{2}+2\left(1-\Sigma^{2}-x^{2}-y^{2}\right)\right],\\
\frac{\ud y}{\ud\alpha} & = & \lambda xy+y\left[3\Sigma^{2}+3x^{2}+2\left(1-\Sigma^{2}-x^{2}-y^{2}\right)\right],\label{system-m0-y}
\end{eqnarray}
together with the Friedmann constraint $\Sigma^{2}+x^{2}+y^{2}+z^{2}=1$. The critical points $(x_{c},y_{c},z_{c},\Sigma_{c})$ are found by
setting $\frac{\ud\Sigma}{\ud\alpha}=\frac{\ud x}{\ud\alpha}=\frac{\ud y}{\ud\alpha}=0$.
There are 3 critical points for this system, as shown below.

%---------------------------------------------------------------------------------------------------------------------
%---------------------------------------------------------------------------------------------------------------------
\subsubsection{The Standard Slow-Roll Solution, $\mathcal{SSR}$.}

The first critical point to be considered
is
\begin{equation}\label{SSR-m0}
\mathcal{SSR}:\qquad(x_{c},y_{c},z_{c},\Sigma_{c})=
\left(-\frac{\lambda}{3},\,\sqrt{1-\left(\frac{\lambda}{3}\right)^{2}},\,0,\,0\right),
\end{equation}
where the eigenvalues of the matrix $\mathcal{M}$ for this solution
are
\begin{equation}\label{SSR-eigen}
m_{1,2}=-3+\frac{\lambda^{2}}{3}\quad\text{and}\quad m_{3}=-4+\frac{2}{3}\lambda\left(\lambda+\Gamma\right).
\end{equation}
Hence the existence and stability of this solution requires
\begin{equation}
\lambda<3\qquad\text{and}\qquad\lambda\left(\lambda+\Gamma\right)<6.\label{SSR-cond}
\end{equation}
This critical point becomes the inflationary standard slow-roll solution
when $\lambda\ll3$ or equivalently $\epsilon\ll3$. The scalar field
slowly rolls down its potential which dominates the energy density
of the Universe $y_c\approx1$. We find that the vector field energy density vanishes
at this critical point $z_c=0$, and consequently so does the anisotropy $\Sigma_c=0$. These are completely diluted by the expansion of space and provides an example of the cosmic no-hair theorem \cite{wald}. The slow-roll parameters Eqs. (\ref{slow-roll-param}) and (\ref{slow-roll-paraHJ}) are now
\begin{eqnarray}\label{SSR-mo-slwParam}
\epsilon_{\textrm{H}} & =\epsilon= & \frac{1}{3}\lambda^{2}\qquad\text{and}\qquad\eta_{\textrm{H}}=\eta-\epsilon,
\end{eqnarray}
and the scalar equation of state is
\begin{equation}\label{SSR-mo-eqofstate}
\gamma_{\phi}\equiv1+w_{\phi}=\frac{2x^{2}}{x^{2}+y^{2}}=\frac{2}{9}\lambda^{2}=\frac{2}{3}\epsilon_{\textrm{H}}.
\end{equation}

The method to establish the stability of the system relies on the fact that the eigenvalues are constant or slowly varying. This may be achieved if the model parameters are constant or slowly varying, but this may also be achieved in the limits $\lambda\ll3$ and $\lambda\left(\lambda+\Gamma\right)\ll6$ where the eigenvalues depend weakly on the model parameters. This therefore ensures the validity of our method in establishing the stability of the critical points. These limits would also guarantee the existence and stability of this critical point.

For slowly varying model parameters we can identify the bifurcations. The bifurcation value $\lambda=3$ determines the boundary between the existence and non-existence of the critical point. And passing through the bifurcation value $\lambda(\lambda+\Gamma)=6$ the stability of the critical point changes. We will see what happens to the other critical points at these bifurcations in the following sections.

%---------------------------------------------------------------------------------------------------------------------
%---------------------------------------------------------------------------------------------------------------------
\subsubsection{The Anisotropic Kination Solution, $\mathcal{AKS}$.}\label{Sec-AKS}

The second critical point of this system is
\begin{equation}\label{AKS}
\mathcal{AKS}:\qquad(x_{c},y_{c},z_{c},\Sigma_{c})=
\left(x,\,0,\,0,\,\sqrt{1-x^{2}}\right).
\end{equation}
This point turns out to be a non-hyperbolic critical point as one
of the eigenvalues of the matrix $\mathcal{M}$ is zero. Therefore
we cannot establish its stability with the linearised system. In any
case we find that this critical point does not correspond to a cosmology
of interest. This is because the scalar potential vanishes at the point $y_c=0$
and therefore cannot correspond to an inflating Universe. We also
notice that for the anisotropy $\Sigma_c$ to be small we require that
$x_c\approx1$, and therefore the Universe is dominated by the scalar
kinetic energy. Otherwise we would have a strongly anisotropic Universe, contradicting
observations. Therefore we will not pursue the analysis of this critical
point any further.

%---------------------------------------------------------------------------------------------------------------------
%---------------------------------------------------------------------------------------------------------------------
\subsubsection{The Vector Scaling Solution, $\mathcal{VSS}$.}\label{Sec-m0-VSS}

The vector scaling critical point is given by
\begin{equation}\label{VSS-m0}
\mathcal{VSS}:\quad(x_{c},y_{c},z_{c},\Sigma_{c})=
\left(-6\Delta_{1}\left(\lambda+\Gamma\right)\,,\:
\Delta_{1}\Delta_{2}\sqrt{\Gamma^{2}+\lambda\Gamma+6}\,,\:
\Delta_{1}\Delta_{2}\sqrt{\lambda^{2}+\lambda\Gamma-6}\,,\:
2\Delta_{1}\left(\lambda^{2}+\lambda\Gamma-6\right)\right),
\end{equation}
\[
\text{where}\quad\Delta_{1}\equiv\left(3\Gamma^{2}+4\Gamma\lambda+\lambda^{2}+12\right)^{-1}\quad\text{and}\quad
\Delta_{2}\equiv\sqrt{3\left(3\Gamma^{2}+2\lambda\Gamma-\lambda^{2}+12\right)}.
\]

Calculating the eigenvalues of the matrix $\mathcal{M}$ for the vector
scaling solution becomes highly complicated, however we will see that
we are interested in the special case where $\Gamma\gg\lambda$ and
$\Gamma\gg1$. In these limits the critical point may be approximated by
\begin{equation}\label{VSS-m0-cp}
\mathcal{VSS}:\qquad(x_{c},y_{c},z_{c},\Sigma_{c})=\left(-\frac{2}{\Gamma}\,,\:1\,,\:\frac{\sqrt{\lambda\Gamma-6}}{\Gamma}\,,\:
\frac{2\left(\lambda\Gamma-6\right)}{3\,\Gamma^{2}}\right).
\end{equation}

Plots are shown in Appendix (\ref{App-VSS-m0}) of the functional dependence of the real parts of the eigenvalues on the model parameters. We can clearly see that in the limits $\Gamma\gg\lambda$ and $\Gamma\gg1$ the real parts
of the eigenvalues are approximated by
\begin{equation}\label{VSS-m0-Remi}
\textrm{Re}[m_{1}]\simeq-3\qquad\text{and}\qquad\textrm{Re}[m_{2,3}]\simeq-\frac{3}{2}.
\end{equation}
The real parts of the eigenvalues are negative, therefore the critical point is stable i.e. a late-time attractor. In the limits considered the eigenvalues depend weakly on the model parameters. This can be seen as plateaus in the plots of Appendix (\ref{App-VSS-m0}). This therefore ensures the validity of our method in establishing the stability of the critical points.

The existence and stability of this solution requires
\begin{equation}
\lambda\Gamma>6\:,\qquad\Gamma\gg\lambda\qquad\text{and}\qquad\Gamma\gg1,\label{VSS-cond}
\end{equation}
which are sufficient but not necessary conditions. We can see that
the energy density is dominated by the scalar potential $y_c\approx1$ with the scalar field in slow-roll, therefore we have an inflating Universe. We note
that as $\Gamma\rightarrow\infty$ the Universe becomes de-Sitter.
The energy density ratio in Eq. (\ref{Ratios}) is given by
\begin{equation}\label{Ratios-VSS-m0}
\mathcal{R}=\frac{z^{2}}{x^{2}+y^{2}}\simeq\frac{\lambda\Gamma-6}{\Gamma^{2}}\ll1.
\end{equation}
The energy density of the vector field tracks that of the scalar field if the dimensionless model parameters are constant. Otherwise the energy density ratio is varying, with time-dependence determined by that of the dimensionless model parameters. Under the limits considered, the vector field energy density contribution is kept subdominant. The backreaction, from Eqs. (\ref{Ratios}) and (\ref{Veff01}) leads to the following effective scalar slope
\begin{equation}\label{VSS-mo-RB}
\mathcal{R_{B}}=-\frac{\Gamma}{\lambda}
\left(\frac{z}{y}\right)^{2}\simeq\frac{6-\lambda\Gamma}{\lambda\Gamma}\qquad\text{and}\qquad
V'_{\textrm{eff}}=\frac{6}{\lambda\Gamma}V'.
\end{equation}
The existence condition $\lambda\Gamma>6$ from Eq. (\ref{VSS-cond}) means that the effective slope seen by the scalar field becomes flatter, and so the field slows down as it evolves along its potential. This effect is quantified by the slow-roll parameters
\begin{equation}\label{VSS-m0-slowrollparam}
\epsilon_{\textrm{H}}\simeq\frac{2\lambda}{\Gamma}\ll1\qquad\text{and}\qquad
\eta_{\textrm{H}}\simeq\frac{2\lambda}{\Gamma}
+\frac{\sqrt{6}m_{P}}{\Gamma}\left(\frac{\lambda'}{\lambda}-\frac{\Gamma'}{\Gamma}\right).
\end{equation}
Comparing to the $\mathcal{SSR}$ slow-roll condition Eq. (\ref{SSR-mo-slwParam}), we find
$\epsilon_{\textrm{H}}(\mathcal{VSS})\simeq \frac{2\lambda}{\Gamma}\epsilon_{\textrm{H}}(\mathcal{SSR})$. The slow-roll parameter is therefore reduced in the vector scaling solution.

The anisotropy in the expansion, $\Sigma$ is given by
\begin{equation}\label{VSS-mo-sigma}
\Sigma_{c}\simeq\frac{2}{3}\left(\frac{\lambda\Gamma-6}{\Gamma^{2}}\right)=\frac{2}{3}\, z_{c}^{2}=\frac{2}{3}\mathcal{R}\ll1.
\end{equation}
The anisotropy is proportional to the energy density ratio and therefore evolves in the same way. This is a generalisation of the result obtained in
Ref.~\cite{anisinf} for any potential and kinetic function that satisfy Eq. (\ref{VSS-cond}). We can see that this vector scaling solution is a new slow-roll inflationary stage which supports small but prolonged anisotropy which would otherwise be completely diluted in the standard slow-roll regime. This prolonged
anisotropy may be useful in generating statistical anisotropy in the
curvature perturbation as described in Ref. \cite{anisinf,anisinf+}.
The non-vanishing anisotropy provides a counter-example to the cosmic no-hair conjecture \cite{wald}.

We can now see how the kinetic function behaves at the critical point. As we have obtained the attractor solution in Eq. (\ref{scalingSol}), the kinetic function scales as
\begin{equation}
f_{\rm att}(\alpha)\propto e^{-4\alpha}.
\end{equation}
The scalar equation of state is
\begin{equation}\label{VSS-mo-eqofstate}
\gamma_{\phi}\equiv1+w_{\phi}=\frac{2x^{2}}{x^{2}+y^{2}}\simeq\frac{8}{\Gamma^{2}}\ll1.
\end{equation}
We notice how the equation of state $\gamma_{\phi}>0$ and therefore
we are not dealing with a phantom field which would otherwise violate
the strong energy condition.

We also note that in the $\mathcal{VSS}$ parameter space where $\lambda\Gamma>6$ and $\Gamma\gg\lambda$ one of the eigenvalues of the $\mathcal{SSR}$, namely $m_3$ of Eq. (\ref{SSR-eigen}) is positive. Therefore the $\mathcal{SSR}$ is an unstable saddle
point as long as the condition $\lambda<3$ is also satisfied, otherwise the $\mathcal{SSR}$ would not exist. This really tells us that the $\mathcal{SSR}$ is unstable because the backreaction is growing and would eventually effect the dynamics of the system. If however the backreaction is subdominant at some early time, the standard slow-roll attractor could be reached prior to the vector scaling solution. This transient period of standard slow-roll inflation could last for a large number of e-folds depending on how subdominant the backreaction is. The $\mathcal{VSS}$ is however the true late-time attractor.

For the $\mathcal{VSS}$ it is more difficult to identify the bifurcations because the eigenvalues are so complicated. However we notice that the bifurcation value $\lambda(\lambda+\Gamma)=6$, observed at the $\mathcal{SSR}$, determines the boundary between the existence and non-existence of the $\mathcal{VSS}$. We also notice that the stability of the $\mathcal{SSR}$ and $\mathcal{VSS}$ critical points are interchanged at the bifurcation. As we are considering the limit $\Gamma\gg\lambda$ for the $\mathcal{VSS}$, then the bifurcation of interest, through which the dynamics of the system changes dramatically is $\lambda\Gamma=6$.

%---------------------------------------------------------------------------------------------------------------------
%---------------------------------------------------------------------------------------------------------------------
\subsection{Time Dependence of the Dimensionless Model Parameters}\label{Sec-modParam}

One important issue that we must deal with is the fact that generally
the model parameters, $\left(\lambda,\Gamma,\Xi\right)$ are time-dependent
$\left(\lambda(\alpha),\Gamma(\alpha),\Xi(\alpha)\right)$ and therefore
the critical points derived above are also time-dependent. The trivial
case corresponds of course to constant model parameters, where the
critical points are the true late time solutions. Otherwise they are
known as the instantaneous critical points. If the solutions approach
the attractors faster than the evolution of the critical points themselves,
then it is reasonable to assume that the attractor solutions are
reached and solutions are dragged along with the critical point. Otherwise the late time solutions are determined by
the asymptotic behaviour of the dimensionless model parameters. In the vicinity
of the stationary points the approach to the attractors is determined by
the eigenvalues of the matrix $\mathcal{M}$. For the critical solutions to be reached, their time-dependence must be smaller than the eigenvalue of smallest magnitude. We may write this as
\begin{equation}
\left|\frac{1}{F_{c}}\frac{\ud F_{c}}{\ud\alpha}\right|=\sqrt{6}m_{P}\left|x_{c}\frac{F_{c}'}{F_{c}}\right|<\left|m_{1,2,3}\right|,\label{cptimedepgen2}
\end{equation}
where $F_{c}=\left(x_{c},y_{c},z_{c},s_{c},\Sigma_{c}\right)$. Therefore
we need to establish which variable has the largest time variation
and then which eigenvalue has the smallest magnitude.

%---------------------------------------------------------------------------------------------------------------------
%---------------------------------------------------------------------------------------------------------------------
\subsubsection{The Standard Slow-Roll Solution, $\mathcal{SSR}$.}

For the $\mathcal{SSR}$ we see from Eq. (\ref{SSR-m0}) that $F_{c}=\left(x_{c},y_{c}\right)$,
therefore we have to establish which of these two variables has the
largest time variation. We find that Eq. (\ref{SSR-m0}) leads
to
\begin{equation}\label{cp-motion-m0-SSR}
\left|\frac{1}{x_{c}}\frac{\ud x_{c}}{\ud\alpha}\right|=\sqrt{\frac{2}{3}}m_{P}\left|\lambda'\right|\quad\text{and}\quad
\left|\frac{1}{y_{c}}\frac{\ud y_{c}}{\ud\alpha}\right|\simeq\sqrt{\frac{2}{3}}m_{P}\left|\lambda'\right|\left(\frac{\lambda}{3}\right)^{2}.
\end{equation}
The existence condition in Eq. (\ref{SSR-cond}) therefore tells us that the $x_{c}$ variable has the largest time dependence.

Next we need to establish which eigenvalue has the smallest magnitude. If the dimensionless model parameters are very slowly varying,
then from the existence and stability conditions Eq. (\ref{SSR-cond}), we can see that in the regime $\Gamma\ll\lambda$,
and considering slow-roll $\epsilon\ll1$, that $|m_{1,2}|<|m_{3}|$. If
we consider the regime $\Gamma\gg\lambda$ then we find two possibilities,
first if $\frac{3}{2}<\lambda\Gamma<6$ then $|m_{1,2}|>|m_{3}|$ otherwise
if $\lambda\Gamma<\frac{3}{2}$ then $|m_{1,2}|<|m_{3}|$. For the standard
slow-roll solution, Eq. (\ref{cptimedepgen2}) with $F_{c}=x_{c}$
leads to the conditions
\begin{eqnarray}\label{SSR-mov-cond}
\mathcal{SSR}:\qquad\sqrt{\frac{2}{3}}m_{P}\left|\lambda'\right|
=\left|2\epsilon-\eta\right| & <
\begin{cases}
\left|m_{1,2}\right| & =\left|-3+\epsilon\right|\\
\left|m_{3}\right| & =\left|-4+2\epsilon+\frac{2}{3}\lambda\Gamma\right|.
\end{cases}
\end{eqnarray}
The slow-roll conditions $\epsilon\ll1$ and $\left|\eta\right|\ll1$
guarantee that the first inequality above is satisfied. Then all we require is that $|\frac{2}{3}\lambda\Gamma-4|>\mathcal{O}(\eta)$, to satisfy the second inequality. This can be readily satisfied as long as we are not too close to the bifurcation value $\lambda\Gamma\not\approx 6$. In fact if $|\frac{2}{3}\lambda\Gamma-4|\gtrsim\mathcal{O}(1)$, then both inequalities can be met even if $\left|\eta\right|\sim1$, violating this slow-roll condition. In other words, the $\mathcal{SSR}$ can also apply to fast-roll inflation, where only one of the slow-roll conditions $\epsilon\ll1$ is met.

If the dimensionless model parameters are varying rapidly, then the validity of our
method in establishing the stability of the critical points requires the strong conditions $\lambda\ll3$ and $\lambda\Gamma\ll6$. These conditions ensures that the eigenvalues depend weakly on the dimensionless model parameters.
In this case the eigenvalues of smallest magnitude are clearly $|m_{1,2}|\simeq3$. Therefore the inequalities above can be readily satisfied even for fast-roll
inflation where $\left|\eta\right|\sim1$.
%---------------------------------------------------------------------------------------------------------------------
%---------------------------------------------------------------------------------------------------------------------
\subsubsection{The Vector Scaling Solution, $\mathcal{VSS}$.}

For the $\mathcal{VSS}$ we see from Eq. (\ref{VSS-m0}) that $F_{c}=\left(x_{c},y_{c},z_{c},\Sigma_{c}\right)$,
therefore we have to establish which of these four variables has the
largest time variation. In the limits $\Gamma\gg\lambda$ and $\Gamma\gg1$
we find that Eq. (\ref{cptimedepgen2}) leads to
\begin{equation}\nonumber
\left|\frac{1}{x_{c}}\frac{\ud x_{c}}{\ud\alpha}\right|=2\sqrt{6}m_{P}\left|\frac{\Gamma'}{\Gamma^{2}}
+\frac{1}{3}\frac{\lambda'}{\Gamma^{2}}\right|\:,\qquad
\left|\frac{1}{y_{c}}\frac{\ud y_{c}}{\ud\alpha}\right|=\sqrt{6}m_{P}\left|\frac{\lambda'}{\Gamma^{2}}\right|\:,
\end{equation}
\begin{equation}\label{cp-motion-m0-VSS}
\left|\frac{1}{z_{c}}\frac{\ud z_{c}}{\ud\alpha}\right|=
2\sqrt{6}m_{P}\left|\frac{\Gamma'}{\Gamma^{2}}+\frac{\lambda'}{\Gamma^{2}}
-\frac{\lambda'\Gamma+\lambda\Gamma'}{\Gamma\left(\lambda\Gamma-6\right)}\right|\quad\text{and}\quad
\left|\frac{1}{\Sigma_{c}}\frac{\ud\Sigma_{c}}{\ud\alpha}\right|\simeq2\sqrt{6}m_{P}\left|
-\frac{2}{3}\frac{\lambda'}{\Gamma^{2}}+\left(\frac{\Gamma'}{\Gamma^{2}}\right)\left(\frac{2}{3}\frac{\lambda}{\Gamma}
-\frac{8}{\Gamma^{2}}\right)\right|.
\end{equation}
Therefore the eigenvalue of smallest magnitude, which from Eq. (\ref{VSS-m0-Remi})
is $\left|m_{2,3}\right|\simeq\frac{3}{2}$ in the limits considered, has to be larger than all of the expressions above, $\left|\frac{1}{F_{c}}\frac{\ud F_{c}}{\ud\alpha}\right|<\mathcal{O}(1)$ (or the strong bound $\ll1$) for the attractor to be reached. Now if we consider a relatively flat potential $\sqrt{\frac23}m_P|\lambda'|=|\eta-2\epsilon|\lesssim\mathcal{O}(1)$ and because $\Gamma\gg1$ then clearly $m_P\left|\frac{\lambda'}{\Gamma^2}\right|\ll1$. Then if
\begin{equation}
\lambda\Gamma\not\approx6\quad{\rm or}\quad
\lambda'\Gamma\approx-\lambda\Gamma',
\label{cond}
\end{equation}
we find that all the expressions above can be reduced to a single (strong) condition
\begin{equation}\label{VSS-param_cond}
m_P\left|\frac{\Gamma'}{\Gamma^2}\right|=\left|\frac{ff''}{f'^2}-1\right|\ll1.
\end{equation}
As expected, if the kinetic function is exponential $f\propto e^{c\phi/m_P}$,
where $c$ is some constant, then the expression above is exactly zero. This is because only exponential functions will give constant dimensionless model parameters.
Potentials and kinetic functions obeying $\Gamma\propto 1/\lambda$ give
$\lambda'\Gamma=-\lambda\Gamma'$. These have been considered in
Ref.~\cite{anisinf,anisinf+} where
\mbox{$f\propto\exp\left(cm_P^{-2}\int\frac{V}{V'}\ud\phi\right)$} and satisfy the condition
in Eq.~(\ref{cond}).
%---------------------------------------------------------------------------------------------------------------------
%---------------------------------------------------------------------------------------------------------------------
\section{The Massive Vector Field }\label{Sec-massiveVFcase}

We now consider a massive vector field whose mass is also modulated by the inflaton, hence $s,\mu,\Xi\neq0$.
Therefore we now have to consider the full, 5-dimensional system of equations Eqs. (\ref{fullsystem})-(\ref{fullsystem-m}).

As we have seen in Sec. \ref{Sec-m0-VSS} for a massless vector field, when the $f$-term backreaction becomes important, the dynamics lead to the
attractor solution $\mathcal{VSS}$ where $f_{\rm att}(\alpha)\propto e^{-4\alpha}$. Such scaling for $f$, results in the generation of a superhorizon spectrum of perturbations for the transverse components of the vector field.
However, if the vector field is massive, we have to consider also the longitudinal component. As shown in Ref.~\cite{varkin}, to obtain a scale invariant spectrum for the latter we need to have \mbox{$m\propto a$}. Thus,
we are also interested in finding out whether the attractor solution is such that $m_{\rm att}(\alpha)\propto e^{\alpha}$ simultaneously.
%These specific scalings for the kinetic function and mass are of interest as they lead to a scale invariant vector field spectrum of superhorizon perturbations. The vector field spectrum may then contribute to the curvature perturbation of the Universe $\zeta$.
But of course there can only be one solution $\phi(\alpha)$ and therefore there must be a connection between $f(\phi)$ and $m(\phi)$ to obtain the desired scaling of $f$ and $m$ simultaneously. To this end we have to impose the following connection
\begin{equation}\label{connetion}
\Xi=-\frac{1}{2}\Gamma
\qquad\Longleftrightarrow\qquad\frac{f'}{f}=-4\frac{m'}{m}.
\end{equation}
This condition is the price we have to pay to obtain the desired behaviour
for both the mass and the kinetic function. The question to be investigated is
whether, with the contribution of the mass backreaction taken into account, the
desired scaling remains an attractor solution.

For the massive vector field case there are now two terms in the backreaction, see Eq. (\ref{baf-decomp}). Their relative magnitude is given by
\begin{equation}
\frac{\mathcal{B}_{A-f}}{\mathcal{B}_{A-m}}=-\frac{\Gamma}{\Xi}\left(\frac{z}{s}\right)^{2}
=2\left(\frac{z}{s}\right)^{2}.
\end{equation}
Where we considered the connection Eq. (\ref{connetion}) in the last equality.

A natural choice of initial conditions for the vector field can be based on energy equipartition grounds.
Energy equipartition states that at some early time,
for example at the onset of inflation, that $\rho_{{\rm {kin}}}\approx V_{A}$. In our expansion normalised set of variables this gives us $z_{i}^{2}\approx s_{i}^{2}$, where the subscript `{\it i}' indicates
the early epoch at the onset of inflation. Considering the connection Eq. (\ref{connetion}) we find that
the initial backreaction ratio is $\frac{\mathcal{B}_{A-f}}{\mathcal{B}_{A-m}}\Big|_{i}\approx2$.
This suggests that the $m$-term backreaction is important from the
onset. Therefore we have to be careful as we cannot neglect the $m$-term backreaction even if we are considering a very light
vector field. Thus, assuming initial equipartition of energy, we cannot simply extrapolate the analysis from
the previous massless case where $\mathcal{B}_{A-m}=0$, to a small but non-zero massive case, because initially $\mathcal{B}_{A-f}\sim\mathcal{B}_{A-m}$.

Calculating the stationary points and the eigenvalues of the linearised system becomes highly complicated
when considering a massive vector field. The problem stems from the mass
function $\mu$, whose first derivative, taken from Eq. (\ref{fullsystem-m}), is
\begin{equation}
\frac{\ud\mu}{\ud\alpha}=\left[\left(\Xi-\Gamma\right)x+\left(3\Sigma^{2}
+3x^{2}+2z^{2}+s^{2}\right)\right]\mu.
\end{equation}

Under the condition that $\Gamma$ has an opposite sign to both $\Xi$ and
$x$ we can see that $\mu$ is a monotonically increasing function.
Asymptotically this function will tend to $\mu\rightarrow\infty$.
The phase space analysis is designed to investigate the asymptotic (final)
behaviour of systems of differential equations.
The problem is that we wish to study the qualitative
behaviour of solutions for a light vector field,
which, since the effects of a non-zero mass are increasing in time,
corresponds to an intermediate phase in the evolution of the system.
Therefore, it is not clear
whether we can understand the system qualitatively in this intermediate
regime where the vector field has a small but non-zero mass.
However, as shown below, it turns out that we can indeed
%proceed forward and
obtain some useful information if we first consider
a very light field $\mu\ll1$.

%---------------------------------------------------------------------------------------------------------------------
%---------------------------------------------------------------------------------------------------------------------
\subsection{The Light Vector Field Case}

In this section we assume that,
if the field is light enough, we may neglect
any terms proportional to the mass function $\mu$ in the system
of differential equations. We can then establish the existence of
any critical points and their stability. If any of the stationary points are
stable, i.e. attractors, and the mass is small enough as to not effect
the dynamics, we should expect the solutions to reach these attractors.
We later confirm the validity of this assumption numerically for different
model examples, (e.g. see Fig.~\ref{fig-light-heavy-2xGamma}).

\begin{figure}[h]
\includegraphics[width=100mm,angle=0]{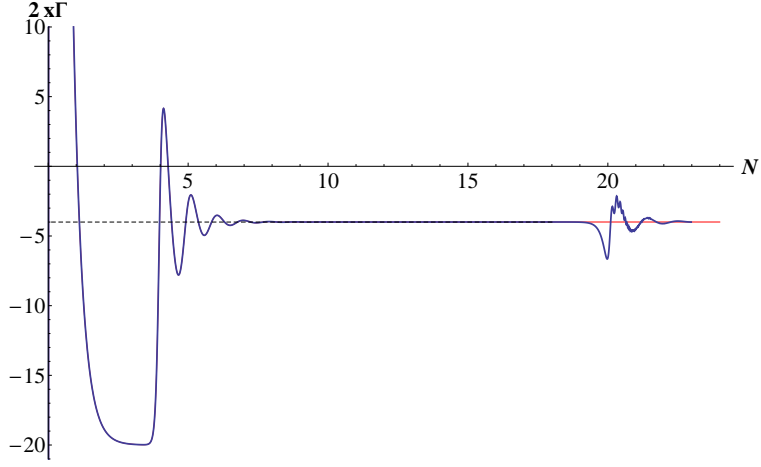}
\caption{This numerical plot demonstrates the validity of our approximation in neglecting terms proportional to $\mu$ when considering a light vector field. The plot shows a numerical solution for the evolution of the kinetic function scaling $\frac{1}{f}\frac{\ud f}{\ud \alpha}=2x\Gamma$, given by Eq. (\ref{scalingSol1}), with respect to the number of elapsed e-folds $N$, for the particular case where the kinetic function, mass and scalar potential are exponential. The numerical solution (in blue-thick line) to the full system of equations, Eqs.~(\ref{fullsystem})-(\ref{fullsystem-m}), is indistinguishable from the numerical solution (in red-thin line) where the terms proportional to $\mu$ have been removed until the vector field becomes heavy ($N\gtrsim 20$ in the plot). The plot is also an example of how the vector scaling solution $f_{\textrm{att}}\propto e^{-4\alpha}$ is attained in the light and heavy vector field case. Note that the attractors take only a handful of e-folds to be reached, corresponding to the oscillating behaviour for $N\sim5$ (light field) and $N\sim20$ (heavy field).
More examples are shown in Sec.~\ref{examples}.
}\label{fig-light-heavy-2xGamma}
\end{figure}

The full system of differential equations, Eqs. (\ref{fullsystem})-(\ref{fullsystem-m}) now reduces to the 4-dimensional sub-system given below
\begin{eqnarray}
\frac{\ud\Sigma}{\ud\alpha} & = & -3\Sigma+2z^{2}-2s^{2}+\Sigma\left(3\Sigma^{2}+3x^{2}+2z^{2}+s^{2}\right), \label{LVF-sigma}\\
\frac{\ud x}{\ud\alpha} & = & -3x-\lambda\left(1-\Sigma^{2}-x^{2}-z^{2}-s^{2}\right)+\Gamma z^{2}-\Xi s^{2}+x\left(3\Sigma^{2}+3x^{2}+2z^{2}+s^{2}\right),\label{LVF-x}\\
\frac{\ud z}{\ud\alpha} & = & \left[-2-2\Sigma-\Gamma x+\left(3\Sigma^{2}+3x^{2}+2z^{2}+s^{2}\right)\right]z,\label{LVF-z} \\
\frac{\ud s}{\ud\alpha} & = & \left[-1+2\Sigma+\Xi x+\left(3\Sigma^{2}+3x^{2}+2z^{2}+s^{2}\right)\right]s.\label{LVF-s}
\end{eqnarray}

We may obtain a qualitative feel of how the variables $z$ and $s$ evolve by looking at Eqs. (\ref{LVF-z}) and (\ref{LVF-s}). These variables determine the backreaction and vector field energy density. If the energy density is dominated by the scalar potential, i.e. $y\approx1$, and considering the model parameter $\Gamma\gg1$, then Eqs. (\ref{LVF-z}) and (\ref{LVF-s}) may be approximated by
\begin{equation}\label{LF-z&s}
\frac{\ud z}{\ud\alpha}  \simeq  \left(-2-\Gamma x\right)z\qquad\text{and}\qquad
\frac{\ud s}{\ud\alpha}  \simeq  \left(-1+\Xi x\right)s=\frac12\left(-2-\Gamma x\right)s,
\end{equation}
where we used the connection Eq. (\ref{connetion}) in the last equality. First we notice that in this regime the $z$ variable evolves twice as fast as the $s$ variable. Hence the $f$-term backreaction evolves two times faster than the $m$-term backreaction. Because of this difference in evolutionary rate of the backreation terms, there is only a small region of parameter space where $\mathcal{B}_{A-f}\sim\mathcal{B}_{A-m}$ at the time where the backreaction becomes important to the dynamics of the system. Most of parameter space will be taken by either one or the other backreaction terms dominating.

Let us consider some early time where the vector field energy density and backreaction, and consequently anisotropy are completely subdominant. Then a period of slow-roll (or indeed fast-roll) inflation where $\epsilon_{\textrm{H}}\ll1$ leads to $\mathcal{B}_{A-f}$ growing two times faster than  $\mathcal{B}_{A-m}$. Remember from Eq. (\ref{slow-roll-paraHJ}) that the term in the system above \mbox{$3\Sigma^{2}+3x^{2}+2z^{2}+s^{2}=-\frac{\ddot{\alpha}}{\dot{\alpha}^{2}}=\epsilon_{\textrm{H}}$} and is therefore very small during inflation. The $f$-term backreaction is therefore very likely to completely dominate the total backreaction and will effect the dynamics of the system before the $m$-term does so. We are effectively considering the same system as that of the massless vector field in which only the $f$-term backreaction existed, see Sec. \ref{Sec-masslessVFcase}.

From the approximations in Eq. (\ref{LF-z&s}) we can directly see that at stationary points where $z,s\not=0$ the solutions
$x_{c}\simeq-\frac{2}{\Gamma}$ and/or $x_{c}\simeq\frac{1}{\Xi}$ emerge, see Eq. (\ref{scalingSol}). Therefore
we observe that both scaling solutions
\begin{equation}
f(\alpha)\propto e^{-4\alpha}\qquad\text{and}\qquad m(\alpha)\propto e^{\alpha}
\end{equation}
are inherent in the equations of motions. But of course we
require the connection $\Xi=-\frac{1}{2}\Gamma$ to ensure that we
obtain these two solutions simultaneously.

We now make these ideas more concrete by looking
for stationary solutions to the full system. There are 5 critical
points for this system. These critical points include the non-hyperbolic anisotropic kination solution, $\mathcal{AKS}$ discussed in Sec. \ref{Sec-AKS}. We will not consider this point again. The remaining stationary points for the system of Eqs. (\ref{LVF-sigma})-(\ref{LVF-s}) are given below.

%---------------------------------------------------------------------------------------------------------------------
%---------------------------------------------------------------------------------------------------------------------
\subsubsection{The Standard Slow-Roll Solution, $\mathcal{SSR}$.}
The standard slow-roll critical point is given by [c.f. Eq. (\ref{SSR-m0})]
\begin{equation}\label{SSR-cp-mh}
\mathcal{SSR}:\qquad
(x_{c}, y_{c}, z_{c}, s_{c},\Sigma_{c})=
\left(-\frac{\lambda}{3},\,\sqrt{1-\left(\frac{\lambda}{3}\right)^{2}},\,
0,\,0,\,0\right).
\end{equation}
The eigenvalues of the matrix $\mathcal{M}$ are
\begin{equation}
m_{1,2}=-3+\frac{1}{3}\lambda^{2}\:,\quad m_{3}=-1+\frac{1}{3}\lambda\left(\lambda-\Xi\right)\:,\quad\text{and}\quad m_{4}=-2+\frac{1}{3}\lambda\left(\lambda+\Gamma\right).
\end{equation}
The existence and stability of this solution therefore requires
\begin{equation}
\lambda<3\,,\quad\lambda\left(\lambda+\Gamma\right)<6\quad\text{and}\quad\lambda\left(\lambda-\Xi\right)<3.
\end{equation}
As seen in the massless vector field case, this critical point corresponds to the
standard slow-roll inflationary solution when $\lambda\ll3$. In this region of parameter space the scalar potential dominates the energy density of the Universe $y_c\approx1$. The
vector field energy density and backreaction completely vanishes at the critical
point $z_c=s_c=0$, and consequently so does any anisotropy $\Sigma_c=0$.
All the results derived from this critical point in the massless case, Eqs. (\ref{SSR-mo-slwParam}) and (\ref{SSR-mo-eqofstate}) apply here. The only difference with the massless case is the stability condition. Considering the connection
$\Xi=-\frac{1}{2}\Gamma$, the existence and the stronger stability condition are now
\begin{equation}\label{SSR-cond-light}
\lambda<3\qquad\text{and}\qquad\lambda\left(2\lambda+\Gamma\right)<6.
\end{equation}

These bounds apply for constant or almost constant dimensionless model parameters. Otherwise we require the stronger bounds $\lambda\ll3$ and $\lambda\Gamma\ll6$ to make certain that the eigenvalues depend weakly on the dimensionless model parameters. This therefore ensures the validity of our method in establishing the stability of the critical points. In these limits the existence and stability of this critical point are guaranteed.

For time-dependent dimensionless model parameters, the motion of the critical point is given by Eq. (\ref{cp-motion-m0-SSR}). In analogy to Eq. (\ref{SSR-mov-cond}), the constraint on the time-dependence of the dimensionless model parameter becomes
\begin{eqnarray}
\mathcal{SSR}:\qquad\left|2\epsilon-\eta\right| & <
\begin{cases}
\left|m_{1,2}\right| & =\left|-3+\epsilon\right|\\
\left|m_{3}\right| & =\left|-1+\epsilon+\frac{1}{6}\lambda\Gamma\right|\\
\left|m_{4}\right| & =\left|-2+\epsilon+\frac{1}{3}\lambda\Gamma\right|
\end{cases}\label{SSR-mh-mov-cond}
\end{eqnarray}
If the slow-roll conditions are met $\epsilon\ll1$ and $|\eta|\ll1$, and $|\lambda\Gamma-6|>\mathcal{O}(\eta)$, then the inequalities above
will be satisfied. This can be achieved as long as we are not too close to the bifurcation value $\lambda\Gamma\not\approx6$.
The solutions then are able to reach and be dragged along with the critical point.

If the dimensionless model parameters are varying more rapidly, then the validity of our
method in establishing the stability of the critical points requires the strong conditions $\lambda\ll3$ and $\lambda\Gamma\ll6$. These conditions ensures that the eigenvalues depend weakly on the dimensionless model parameters.
In this case the eigenvalue of smallest magnitude is clearly $|m_{3}|\simeq1$. Therefore the
inequalities above are readily satisfied if the slow-roll conditions $\epsilon\ll1$ and $\left|\eta\right|\ll1$ are met.
%---------------------------------------------------------------------------------------------------------------------
%---------------------------------------------------------------------------------------------------------------------
\begin{comment}
\subsubsection{The Anisotropic Kination Solution, $\mathcal{AKS}$.}

\[
\mathcal{AKS}:\qquad\left(x,\,0,\,0,\,0,\,\sqrt{1-x^{2}}\right)
\]
As seen in the massless case this is a non-hyperbolic point because
one of the eigenvalues of the matrix $\mathcal{M}$ is zero and therefore
requires more advanced methods such as the center manifold theorems
to establish it's stability. In any case we find that this critical
point does not correspond to any cosmologies of interest, primarily
as the scalar potential vanishes at the point and therefore cannot
correspond to an inflating Universe. We will not pursue the analysis
of this point any further.
\end{comment}
%---------------------------------------------------------------------------------------------------------------------
%---------------------------------------------------------------------------------------------------------------------
\subsubsection{The Vector Scaling Solutions, $\mathcal{VSS}$.}

When considering a massive vector field we find two additional vector scaling solutions to the one found in the massless field case. Different vector scaling solutions appear because there are now two terms in the backreaction Eq. (\ref{baf-decomp}) which may effect the dynamics of the system. These critical points arise from
the effect of $f$-term backreaction only, $m$-term backreaction
only and a combination of the two. We will analyse these points separately
as follows.

%---------------------------------------------------------------------------------------------------------------------
%---------------------------------------------------------------------------------------------------------------------
%\subsubsection

%\bigskip
%\noindent
\subparagraph{\bf Vector Scaling Solution 1, $\mathcal{VSS}_{{\it 1}}$.}
At this critical point the vector potential term and therefore the $m$-term
backreaction vanishes due to $s_{c}=0$. This is the same critical point as the $\mathcal{VSS}$ of the massless case [c.f. Eq. (\ref{VSS-m0})],
\begin{equation}\label{VSS1}
\!\;
\mathcal{VSS}_{{\it 1}}:\;(x_{c},y_{c},z_{c},s_c, \Sigma_{c})=\!
\left(-6\Delta_{1}\left(\lambda+\Gamma\right),\:
\Delta_{1}\Delta_{2}\sqrt{\Gamma^{2}+\lambda\Gamma+6}\,,\:
\Delta_{1}\Delta_{2}\sqrt{\lambda^{2}+\lambda\Gamma-6}\,,\:
0\,,
2\Delta_{1}\left(\lambda^{2}+\lambda\Gamma-6\right)\right)\!,
\hspace{-1cm}\end{equation}
\[
\text{where}\quad\Delta_{1}\equiv\left(3\Gamma^{2}+4\Gamma\lambda+\lambda^{2}+12\right)^{-1}\quad\text{and}\quad
\Delta_{2}\equiv\sqrt{3\left(3\Gamma^{2}+2\lambda\Gamma-\lambda^{2}+12\right)}.
\]
The eigenvalues of the matrix $\mathcal{M}$ and thus the stability conditions are however different.
\begin{comment}
The vector scaling critical point
$(x_{c}, y_{c}, z_{c}, s_{c},\Sigma_{c})$ is given by
\[
\mathcal{VSS}_{{\it 1}}:\quad\left(-6\Delta_{1}\left(\lambda+\Gamma\right)\,,\:\Delta_{1}\Delta_{2}\sqrt{\Gamma^{2}+\lambda\Gamma+6}\,,\:
\Delta_{1}\Delta_{2}\sqrt{\lambda^{2}+\lambda\Gamma-6}\,,\:0\,,\:2\Delta_{1}\left(\lambda^{2}+\lambda\Gamma-6\right)\right)
\]
\[
\text{where}\quad\Delta_{1}\equiv\left(3\Gamma^{2}+4\Gamma\lambda+\lambda^{2}+12\right)^{-1}\quad\text{and}\quad
\Delta_{2}\equiv\sqrt{3\left(3\Gamma^{2}+2\lambda\Gamma-\lambda^{2}+12\right)}
\]
\end{comment}
Considering the connection $\Xi=-\frac{1}{2}\Gamma$ in Eq.~(\ref{connetion}), the eigenvalues are given by
\[
m_{1}=-\Delta_{1}\Delta_{2}^{2}\:,\qquad m_{4}=9\Delta_{1}\left(\lambda^{2}+\lambda\Gamma-4\right)\quad\text{and}
\]
\begin{equation}\label{VSS1-fulleigen}
m_{2,3}=-\frac{1}{2}\Delta_{1}\Delta_{2}\left[\Delta_{2}\pm\sqrt{-8\left(\lambda^{3}\Gamma
+\left(2\Gamma^{2}+\frac{51}{8}\right)\lambda^{2}
+\left(\Gamma^{3}-\frac{3}{4}\Gamma\right)\lambda-\frac{81}{2}-\frac{57}{8}\Gamma^{2}\right)}\right].
\end{equation}
The limits $\Gamma\gg\lambda$ and $\Gamma\gg1$ ensures that we have
an inflating Universe $y_c\approx1$ with small anisotropy $\Sigma_c\ll1$, see Eq. (\ref{VSS-m0-cp}). The critical point is approximated by
\begin{equation}\label{VSS1-cp}
\mathcal{VSS}_{{\it 1}}:\qquad(x_{c},y_{c},z_{c},s_c, \Sigma_{c})=\left(-\frac{2}{\Gamma}\,,\:1\,,\:\frac{\sqrt{\lambda\Gamma-6}}{\Gamma}\,,\:
0\,,
\frac{2\left(\lambda\Gamma-6\right)}{3\,\Gamma^{2}}\right).
\end{equation}
The existence of the critical point requires $\lambda\Gamma>6$. Therefore we are considering the same conditions as those of the massless case in Eq. (\ref{VSS-cond}).
In these limits the real parts of the eigenvalues are approximated by
\begin{equation}\label{VSS1-eigen}
\textrm{Re}[m_{1}]\simeq-3\:,\qquad \textrm{Re}[m_{2,3}]\simeq-\frac{3}{2}\:,\quad\text{and}\quad \textrm{Re}[m_{4}]\simeq\frac{3\lambda\Gamma-12}{\Gamma^{2}}.
\end{equation}
The real parts of eigenvalues $m_{1,2,3}$ in these limits are the same as for the massless case, see Eq. (\ref{VSS-m0-Remi}). These are negative and lead to solutions being attracted to this point. Having now considered a massive vector field we obtain an additional eigenvalue $m_{4}$ which is very small but positive. We can see this through the existence
condition $\lambda\Gamma>6$ that $0<\textrm{Re}[m_{4}]\ll1$.  The critical point
is therefore technically an unstable saddle point, however because $m_4$ is so small the critical point is said to be quasi-stable. If a solution approaches this critical point it will remain there for a long period of time. At the vicinity of the critical point, solutions evolve very slowly. In these limits the eigenvalues depend weakly on the model parameters, see Appendix (\ref{App-VSS1-mh-light}). This therefore ensures the validity of our method in establishing the stability of the critical points.

All the results found in Sec. \ref{Sec-m0-VSS} for the energy density ratio $\mathcal{R}$, backreaction ratio $\mathcal{R_B}$, effective scalar slope $V'_{\textrm{eff}}$, slow-roll conditions $\epsilon_{\textrm{H}}$ and $\eta_{\textrm{H}}$, anisotropic stress $\Sigma$ and scalar equation of state $\gamma_{\phi}$, in the massless vector field case also apply here, see Eqs. (\ref{Ratios-VSS-m0})-(\ref{VSS-mo-sigma}) and (\ref{VSS-mo-eqofstate}) respectively. However as we have now considered the connection in Eq. (\ref{connetion}), we obtain the attractor solutions simultaneously, see Eq. (\ref{scalingSol})
\begin{equation}\label{VSS1-fnm}
f_{\rm att}(\alpha)\propto e^{-4\alpha}\qquad\text{and}\qquad m_{\rm att}(\alpha)\propto e^{\alpha}.
\end{equation}

To reach this critical point the $f$-term backreaction must dominate the total backreaction at the point where it becomes dynamically important $z^2\gg s^2$. As described above, this is the most likely scenario because the $f$-term backreaction grows twice as fast as the $m$-term backreaction if we consider a prior period of standard slow-roll inflation.

For time-dependent model parameters, as seen in the massless case, the motion of the critical point is given by  Eq. (\ref{cp-motion-m0-VSS}). Therefore, for solutions to be able to reach and be dragged along with the critical point, the eigenvalue of smallest magnitude in Eq. (\ref{VSS1-eigen}) must be larger than all of the expressions in Eq. (\ref{cp-motion-m0-VSS}). We can see from Eq. (\ref{VSS1-eigen}) that $\textrm{Re}[m_{4}]\ll1$. However, when considering the motion of the critical point we do not need to consider this small eigenvalue. The reason why is discussed at the end of Sec.~\ref{Sec-mh-VSS3}.
Therefore the eigenvalues of smallest magnitude that we need to consider are $|\textrm{Re}[m_{2,3}]|\simeq\frac32$. Then, if we consider a relatively flat scalar potential under the constraints in Eq.~(\ref{cond}), the conditions of Eq. (\ref{cp-motion-m0-VSS}), can be reduced to a single condition, see Eq. (\ref{VSS-param_cond}).

%---------------------------------------------------------------------------------------------------------------------
%---------------------------------------------------------------------------------------------------------------------
\subparagraph{Vector Scaling Solution 2, $\mathcal{VSS}_{{\it 2}}$.}

At this critical point the vector kinetic term and therefore the $f$-term
backreaction vanishes due to $z_c=0$. The critical point is given by
\begin{equation}
\mathcal{VSS}_{{\it 2}}:\quad(x_{c}, y_{c}, z_{c}, s_{c},\Sigma_{c})=
\left(3\Delta_{3}\left(\Xi-2\,\lambda\right),\:\Delta_{3}\Delta_{4}\sqrt{\left(\Xi^{2}-\Xi\,\lambda+6\right)},\:0,\:
\Delta_{3}\Delta_{4}\sqrt{\left(\lambda^{2}-\Xi\,\lambda-3\right)},\:2\Delta_{3}\left(\Xi\,\lambda-\lambda^{2}+3\right)\right)\label{VSS2}
\end{equation}
\[
\text{where}\quad\Delta_{3}\equiv\left(3\,\Xi^{2}-5\,\Xi\,\lambda+2\,\lambda^{2}+12\right)^{-1}\quad\text{and}\quad
\Delta_{4}\equiv3\sqrt{\left(\Xi^{2}-\frac{4}{3}\Xi\,\lambda+4\right)}.
\]
The case of interest occurs in the limits $\left|\Xi\right|\gg\lambda$
and $\left|\Xi\right|\gg1$, this critical point may then be approximated
by
\begin{equation}\label{VSS2-cp}
\mathcal{VSS}_{{\it 2}}:\qquad(x_{c}, y_{c}, z_{c}, s_{c},\Sigma_{c})=
\left(\frac{1}{\Xi},\:1,\:0,\:-\frac{\sqrt{-3-\lambda\Xi}}{\Xi},\:\frac{6+2\lambda\Xi}{3\Xi^{2}}\right).
\end{equation}

The existence of this solution in these limits requires $|\lambda\Xi|>3$. We can see that these limits ensure that the scalar potential dominates the energy density $y_c\approx1$ and therefore leads to an inflating Universe. However these limits do not guarantee stability. In Appendix (\ref{App-VSS2-mh-light}) plots are shown of the functional dependence of the real parts of the eigenvalues on the model parameters. The real parts of the eigenvalues of the matrix $\mathcal{M}$ for this critical point in these limits are ($m_4$ is obtained by the analytic expression)
\begin{equation}\label{VSS2-eigen}
\textrm{Re}[m_{1}]\simeq-3,\quad
\textrm{Re}[m_{2,3}]\simeq-\frac{3}{2},\quad
\textrm{Re}[m_{4}]\simeq-\frac{3\left(\lambda\Xi+4\right)}{\Xi^{2}}
=\frac{6\lambda\Gamma-48}{\Gamma^{2}},
\end{equation}
%m_{4}=3\Delta_{3}\left(2\lambda^{2}-3\lambda\Xi-12\right)
where we considered the connection Eq. (\ref{connetion}) in the last equality. With this connection, the critical point in Eq. (\ref{VSS2-cp}) is given by
\begin{equation}\label{VSS2-eigen+}
\mathcal{VSS}_{{\it 2}}:\qquad(x_{c}, y_{c}, z_{c}, s_{c},\Sigma_{c})=
\left(-\frac{2}{\Gamma},\:1,\:0,\:\frac{\sqrt{2(\lambda\Gamma-6)}}{\Gamma}
,\:\frac{4(6-\lambda\Gamma)}{3\Gamma^{2}}\right).
\end{equation}

These limits are now $\Gamma\gg\lambda$ and $\Gamma\gg1$. We then find that the real parts of the first three eigenvalues $m_{1,2,3}$ are identical to those of $\mathcal{VSS}_{\it 1}$ above and $\mathcal{VSS}$ of the massless case. These eigenvalues drive solutions towards the critical point. However, as in $\mathcal{VSS}_{\it 1}$ above we find an additional eigenvalue $m_4$ which is very small in the limits considered. There is a window $6<\lambda\Gamma<8$ where $\textrm{Re}[m_{4}]<0$ and the critical point is stable. Otherwise for $\lambda\Gamma>8$
the critical point is quasi-stable. Solutions that approach this critical point evolve very slowly, and therefore remain at the critical point for a long period of time. In these limits the eigenvalues depend weakly on the model parameters, seen as plateaus in the plots of Appendix (\ref{App-VSS2-mh-light}). This therefore ensures the validity of our method in establishing the stability of the critical points.

Having considered the connection Eq. (\ref{connetion}) we find the following result for the energy density ratio
\begin{equation}\label{VSS2-R}
\mathcal{R}\simeq\frac{2(\lambda\Gamma-6)}{\Gamma^{2}}\ll1,
\end{equation}
which is two times larger than the ratio of the massless and $\mathcal{VSS}_{\it1}$ case given in Eq. (\ref{Ratios-VSS-m0}). The energy density of the vector field tracks that of the scalar field if the dimensionless model parameters are constant. Otherwise the energy density ratio is varying with time-dependence determined by that of the dimensionless model parameters. Under the limits considered the vector field energy density contribution is kept subdominant.

We notice that this critical point leads to the same backreaction ratio $\mathcal{R_B}$, effective potential slope $V'_{\textrm{eff}}$, slow-roll parameters $\epsilon_{\textrm{H}}$ and $\eta_{\textrm{H}}$ and equation of state $\gamma_{\phi}$ as the $\mathcal{VSS}_{\it 1}$ above and the $\mathcal{VSS}$ of the massless case, see Eqs. (\ref{VSS-mo-RB}), (\ref{VSS-m0-slowrollparam}) and (\ref{VSS-mo-eqofstate}) respectively.

The anisotropy in the expansion, $\Sigma$ is given by
\begin{equation}\label{VSS2-sigma}
\Sigma_{c}\simeq-\frac{4}{3}\left(\frac{\lambda\Gamma-6}{\Gamma^{2}}\right)=-\frac{2}{3}\mathcal{R}\quad\Rightarrow\quad|\Sigma_{c}|\ll1.
\end{equation}
The anisotropy is proportional to the energy density ratio and therefore evolves in the same way.

As in the $\mathcal{VSS}_{\it1}$ case above, having used the connection in Eq. (\ref{connetion}), we can simultaneously obtain the desired attractor behaviour for the kinetic function and mass given in Eq. (\ref{VSS1-fnm}). To reach this critical point the $m$-term backreaction must dominate the total backreaction at the point where it becomes dynamically important $s^2\gg z^2$.

For time-dependent model parameters, the motion of the critical point is given by
\begin{equation}
\left|\frac{1}{x_{c}}\frac{\ud x_{c}}{\ud\alpha}\right|
=2\sqrt{6}m_{P}\left|\frac{\Gamma'}{\Gamma^{2}}-\frac{2}{3}\frac{\lambda'}{\Gamma^{2}}\right|\:,\qquad
\left|\frac{1}{y_{c}}\frac{\ud y_{c}}{\ud\alpha}\right|
=2\sqrt{6}m_{P}\left|\frac{\lambda'}{\Gamma^{2}}\right|\:,
\qquad\left|\frac{1}{z_{c}}\frac{\ud z_{c}}{\ud\alpha}\right|=0\nonumber
\end{equation}
\begin{equation}\label{cp-motion-mh-VSS2}
\left|\frac{1}{s_{c}}\frac{\ud s_{c}}{\ud\alpha}\right|
=2\sqrt{6}m_{P}\left|\frac{\Gamma'}{\Gamma^{2}}+2\frac{\lambda'}{\Gamma^{2}}
-\frac12\frac{\lambda'\Gamma+\lambda\Gamma'}{\Gamma\left(\lambda\Gamma-6\right)}\right|
\quad\text{and}\quad
\left|\frac{1}{\Sigma_{c}}\frac{\ud\Sigma_{c}}{\ud\alpha}\right|
=4\sqrt{6}m_{P}\left|\frac{\Gamma'}{\Gamma^{2}}+\frac53\frac{\lambda'}{\Gamma^{2}}
-\frac12\frac{\lambda'\Gamma+\lambda\Gamma'}{\Gamma\left(\lambda\Gamma-6\right)}\right|.
\end{equation}
For solutions to be able to reach this critical point and be dragged along with it, the eigenvalue of smallest magnitude must be larger than all of the expressions above. As in the $\mathcal{VSS}_{\it1}$ case, the eigenvalue of smallest magnitude which drives solutions to the vector scaling solutions, from Eq. (\ref{VSS2-eigen}) is $\left|m_{2,3}\right|\simeq\frac{3}{2}$. Again, we ignore $\left|m_{4}\right|$. The reason why is discussed at the end of Sec.~\ref{Sec-mh-VSS3}.
Then if we consider a relatively flat scalar potential under the constraints in Eq.~(\ref{cond}), the conditions in Eq. (\ref{cp-motion-mh-VSS2}), can be reduced to a single condition, see Eq. (\ref{VSS-param_cond}).

%---------------------------------------------------------------------------------------------------------------------
%---------------------------------------------------------------------------------------------------------------------
\subparagraph{Vector Scaling Solution 3, $\mathcal{VSS}_{{\it 3}}$.}\label{Sec-mh-VSS3}

At this stationary point the kinetic and potential terms of the vector
field energy density are comparable. Therefore the $f$-term
and $m$-term backreaction are both important to the dynamics of the system.
The critical point  is given by
\begin{eqnarray}\label{VSS3}
\mathcal{VSS}_{{\it 3}}:\quad(x_{c}, y_{c}, z_{c}, s_{c},\Sigma_{c})=
\Big(-6\Delta_{5}\,,\:\Delta_{5}\sqrt{6\Gamma^{2}+3\lambda\left(2\Gamma-3\Xi\right)-3\Xi\left(\Gamma-2\Xi\right)+36}\,,\nonumber\\
\Delta_{5}\sqrt{6\lambda^{2}+\left(3\Xi+6\Gamma\right)\lambda-3\Xi\left(\Gamma+2\Xi\right)-36}\,,\:
\Delta_{5}\sqrt{9\lambda^{2}+\left(-6\Xi+6\Gamma\right)\lambda-3\Gamma\left(\Gamma+2\Xi\right)-36}\,,\:
-\Delta_{5}\left(-2\Xi+\lambda-\Gamma\right)\Big),
\end{eqnarray}
\[
\text{where}\quad\Delta_{5}\equiv-\left[2\left(\Xi-2\lambda-\Gamma\right)\right]^{-1}.
%\quad\text{and}\quad
%\Delta_{6}\equiv\left(3\Gamma+4\lambda\right)^{-1}.
\]
Considering the connection in Eq. (\ref{connetion}), and the limits $\Gamma\gg\lambda$ and $\Gamma\gg1$ we can approximate the critical point as
\begin{comment}
\[
\left(-6\Delta_{6}\,,\:-\Delta_{6}\sqrt{\frac{21}{2}\lambda\Gamma+9\Gamma^{2}+36}\,,\:
-\Delta_{6}\sqrt{6\lambda^{2}+\frac{9}{2}\lambda\Gamma-36}\,,\:-3\Delta_{6}\sqrt{\lambda\Gamma+\lambda^{2}-4}\,,\:
-\lambda\Delta_{6}\right)
\]
\end{comment}
\begin{equation}\label{VSS3-cp}
\mathcal{VSS}_{{\it 3}}:\quad(x_{c}, y_{c}, z_{c}, s_{c},\Sigma_{c})=
\left(-\frac{2}{\Gamma}\,,\:1\,,\:\frac{\sqrt{\lambda\Gamma-8}}{\sqrt2\,\Gamma}\,,\:
\frac{\sqrt{\lambda\Gamma-4}}{\Gamma}\,,\:-\frac{\lambda}{3\Gamma}\right).
\end{equation}
In Appendix (\ref{App-VSS3-mh-light}) we show plots of the eigenvalues as functions of the model parameters. We can clearly see that in these limits the real parts of the eigenvalues of the matrix $\mathcal{M}$ are approximated by ($m_4$ is obtained by the analytic expression)
\begin{equation}\label{VSS3-eigen}
 \textrm{Re}[m_{1}]\simeq-3
 \,,\qquad
 \textrm{Re}[m_{2,3}]\simeq-\frac{3}{2}\qquad\text{and}\qquad
 \textrm{Re}[m_4]\simeq -12\,{\frac { \left(\Gamma\lambda-8 \right)
 \left(\Gamma\lambda-4 \right) }{ \left( 3 \Gamma \lambda-20 \right)
 {\Gamma }^{2}}}.
\end{equation}

Three eigenvalues are found to be identical in all the vector scaling solutions for a massless and light vector field, namely $m_{1,2,3}$. These eigenvalues are negative and therefore drive solutions towards the $\mathcal{VSS}$. In the same way as the other two vector scaling solutions we find an additional eigenvalue $m_4$ of very small magnitude. For this critical point however we find $\textrm{Re}[m_4]<0$. Therefore the stationary point is stable
and the late-time attractor of the system. All solutions will eventually converge
to this point. In the limits considered $\Gamma\gg\lambda$ and $\Gamma\gg1$, the eigenvalues depend weakly on the dimensionless model parameters, as seen by plateaus in the eigenvalue plots of Appendix (\ref{App-VSS3-mh-light}). This therefore ensures the validity of our method in establishing the stability of the critical points. The existence and stability of this solution requires
\begin{equation}
\lambda\Gamma>8\:,\qquad\Gamma\gg\lambda\qquad\text{and}\qquad\Gamma\gg1,\label{VSS3-cond}
\end{equation}
which are sufficient but not necessary conditions.

The energy density ratio is now given by
\begin{equation}\label{VSS3-R}
\mathcal{R}=\frac{z^{2}+s^{2}}{x^{2}+y^{2}}\simeq\frac{\frac{3}{2}\lambda\Gamma-8}{\Gamma^{2}}\ll1.
\end{equation}
The energy density of the vector field tracks that of the scalar field if the dimensionless model parameters are constant. Otherwise the energy density ratio is varying with time-dependence determined by that of the dimensionless model parameters. Under the limits considered the vector field energy density contribution is kept subdominant. We notice again that this critical point leads to the same backreaction ratio $\mathcal{R_B}$, effective scalar slope $V'_{\textrm{eff}}$, slow-roll parameters $\epsilon_{\textrm{H}}$ and $\eta_{\textrm{H}}$ and scalar equation of state $\gamma_{\phi}$ as the $\mathcal{VSS}_{\it 1,2}$ above and the $\mathcal{VSS}$ of the massless case, see Eqs. (\ref{VSS-mo-RB}), (\ref{VSS-m0-slowrollparam}) and (\ref{VSS-mo-eqofstate}) respectively. The anisotropic stress for the $\mathcal{VSS}_{\it3}$ is however given by
\begin{equation}
\Sigma_c\simeq-\frac{\lambda}{3\Gamma}\quad\Rightarrow\quad|\Sigma_{c}|\ll1.
\end{equation}
As in the $\mathcal{VSS}_{\it{1,2}}$ cases above, having used the connection in Eq. (\ref{connetion}), we can simultaneously obtain the desired attractor behaviour for the kinetic function and mass given in Eq. (\ref{VSS1-fnm}).

For time-dependent model parameters, the motion of the $\mathcal{VSS}_{\it 3}$ critical point is given by
\begin{equation}\nonumber
\left|\frac{1}{x_{c}}\frac{\ud x_{c}}{\ud\alpha}\right|
=2\sqrt{6}m_{P}\left|\frac{\Gamma'}{\Gamma^{2}}+\frac{4}{3}\frac{\lambda'}{\Gamma^{2}}\right|
\:,\qquad
\left|\frac{1}{y_{c}}\frac{\ud y_{c}}{\ud\alpha}\right|=\frac32\sqrt{6}m_{P}\left|\frac{\lambda'}{\Gamma^{2}}\right|\:,
\end{equation}
\begin{equation}\nonumber
\left|\frac{1}{z_{c}}\frac{\ud z_{c}}{\ud\alpha}\right|
=2\sqrt{6}m_{P}\left|\frac{\Gamma'}{\Gamma^{2}}+\frac43\frac{\lambda'}{\Gamma^{2}}
-\frac12\frac{\lambda'\Gamma+\lambda\Gamma'}{\Gamma\left(\lambda\Gamma-8\right)}\right|
\:,\qquad
\left|\frac{1}{s_{c}}\frac{\ud s_{c}}{\ud\alpha}\right|
=2\sqrt{6}m_{P}\left|\frac{\Gamma'}{\Gamma^{2}}+\frac43\frac{\lambda'}{\Gamma^{2}}
-\frac12\frac{\lambda'\Gamma+\lambda\Gamma'}{\Gamma\left(\lambda\Gamma-4\right)}\right|
\end{equation}
\begin{equation}\label{cp-motion-mh-VSS3}
\text{and}\qquad\left|\frac{1}{\Sigma_{c}}\frac{\ud\Sigma_{c}}{\ud\alpha}\right|\simeq2\sqrt{6}m_{P}\left|
\frac{\Gamma'}{\Gamma^{2}}+\frac{4}{3}\frac{\lambda'}{\Gamma^{2}}
-\frac{\lambda'}{\lambda\Gamma}
\right|.
\end{equation}

For solutions to be able to reach this critical point and be dragged along with it, the eigenvalue of smallest magnitude must be larger than all of the expressions above. As in the $\mathcal{VSS}_{\it{1,2}}$ cases above, the appropriate eigenvalue of smallest magnitude, which drives solutions to the vector scaling solutions, from Eq. (\ref{VSS3-eigen}), is not $\left|m_{4}\right|$ but $\left|m_{2,3}\right|\simeq\frac{3}{2}$. We discuss why this is so below. As in previous vector scaling solutions we may approximate the conditions above to a single condition if we consider a relatively flat scalar potential under the constraints in Eq.~(\ref{cond}), see Eq. (\ref{VSS-param_cond}).

%---------------------------------------------------------------------------------------------------------------------
%---------------------------------------------------------------------------------------------------------------------

We have found that all three vector scaling solutions in the limits $\Gamma\gg\lambda$ and $\Gamma\gg1$, and considering the connection in Eq. (\ref{connetion}), exhibits the same backreaction ratio $\mathcal{R_B}$, effective scalar potential slope $V'_{\textrm{eff}}$, slow-roll parameters $\epsilon_{\textrm{H}}$ and $\eta_{\textrm{H}}$ and equation of state $\gamma_{\phi}$. We also find that all three vector scaling critical points gives us the attractor behaviour desired for the kinetic function and mass of the vector field, from Eq. (\ref{scalingSol})
\begin{equation}
f_{\rm att}(\alpha)\propto e^{-4\alpha}\qquad\text{and}\qquad m_{\rm att}(\alpha)\propto e^{\alpha}.
\end{equation}

To which vector scaling critical point will a solution evolve to depends on the relative magnitude of the $f$-term and $m$-term backreaction at the point where the backreaction becomes dynamically important. If the $f$-term backreaction dominates the total backreaction at the point where it becomes dynamically important, then solutions will approach the $\mathcal{VSS}_{\it1}$ critical point. In this case we may approximate the system with $z^2\gg s^2$. We are then dealing with effectively the same system as that for the massless case, Eqs. (\ref{System-m0})-(\ref{system-m0-y}) for which we know the $\mathcal{VSS}_{\it1}$ is an attractor.

For constant or very slowly varying dimensionless model parameters, the small but positive eigenvalue $m_4$ of Eq. (\ref{VSS1-eigen}) corresponds to flows from the $\mathcal{VSS}_{\it1}$ to $\mathcal{VSS}_{\it3}$, where $s$ increases and $z$ decreases. Throughout the flow the $x_c$ and $y_c$ solutions are approximately conserved, and therefore so does an inflating Universe with the desired scaling solutions for the kinetic function and mass. In the limit $\Gamma\gg1$ this flow is very slow, it takes many e-folds to deviate significantly from the critical point. For constant or very slowly varying dimensionless model parameters the late-time attractor solution is the $\mathcal{VSS}_{\it3}$ (see Fig.~\ref{fig-VSS-1-2-3}).
\begin{figure}[h]
\includegraphics[width=90mm,angle=0]{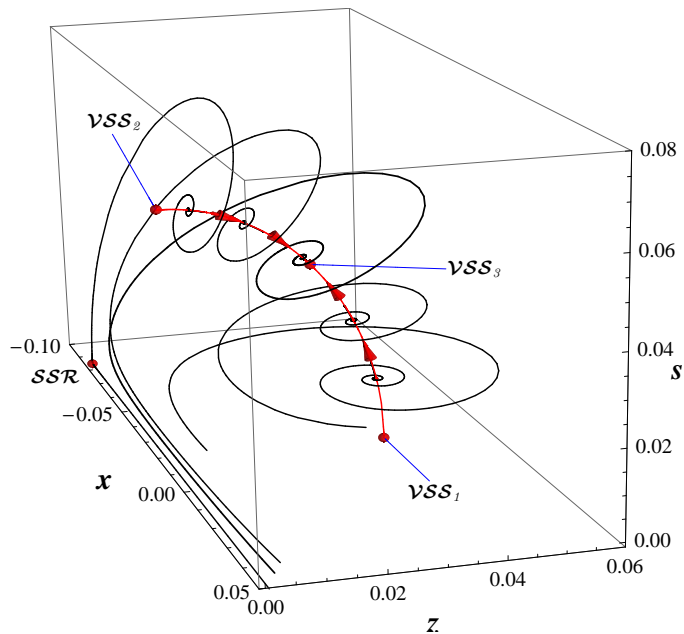}
\caption{This numerical plot demonstrates the flow between the three vector scaling solutions found in the light vector field case, for the particular example where the kinetic function, mass and scalar potential are exponential. The plot shows how different initial conditions can give rise to solutions approaching different vector scaling critical points, depending on the relative magnitudes of $z$ and $s$. The model parameters are chosen so that the $\mathcal{VSS}_{\it3}$ is the late-time attractor. If a solution does not directly fall on this critical point, the small but positive eigenvalues $m_4$ for the $\mathcal{VSS}_{\it{1,2}}$ in Eqs.~(\ref{VSS1-eigen}) and (\ref{VSS2-eigen}), and the small but negative eigenvalue $m_4$ for the $\mathcal{VSS}_{\it3}$ in Eq.~(\ref{VSS3-eigen}), drive very weak flow of solutions along the arc (shown in red) towards the $\mathcal{VSS}_{\it3}$. However, throughout the flows, the solution $x\simeq-2/\Gamma$ is conserved and therefore so is the scaling solution attractor $f_{\textrm{att}}\propto a^{-4}$. The figure also depicts the $\mathcal{SSR}$ critical point which is an unstable saddle point for the parameter space considered.}\label{fig-VSS-1-2-3}
\end{figure}

In the same way, if the $m$-term backreaction dominates the total backreaction at the point where it becomes dynamically important $z^2\ll s^2$, then solutions will approach the $\mathcal{VSS}_{\it2}$ critical point. However, as for $\mathcal{VSS}_{\it1}$, this critical point also has an additional positive eigenvalue $m_4$ of Eq. (\ref{VSS2-eigen}) which drives solutions very slowly towards $\mathcal{VSS}_{\it3}$. Eigenvalue $m_4$ is positive if the dimensionless model parameters do not land in the stable window as discussed after Eq. (\ref{VSS2-eigen+}), otherwise the $\mathcal{VSS}_{\it2}$ is itself stable and $\mathcal{VSS}_{\it3}$ does not exist. Again, throughout the flow, the $x_c$ and $y_c$ solutions are approximately conserved, and therefore so does an inflating universe with the desired scaling solutions for the kinetic function and mass. However, the anisotropy varies along the flow, and it can change from positive to negative.

If a solution is driven near $\mathcal{VSS}_{\it{1,2}}$ then the small but positive eigenvalues, $m_4$ for both points, means that the solutions are then slowly attracted towards the $\mathcal{VSS}_{\it3}$, which is the late-time attractor (see Fig.~\ref{fig-VSS-1-2-3}).

For varying dimensionless model parameters, the motion of the critical point is generally faster than the flow to $\mathcal{VSS}_{\it3}$. We have argued that flows to the $\mathcal{VSS}_{\it3}$ from $\mathcal{VSS}_{\it{1,2}}$ are characterised by the small eigenvalues $m_4$. However the approach to any of the $\mathcal{VSS}$ critical points is characterised by the other eigenvalues $m_{1,2,3}$. As long as the time dependence of the critical point is smaller than the approach of the solutions towards the attractor then the solutions are able to reach and be dragged along with the critical points for a large number of e-folds. As described in Sec.~\ref{Sec-modParam}, in the vicinity of the stationary points the approach of solutions to the attractors is given by
the eigenvalues of the matrix $\mathcal{M}$. Therefore for
the critical solutions to be reached, their time-dependence must be
smaller than the eigenvalue of smallest magnitude. We will later confirm these arguments numerically for a number of model examples.

The eigenvalues of smallest magnitude for $\mathcal{VSS}_{\it{1,2,3}}$ from Eqs. (\ref{VSS1-eigen}), (\ref{VSS2-eigen}) and (\ref{VSS3-eigen}) respectively, are then \mbox{$\left|\textrm{Re}[m_{2,3}]\right|\simeq\frac{3}{2}$} (We have discounted the eigenvalue $m_4$ as it only corresponds to flows between the vector scaling solutions). These eigenvalues have to be larger than all of the expressions describing the motion of the critical points Eqs. (\ref{cp-motion-m0-VSS}), (\ref{cp-motion-mh-VSS2}) and (\ref{cp-motion-mh-VSS3}).

To directly reach the $\mathcal{VSS}_{\it3}$ critical point with varying dimensionless model parameters the $f$-term and $m$-term backreaction must be comparable at the point where they become dynamically important. This corresponds to a small area in parameter space as discussed previously.

We notice that all three vector scaling solutions in the limits $\Gamma\gg\lambda$ and $\Gamma\gg1$ and considering the connection Eq. (\ref{connetion}) have
the same $x_c$ and $y_c$ solutions, namely $x_c\simeq-\frac{2}{\Gamma}$ and $y_c\simeq1$. This is good news for us as no matter which critical point a solutions approaches we obtain an inflating universe with the desired scaling solutions for the kinetic function and mass, i.e. the ones which result in scale-invariant spectra for the perturbations of all the components of the vector field.

%---------------------------------------------------------------------------------------------------------------------
%---------------------------------------------------------------------------------------------------------------------
\subsection{The Heavy Vector Field Case}

Once we consider a heavy vector field the dimension of the problem
increases because terms in the full system Eqs.~(\ref{fullsystem})-(\ref{fullsystem-m}) proportional to $\mu$ cannot be ignored. The dynamical analysis becomes very complicated, however
we can once again reduce the dimension of the system of differential
equations by considering a certain approximation. Once the vector field
becomes heavy it starts to oscillate rapidly and its equation of
motion Eq. (\ref{veceq}) is approximated by
\begin{equation}
\ddot{A}_{z}+\frac{m^{2}}{f}A_{z}\simeq0,
\end{equation}
whose solution gives
\begin{equation}\label{vsol-heavy}
\langle\dot{A}_{z}^{2}\rangle=\frac{m^{2}}{f}\langle A_{z}^{2}\rangle,
\end{equation}
where ``$\langle\,\rangle$'' denotes average over many oscillations.
This leads to $\left\langle \rho_{\textrm{kin}}\right\rangle =\left\langle V_{A}\right\rangle $
and therefore $\left\langle z^{2}\right\rangle =\left\langle s^{2}\right\rangle $, see Fig.~\ref{fig-Heavy-zs}.
As was shown in Ref. \cite{vecurv}, when a vector field becomes heavy it oscillates rapidly
and behaves like a pressureless isotropic fluid. This is confirmed by
analysing the stationary solution determined by $\frac{\ud\Sigma}{\ud\alpha}=0$, from Eq. (\ref{fullsystem})
\begin{equation}
-3\Sigma+2\left\langle z^{2}\right\rangle -2\left\langle s^{2}\right\rangle
 +\Sigma\left(3\Sigma^{2}+3x^{2}+2\left\langle z^{2}\right\rangle +\left\langle s^{2}\right\rangle \right)=0.
\end{equation}
We can clearly see that the solution with $\Sigma_c=0$ then implies $\left\langle z^{2}\right\rangle =\left\langle s^{2}\right\rangle $ which is satisfied by an oscillating field, confirmed numerically in Figs.~\ref{fig-Heavy-zs} and \ref{fig-VSS-light-heavy-3d}.
Using this result, the full system Eqs. (\ref{fullsystem})-(\ref{fullsystem-m}) reduces
to the following 3-dimensional system
\begin{eqnarray}
\frac{\ud\Sigma}{\ud\alpha} & = & -3\Sigma+3\Sigma\left[\Sigma^{2}+x^{2}
+\frac{1}{2}\left(1-\Sigma^{2}-x^{2}-y^{2}\right)\right], \\
\frac{\ud x}{\ud\alpha} & = & -3x-\lambda y^{2}+\frac{1}{2}\left(\Gamma-\Xi\right)
\left(1-\Sigma^{2}-x^{2}-y^{2}\right)+3x\left[\Sigma^{2}+x^{2}+\frac{1}{2}\left(1-\Sigma^{2}-x^{2}-y^{2}\right)\right],\\
\frac{\ud y}{\ud\alpha} & = & \lambda xy+3y\left[\Sigma^{2}+x^{2}+\frac{1}{2}\left(1-\Sigma^{2}-x^{2}-y^{2}\right)\right]. \end{eqnarray}
The Friedmann constraint is now $\Sigma^{2}+x^{2}+y^{2}+2\left\langle z^{2}\right\rangle =1$.
There are 4 critical points for this system. These critical points include the non-hyperbolic anisotropic kination solution, $\mathcal{AKS}$ discussed in Sec. \ref{Sec-AKS}. We will therefore not consider this point again. The remaining stationary points $(x_{c},y_{c},z_{c},\Sigma_{c})$
are given below.

\begin{figure}[h]
\includegraphics[width=95mm,angle=0]{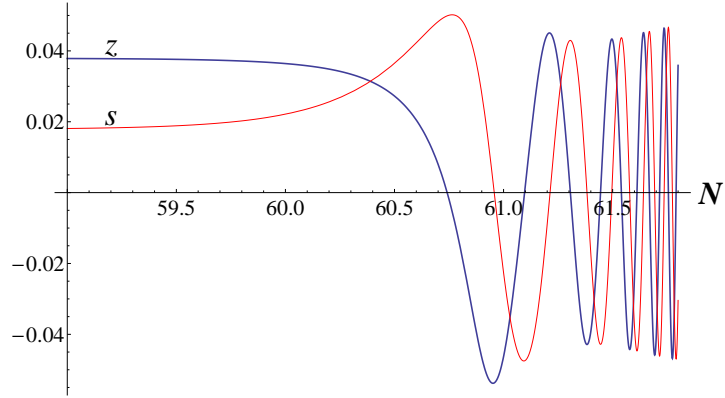}
\caption{This numerical plot shows the evolution of the dimensionless variables $z$ and $s$ as the vector field becomes heavy. The plot demonstrates the validity of our approximation $\left\langle z^{2}\right\rangle =\left\langle s^{2}\right\rangle$, used when considering a heavy vector field. }\label{fig-Heavy-zs}
\end{figure}

%---------------------------------------------------------------------------------------------------------------------
%---------------------------------------------------------------------------------------------------------------------
\subsubsection{The Standard Slow-Roll Solution, $\mathcal{SSR}$.}

The standard slow-roll critical point [c.f. Eq. (\ref{SSR-cp-mh})], is again one of the stationary solutions. It belongs to the sub-set where $z,s,\Sigma=0$. The eigenvalues of the matrix $\mathcal{M}$ for the heavy field case are now
\begin{equation}
m_{1,2}=-3+\frac{\lambda^{2}}{3}\qquad\text{and}\qquad m_{3}=-3+\frac{1}{3}\lambda\left(2\lambda+\Gamma-\Xi\right).
\end{equation}
The existence and stability of this solution requires
\begin{equation}
\lambda<3\qquad\text{and}\qquad\lambda\left(2\lambda+\Gamma-\Xi\right)<9.
\end{equation}
As previously seen, this critical point corresponds to an inflationary solution when $\lambda\ll3$. The results derived from this critical point in the massless case, Eqs. (\ref{SSR-mo-slwParam}) and (\ref{SSR-mo-eqofstate}) apply here. The only difference with the massless case are the stability conditions.
Considering the connection in Eq. (\ref{connetion}), the existence and stability conditions reduce to
\begin{equation}\label{SSR-cond-Heavy}
\lambda<3\qquad\text{and}\qquad\lambda\left(\frac43\lambda+\Gamma\right)<6.
\end{equation}

For time-dependent dimensionless model parameters, the motion of the critical point is given by Eq. (\ref{cp-motion-m0-SSR}). In analogy to Eq. (\ref{SSR-mov-cond}), the constraint on the time-dependence of the dimensionless model parameters become
\begin{eqnarray}
\mathcal{SSR}:\qquad\left|2\epsilon-\eta\right| & <
\begin{cases}
\left|m_{1,2}\right| & =\left|-3+\epsilon\right|\\
\left|m_{3}\right| & =\left|-3+2\epsilon+\frac{1}{2}\lambda\Gamma\right|
\end{cases}\label{SSR-heavy-mov-cond}
\end{eqnarray}
If the slow-roll conditions are met $\epsilon\ll1$ and $|\eta|\ll1$, and $|\lambda\Gamma-6|>\mathcal{O}(\eta)$, then the inequalities above
are satisfied. This can be achieved as long as we are not too close to the bifurcation value $\lambda\Gamma\not\approx6$. Solutions will therefore be able to reach and be dragged along with the critical point. If the dimensionless model parameters are varying more rapidly, then the validity of our
method in establishing the stability of the critical points requires the strong conditions $\lambda\ll3$ and $\lambda\Gamma\ll6$. These conditions ensures that the eigenvalues depend weakly on the dimensionless model parameters.
In this case the eigenvalues have approximately the same magnitude $|m_{1,2,3}|\simeq3$. Therefore the
inequalities above are readily satisfied if the slow-roll conditions $\epsilon\ll1$ and $\left|\eta\right|\ll1$ are met,
but they can also be satisfied even if
%the $\eta$ slow-roll condition
$\left|\eta\right|\sim1$, corresponding then to fast-roll inflation. %is violated.
%---------------------------------------------------------------------------------------------------------------------
%---------------------------------------------------------------------------------------------------------------------
\subsubsection{The Vector Kination Solution, $\mathcal{VKS}$.}

Let us investigate this new critical point given by
\begin{equation}
\mathcal{VKS}:\qquad(x_{c},y_{c},z_{c},\Sigma_{c})=
\left(\frac{1}{3}\left(\Gamma-\Xi\right)\,,\:0\,,\:\frac{1}{3\sqrt{2}}\sqrt{9-\left(\Gamma-\Xi\right)^{2}}\,,\:0\right),
\end{equation}
whose existence is valid for $\left|\Gamma-\Xi\right|<3$. The
eigenvalues of the matrix $\mathcal{M}$ are
\begin{equation}
m_{1,2}=-\frac{3}{2}+\frac{1}{6}\left(\Gamma-\Xi\right)^{2}\quad\text{and}\quad m_{3}=\frac{3}{2}+\frac{\lambda}{3}\left(\Gamma-\Xi\right)+\frac{1}{6}\left(\Gamma-\Xi\right)^{2}.
\end{equation}

Considering that the parameters $\lambda$ and $\Gamma$ have the same sign and opposite to that of $\Xi$, as discussed at the end of Sec. \ref{Sec-EqofMotion}, it is clear that $m_3$ is positive. Therefore the critical point is an unstable saddle point.

For constant or very slowly varying dimensionless model parameters the energy density is dominated by either the scalar kinetic term $x_c$ or the oscillating vector field depending on the value of $\left|\Gamma-\Xi\right|$. However for varying dimensionless model parameters we require the stronger condition $\left|\Gamma-\Xi\right|\ll3$ to ensure that the eigenvalues have a small time-dependence. This therefore guarantees the validity of our method in establishing the stability of the critical points.
In this case the critical point gives us a vector field dominated energy density where
the scalar potential and anisotropy vanishes. In any case the point
cannot correspond to an inflating Universe $y_c=0$, and is therefore of no
interest to us.

%---------------------------------------------------------------------------------------------------------------------
%---------------------------------------------------------------------------------------------------------------------
\subsubsection{The Vector Scaling Solution, $\mathcal{VSS}$.}

The vector scaling critical point, in the heavy vector field case is given by
\begin{equation}\label{VSS-Heavy-cp}
\mathcal{VSS}:\qquad(x_{c},y_{c},z_{c},\Sigma_{c})=
\left(-3\Delta_6\,\text{,}\:\Delta_6\sqrt{9+\left(\Gamma-\Xi\right)^{2}+2\,\lambda\,\left(\Gamma-\Xi\right)}\,,\:
\Delta_6\sqrt{2\,\lambda^{2}+\lambda\,\left(\Gamma-\Xi\right)-9},\,0\right),
\end{equation}
where $\Delta_6\equiv\left(2\,\lambda+\Gamma-\Xi\right)^{-1}$. We notice
that when the vector field becomes heavy and starts oscillating the
anisotropy vanishes $\Sigma_c=0$, as expected from Ref. \cite{vecurv} (see Figs.~\ref{fig-VSS-light-heavy-3d} and \ref{fig-PhaseFlow-x-y-SSR-VSS-light-heavy}). The eigenvalues of the matrix $\mathcal{M}$
are
\begin{equation}
m_{1}=-3\Delta_6\left(\lambda+\Gamma-\Xi\right)\qquad\text{and}\nonumber
\end{equation}
\begin{eqnarray}
m_{2,3} & = & -\frac{1}{2}\Delta_6\bigg\{3\left(\Gamma-\Xi\right)+3\lambda\pm\big[\lambda\left(54-16\lambda^{2}\right)\left(\Gamma-\Xi\right)
+\left(\Gamma-\Xi\right)^{2}\left(45-16\lambda^{2}\right)\nonumber\\
 &  & \qquad\qquad\qquad\qquad\quad
 -4\lambda\left(\Gamma-\Xi\right)^{3}-63\lambda^{2}+324\big]^{\frac{1}{2}}\bigg\}\label{VSS-Heavy-eigen}.
\end{eqnarray}

Let us consider the particular case where we set the connection
in Eq. (\ref{connetion}), and the limits $\Gamma\gg\lambda$ and $\Gamma\gg1$.
The vector scaling solution, Eq. (\ref{VSS-Heavy-cp}) is then approximated by
\begin{equation}\label{VSSHeavy-cp}
\mathcal{VSS}:\qquad(x_{c},y_{c},z_{c},\Sigma_{c})=
\left(-\frac{2}{\Gamma}\,\text{,}\:1\,\text{,}\:
\sqrt{\frac23}\frac{\sqrt{\lambda\Gamma-6}}{\Gamma}\,\text{,}\:0\right).
\end{equation}

We can clearly see from Eq. (\ref{VSS-Heavy-eigen}), that in the limits and connection considered, the real parts of the eigenvalues are approximated by
\begin{equation}
\textrm{Re}[m_{1}]\simeq-3\qquad\text{and}\qquad\textrm{Re}[m_{2,3}]\simeq-\frac{3}{2}
\end{equation}
Therefore the critical point is stable in this asymptotic regime. In these limits the eigenvalues depend weakly on the dimensionless model parameters. This therefore ensures the validity of our method in establishing the stability of the critical points. The sufficient but not necessary conditions for the existence and
stability are
\begin{equation}
\lambda\Gamma>6\,,\quad\Gamma\gg\lambda\quad\text{and}\quad\Gamma\gg1.
\end{equation}

In these limits the energy density is dominated by the scalar potential $y_c\approx1$, therefore we have an inflating Universe. We note
that as $\Gamma\rightarrow\infty$ the Universe becomes de-Sitter.
The energy density ratio Eq. (\ref{Ratios}) is now given by
\begin{equation}\label{VSS-Heavy-RBratios}
\mathcal{R}\simeq\frac{4\left(\lambda\Gamma-6\right)}{3\Gamma^{2}}\ll1
\end{equation}
The energy density of the vector field tracks that of the scalar field if the dimensionless model parameters are constant. Otherwise the energy density ratio is varying with time-dependence determined by that of the dimensionless model parameters. Under the limits considered the vector field energy density contribution is kept subdominant. We notice that this critical point leads to the same backreaction ratio $\mathcal{R_B}$, effective potential slope $V'_{\textrm{eff}}$, slow-roll parameters $\epsilon_{\textrm{H}}$ and $\eta_{\textrm{H}}$ and equation of state $\gamma_{\phi}$ as all of the vector scaling solutions of the massless and light vector field cases, see Eqs. (\ref{VSS-mo-RB}), (\ref{VSS-m0-slowrollparam}) and (\ref{VSS-mo-eqofstate}) respectively. We have also obtained the solutions of Eq. (\ref{scalingSol}), which lead to the desired attractor behaviour for the kinetic function and mass, given in Eq. (\ref{VSS1-fnm}) for the light vector field case. Therefore the desired attractive behaviour given by
\begin{equation}
f_{\rm att}(\alpha)\propto e^{-4\alpha}\quad\text{and}\quad m_{\rm att}(\alpha)\propto e^{\alpha},
\end{equation}
is conserved as the vector field evolves from being light to heavy (see Figs.~\ref{fig-VSS-light-heavy-3d} and \ref{fig-PhaseFlow-x-y-SSR-VSS-light-heavy}).

\begin{figure}[h]
\includegraphics[width=85mm,angle=0]{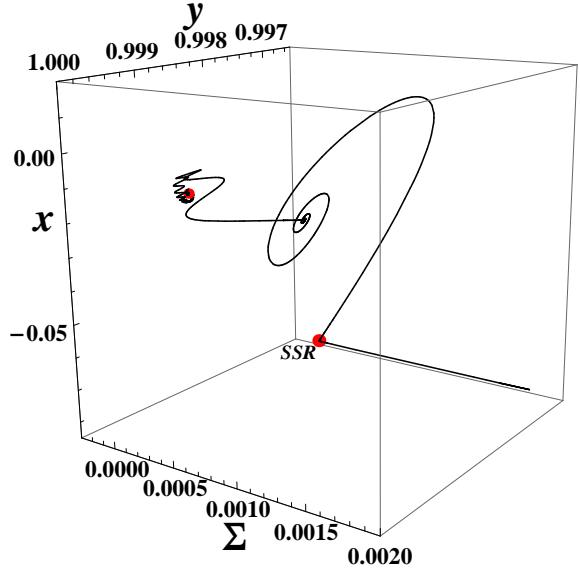}
\caption{This numerical plot shows an example of the evolution of a solution for a massive vector field. As the backreaction is initially subdominant the $\mathcal{SSR}$ critical point is reached, where the anisotropy $\Sigma$ vanishes. Solutions may remain there for a long period of time until the backreaction becomes important. Solutions will then move to the $\mathcal{VSS}$ where $\Sigma\neq0$ and the attractor solution $x_{\textrm{att}}=-2/\Gamma$ is obtained. The vector field may then become heavy. As it does so the anisotropy once again vanishes but the solution $x_{\textrm{att}}=-2/\Gamma$ is conserved. This is clearly demonstrated by the plots in Fig.~\ref{fig-PhaseFlow-x-y-SSR-VSS-light-heavy}.}\label{fig-VSS-light-heavy-3d}
\end{figure}

For time-dependent dimensionless model parameters, the motion of this critical point is given by
\[
\left|\frac{1}{x_{c}}\frac{\ud x_{c}}{\ud\alpha}\right|
=2\sqrt{6}m_{P}\left|\frac{\Gamma'}{\Gamma^{2}}+\frac{4}{3}\frac{\lambda'}{\Gamma^{2}}\right|\:,\qquad
\left|\frac{1}{y_{c}}\frac{\ud y_{c}}{\ud\alpha}\right|=\frac43\sqrt{6}m_{P}\left|\frac{\lambda'}{\Gamma^{2}}\right|\:,
\]
\begin{equation}\label{cp-motion-heavy-VSS}
\left|\frac{1}{z_{c}}\frac{\ud z_{c}}{\ud\alpha}\right|=2\sqrt{6}m_{P}\left|\frac{\Gamma'}{\Gamma^{2}}
+\frac43\frac{\lambda'}{\Gamma^{2}}
-\frac12\frac{\lambda'\Gamma+\lambda\Gamma'}{\Gamma\left(\lambda\Gamma-6\right)}\right|
\quad\text{and}\quad
\left|\frac{1}{\Sigma_{c}}\frac{\ud\Sigma_{c}}{\ud\alpha}\right|=0.
\end{equation}
For solutions to be able to reach this critical point and be dragged along with it, the eigenvalue of smallest magnitude, namely $|\textrm{Re}[m_{2,3}]|\simeq\frac{3}{2}$,
must be larger than all of the expressions above. Now if we consider a relatively flat scalar potential under the constraints in Eq.~(\ref{cond}), the conditions in Eq. (\ref{cp-motion-heavy-VSS}), can be reduced to a single condition, see Eq.~(\ref{VSS-param_cond}).

\begin{figure}[h]
\includegraphics[width=80mm,angle=0]{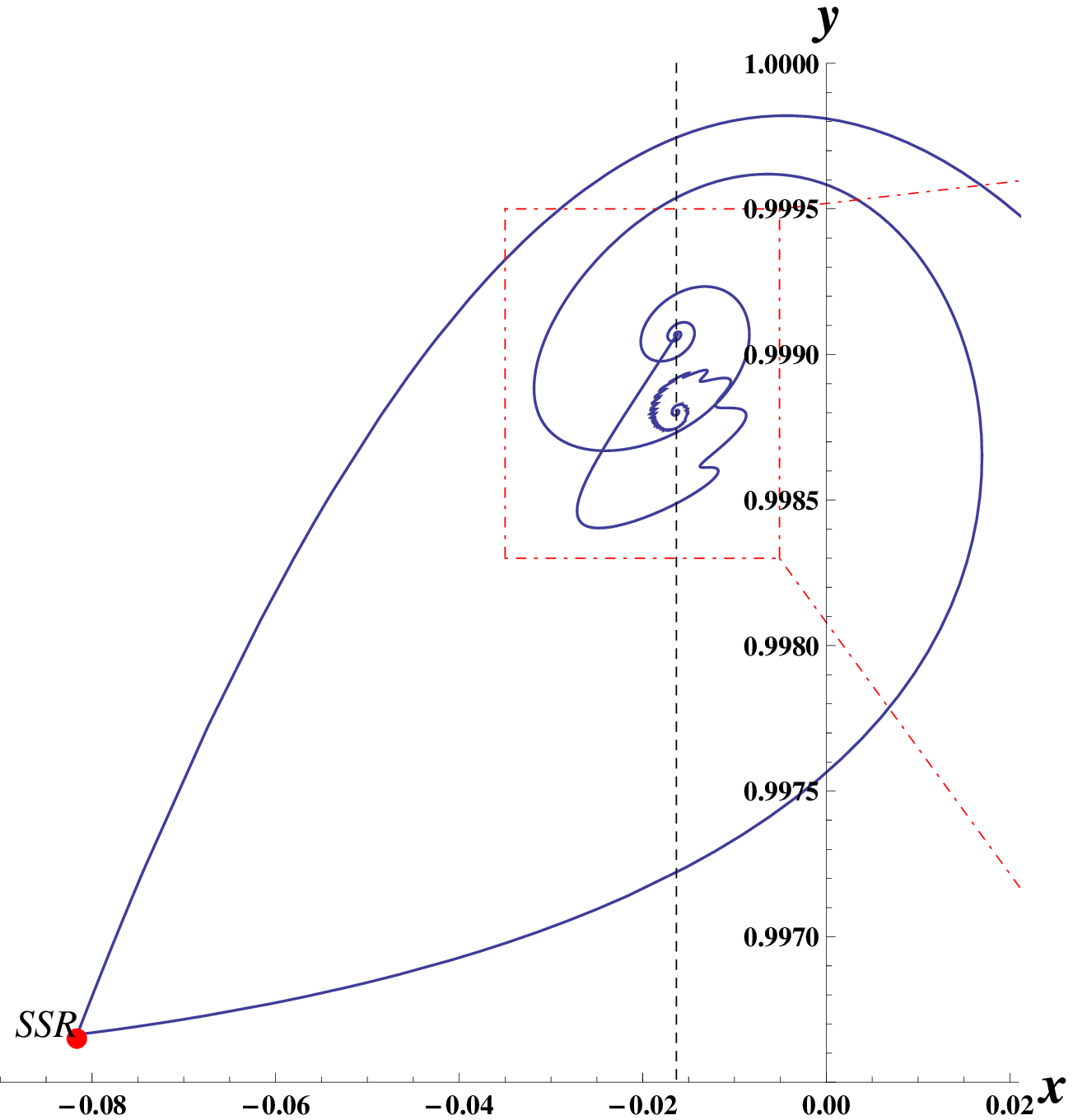}
\includegraphics[width=80mm,angle=0]{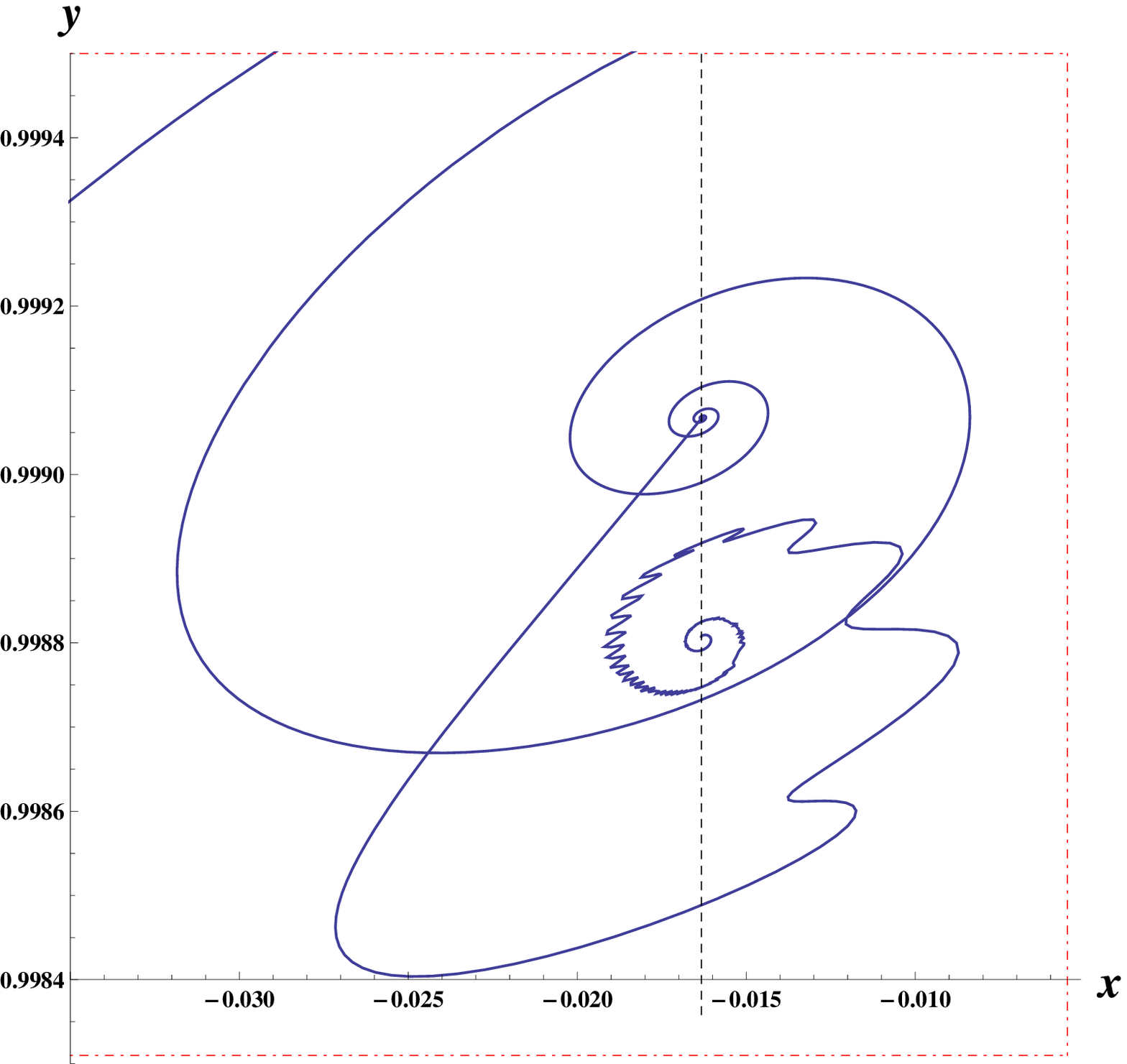}
\caption{These plots show a projection of Fig.~\ref{fig-VSS-light-heavy-3d} in the $x-y$ plane. They show in increasing detail how the attractor solution $x_{\textrm{att}}=-2/\Gamma$ is conserved as the solution jumps from the light to the heavy vector scaling solution.}\label{fig-PhaseFlow-x-y-SSR-VSS-light-heavy}
\end{figure}

%---------------------------------------------------------------------------------------------------------------------
%---------------------------------------------------------------------------------------------------------------------

\section{Summary of results}\label{summary}

In this paper we have investigated the qualitative behaviour of solutions to a system of coupled non-linear differential equations. This system of equations describe an Abelian vector field whose kinetic function and mass are modulated by the scalar field which is driving a period of inflation. The only model dependence of the analysis comes from the dimensionless model parameters $\lambda$, $\Gamma$ and $\Xi$. We have investigated a massless, light and heavy vector field separately. Generally we have found two sorts of stationary points of interest in the phase-space. These are the so-called standard slow-roll $\mathcal{SSR}$ and vector scaling solutions $\mathcal{VSS}$. We will summarize our results in the following tables.

\subsection{The Standard Slow-Roll Solutions, $\mathcal{SSR}$.}
Let us first consider the standard slow-roll critical point. This point is observed in the massless, light and heavy vector field cases, it is given by
\begin{equation}
\mathcal{SSR}:\qquad(x_{c},y_{c},z_{c},s_{c},\Sigma_{c})=
\left(-\frac{\lambda}{3},\,\sqrt{1-\left(\frac{\lambda}{3}\right)^{2}},\,
0,\,0,\,0\right).
\end{equation}

\begin{table}[h]
\begin{center}
\begin{tabular*}{0.48\textwidth}{@{\extracolsep{\fill}} c||c|c}
  \rule[-2mm]{0pt}{4ex}$\mathcal{SSR}$ & Existence$\qquad$ & Stability$\quad$ \\
  \hline
  \hline
  \rule[-2.5mm]{0pt}{5.5ex}Massless & $\lambda<3\qquad$ & $\lambda<3$\,,\quad$\lambda(\lambda+\Gamma)<6\quad$ \\

  \hline
  \rule[-2.5mm]{0pt}{5.5ex}Light & $\lambda<3\qquad$ & $\lambda<3$\,, \,$\lambda(\lambda+\Gamma)<6$\,,\,$2\lambda(\lambda-\Xi)<6\quad$ \\

  \hline
  \rule[-2.5mm]{0pt}{5.5ex}Heavy & $\lambda<3\qquad$ & $\lambda<3$\,,\quad$\lambda(2\lambda+\Gamma-\Xi)<9\quad$ \\
\end{tabular*}
\end{center}
\caption{The Standard Slow-Roll Solutions}\label{Table-SSR}
\end{table}

At this critical point the energy density and backreaction of the vector field vanish $z_c=s_c=0$. We also notice that the anisotropic stress also vanishes, $\Sigma_c=0$, providing an example of the cosmic no-hair theorem \cite{wald} (see Fig.~\ref{fig-VSS-light-heavy-3d}). The slow-roll parameters and scalar equation of state parameter are the same for all cases considered. These are given by
\begin{eqnarray}\label{SSR-gen-slwParam}
\epsilon_{\textrm{H}} & =\epsilon= & \frac{1}{3}\lambda^{2}\,,\qquad
\eta_{\textrm{H}}=\eta-\epsilon\qquad\text{and}\qquad
\gamma_{\phi}=\frac{2}{9}\lambda^{2}=\frac{2}{3}\epsilon_{\textrm{H}}.
\end{eqnarray}
At the $\mathcal{SSR}$ critical point
the energy density of the Universe is dominated by the scalar potential $y_c\approx1$ or equivalently $3m_PH^2\simeq V(\phi)$, if $\lambda\ll3$. In this case, all we require is that $\lambda\Gamma<6$ and $\lambda|\Xi|<3$ for stability of the critical point. Equivalently we may write these conditions as
\begin{equation}
\epsilon\ll3\,,\qquad m_P^2\frac{V'}{V}\frac{f'}{f}<4
\qquad\text{and}\qquad m_P^2\frac{V'}{V}\Big|\frac{m'}{m}\Big|<1.
\end{equation}
When considering varying dimensionless model parameters i.e. time-dependent critical points, an additional condition arises. This condition ensures that the approach of solutions  to the critical point is faster than their evolution. For the $\mathcal{SSR}$, the slow-roll condition $|\eta|\ll1$ ensures this strongly, see Eqs. (\ref{SSR-mov-cond}), (\ref{SSR-mh-mov-cond}) and (\ref{SSR-heavy-mov-cond}) for the massless, light and heavy vector field cases respectively. These conditions may even be satisfied if the $\eta$ slow-roll condition is violated $|\eta|\sim1$. In other words, the $\mathcal{SSR}$ can also apply to fast-roll inflation \cite{FR}.

\subsection{The Vector Scaling Solutions, $\mathcal{VSS}$.}
The vector scaling solutions $\mathcal{VSS}$ are however more interesting. The critical points are too complicated to write again here, c.f. Eqs. (\ref{VSS-m0-cp}), (\ref{VSS1-cp}), (\ref{VSS2-cp}), (\ref{VSS3-cp}) and (\ref{VSSHeavy-cp}) for the massless, light ($\mathcal{VSS}_{\it{1,2,3}}$) and heavy vector field cases respectively.
Note that, in the case of a light vector field,
$\mathcal{VSS}_{\it{3}}$ corresponds to both
the mass and the kinetic function backreacting to the inflaton variation, while
$\mathcal{VSS}_{\it{2}}$ corresponds to only the mass backreaction being
important and $\mathcal{VSS}_{\it{1}}$ corresponds to only the kinetic function
backreaction being important. As such the latter is equivalent to the
$\mathcal{VSS}$ point of the massless vector field.
Throughout the results described below, we have used the connection from Eq. (\ref{connetion}). This connection may be written as $f'/f=-4m'/m$.

When considering the limits $\Gamma\gg\lambda$ and $\Gamma\gg1$, the energy density of the Universe is dominated by the scalar potential for all the vector scaling solutions, $y_c\simeq1$ or equivalently $3m_PH^2\simeq V(\phi)$. These limits correspond to inflationary solutions with small but non-vanishing anisotropy $\Sigma_c\ll1$, providing counter examples to the cosmic no-hair theorem \cite{wald}. The anisotropy does however vanish if the vector field becomes heavy (see Fig.~\ref{fig-VSS-light-heavy-3d}). It turns out that these limits also guarantee the stability of the critical point. Table \ref{Table-VSS} summarises our results for the vector scaling solutions
\begin{table}[h]
\begin{center}
\begin{tabular*}{0.62\textwidth}{@{\extracolsep{\fill}} c||c|c|c|c}
  \rule[-2mm]{0pt}{4ex}$\mathcal{VSS}$ & Existence$\quad$ & Stability$\quad$ & $\mathcal{R}\quad$ & $\Sigma_c\quad$ \\
  \hline
  \hline
  \rule[-2.5mm]{0pt}{5.5ex}Massless & $\lambda\Gamma>6\quad$ & $\Gamma\gg\lambda,\,\Gamma\gg1\quad$ & \large $\frac{\lambda\Gamma-6}{\Gamma^2}\quad$ & $\frac23\mathcal{R}\quad$ \\
  \hline
  \rule[-2.5mm]{0pt}{5.5ex}Light: $\mathcal{VSS}_{\it{1}}$ & $\lambda\Gamma>6\quad$ & $\Gamma\gg\lambda,\,\Gamma\gg1\quad$ & \large$\frac{\lambda\Gamma-6}{\Gamma^2}\quad$ & $\frac23\mathcal{R}\quad$ \\
  \hline
  \rule[-2.5mm]{0pt}{5.5ex}Light: $\mathcal{VSS}_{\it{2}}$ & $\lambda\Gamma>6\quad$ & $\Gamma\gg\lambda,\,\Gamma\gg1\quad$ & 2\large$\left(\frac{\lambda\Gamma-6}{\Gamma^2}\right)\quad$  & $-\frac23\mathcal{R}\quad$ \\
  \hline
  \rule[-2.5mm]{0pt}{5.5ex}Light: $\mathcal{VSS}_{\it{3}}$ & $\lambda\Gamma>8\quad$ & $\Gamma\gg\lambda,\,\Gamma\gg1\quad$ & \large$\frac{3\lambda\Gamma-16}{2\Gamma^2}\quad$ & $-\frac{\lambda}{3\Gamma}\quad$ \\
  \hline
  \rule[-2.5mm]{0pt}{5.5ex}Heavy & $\lambda\Gamma>6\quad$ & $\Gamma\gg\lambda,\,\Gamma\gg1\quad$ & \large$\frac43\left(\frac{\lambda\Gamma-6}{\Gamma^2}\right)\quad$ & $0\quad$ \\

\end{tabular*}
\end{center}
\caption{The Vector Scaling Solutions}\label{Table-VSS}
\end{table}

It is important to note that the stability conditions for the $\mathcal{VSS}$
critical points in Table~\ref{Table-VSS} correspond only to three out of the
four eigenvalues of the linearised system. The fourth eigenvalue is tiny and
its sign determines flows between the $\mathcal{VSS}$ critical points (see
Fig.~\ref{fig-VSS-1-2-3}). Indeed, as explained at the end of Sec.~\ref{Sec-mh-VSS3}, if
$\lambda\Gamma>8$, the
$\mathcal{VSS}_{\it{3}}$ is completely stable i.e. all eigenvalues are negative
in Eq. (\ref{VSS3-eigen}), and therefore $\mathcal{VSS}_{\it3}$ is the absolute late-time attractor, whereas $\mathcal{VSS}_{\it{1,2}}$ are quasi-stable. They have three negative
eigenvalues and one very small but positive eigenvalue, see
Eqs. (\ref{VSS1-eigen}) and (\ref{VSS2-eigen}). However if $6<\lambda\Gamma<8$,
then $\mathcal{VSS}_{\it{3}}$ does not exist, but $\mathcal{VSS}_{\it{2}}$
becomes completely stable, i.e. the late-time attractor, see
Eq. (\ref{VSS2-eigen}). Therefore, as long as $\lambda\Gamma>6$, there is
always a completely stable $\mathcal{VSS}$ critical point that solutions will
eventually converge to.

The backreaction, from Eqs. (\ref{Ratios}) and (\ref{Veff01}) leads to the same effective slope for all cases considered
\begin{equation}\label{RatiosBR-gen}
\mathcal{R_{B}}\simeq\frac{6-\lambda\Gamma}{\lambda\Gamma}\qquad\text{and}\qquad
V'_{\textrm{eff}}=\frac{6}{\lambda\Gamma}V'.
\end{equation}
The existence conditions means that the effective slope seen by the scalar field becomes flatter, and so the field slows down as it evolves along its potential.
The slow-roll parameters and scalar equation of state parameter are also the same for all cases considered, and are given by
\begin{eqnarray}\label{VSS-gen-slwrollparam}
\epsilon_{\textrm{H}} \simeq \frac{2\lambda}{\Gamma}\,,\qquad
\eta_{\textrm{H}}\simeq\frac{2\lambda}{\Gamma}
+\frac{\sqrt{6}m_{P}}{\Gamma}\left(\frac{\lambda'}{\lambda}-\frac{\Gamma'}{\Gamma}\right)
\qquad\text{and}\qquad
\gamma_{\phi}\simeq \frac{8}{\Gamma^2}.
\end{eqnarray}

All the vector scaling solutions for the massless, light and heavy vector field cases, in the limits considered, have the same solution
namely $x_c=-\Gamma/2$ (see Figs.~\ref{fig-VSS-light-heavy-3d} and \ref{fig-PhaseFlow-x-y-SSR-VSS-light-heavy}). As seen from Eq. (\ref{scalingSol}), this solution together with the connection considered Eq. (\ref{connetion}), lead to the desired attractor behaviour for the kinetic function and mass
\begin{equation}
f_{\rm att}(\alpha)\propto e^{-4\alpha}\quad\text{and}\quad m_{\rm att}(\alpha)\propto e^{\alpha}.
\end{equation}
This particular scaling for the kinetic function enables the generation of a
scale-invariant perturbation spectrum for the transverse component of the
vector field.
The additional scaling for the mass further enables the generation of a
scale-invariant perturbation spectrum for the longitudinal component of the
vector field, as shown in Ref.~\cite{varkin}.
Therefore by imposing this connection,
we obtain the scaling of the kinetic function and mass which generates
scale invariant spectra for the perturbations of the vector field.
%which can contribute to the curvature perturbation.
In general, we have shown that the attractor behaviour is not destabilised by
the mass backreaction and it can be such that leads to scale invariant spectra
for all vector field components.

Except for the exponential functional forms of the scalar potential,
kinetic function and mass, where the dimensionless model parameters are constants,
we generally find time-dependent critical points. We then require additional
conditions which ensure that the approach of solutions to the critical
point is faster than their evolution, see Eqs. (\ref{cp-motion-m0-VSS}), (\ref{cp-motion-mh-VSS2}), (\ref{cp-motion-mh-VSS3}) and (\ref{cp-motion-heavy-VSS}) for the massless $\mathcal{VSS}$ and light $\mathcal{VSS}_{\it1}$, light $\mathcal{VSS}_{\it{2,3}}$ and heavy $\mathcal{VSS}$ respectively. Now if we consider a relatively flat inflaton potential $\sqrt{\frac23}m_P|\lambda'|=|\eta-2\epsilon|\lesssim\mathcal{O}(1)$ and if $\lambda\Gamma\not\approx6$ (or $\lambda\Gamma\not\approx8$ for $\mathcal{VSS}_{\it3}$)  or $\lambda'\Gamma\approx-\lambda\Gamma'$, we find that all the conditions can be reduced to a single (strong) condition,
\begin{equation}\label{VSS-param_cond-gen}
m_P\left|\frac{\Gamma'}{\Gamma^2}\right|=\left|\frac{ff''}{f'^2}-1\right|\ll1.
\end{equation}
As expected, if the kinetic function is exponential $f\propto e^{c\phi/m_P}$,
where $c$ is some constant, then the expression above is exactly zero. This is because only exponential functions give constant dimensionless model parameters.
Potentials and kinetic functions obeying $\Gamma\propto 1/\lambda$ result in
$\lambda'\Gamma=-\lambda\Gamma'$. These have been considered in
Refs.~\cite{anisinf,anisinf+} where
%$f\propto e^{\frac{c}{m_P^2}\int\frac{V}{V'}\ud\phi}$
\mbox{$f\propto\exp\left(cm_P^{-2}\int\frac{V}{V'}\ud\phi\right)$}
and satisfy the conditions in Eq.~(\ref{cond}).
%---------------------------------------------------------------------------------------------------------------------
%---------------------------------------------------------------------------------------------------------------------
\section{Specific Model Examples}\label{examples}

We now look at specific model examples choosing the functional forms of
the scalar potential and the vector kinetic function and mass. We obtain
constraints on the dimensionless model parameters and their time-dependence for the different
attractor solutions of interest, namely the standard slow-roll and the vector
scaling solutions. We carry out the analysis for the different cases
separately, i.e. massless, light and heavy vector field cases.

%---------------------------------------------------------------------------------------------------------------------
%---------------------------------------------------------------------------------------------------------------------
\subsection{Constant Dimensionless Model Parameters}

In string theory %the modulation of
parameters such as the masses or kinetic functions of vector boson fields
are determined by the compactification scheme, i.e. their values depend on
so-called moduli fields. The moduli are scalar fields which parameterize the
size and shape of extra dimensions. In that sense they are not fundamental
scalar fields, but appear so from the viewpoint of the 4-dimensional observer.
Typically the dependence of masses and couplings on canonically normalised
(i.e. with canonical kinetic terms) moduli fields is exponential.
And exponential potentials for the moduli fields are also reasonable. One such
modulus can play the role of the inflaton field.

Let us assume the following functional forms for the scalar potential and the
vector field kinetic function
\begin{equation}
V(\phi)=V_{0}e^{\frac{q\phi}{m_{P}}}\qquad\text{and}\qquad f(\phi)=f_{i}e^{\frac{4c}{q m_{P}}\left(\phi-\phi_{i}\right)},
\label{Vf1}
\end{equation}
where $c$ and $q$ are real positive constants.
Our dimensionless model parameters in Eq. (\ref{model_param}) are given by $\lambda=\sqrt{\frac{3}{2}}q$ and $\Gamma=4\sqrt{\frac{3}{2}}\frac{c}{q}$, and the slow-roll parameters defined in Eq. (\ref{slow-roll-param}) are $\epsilon=\frac{1}{2}\eta=\frac{1}{2}q^{2}$.

When we come to consider a massive vector field we need to give a functional form for the mass $m(\phi)$. Considering the connection Eq. (\ref{connetion}), the model parameter $\Xi$ now becomes
$\Xi=-2\sqrt{\frac{3}{2}}\frac{c}{q}$. Equivalently we consider the function
\begin{equation}
m(\phi)=m_{i}e^{-\frac{c}{q m_{P}}\left(\phi-\phi_{i}\right)}.
\label{ex1m}
\end{equation}
Constant parameters $\lambda$ and $\Gamma$ are only achieved by
exponential functions of the form shown in Eqs.~(\ref{Vf1}) and (\ref{ex1m}).

The results for the standard slow-roll solution in this model are given in Table \ref{Table-SSR-Ex1}.
\begin{table}[h]
\begin{center}
\begin{tabular*}{0.46\textwidth}{@{\extracolsep{\fill}} c||c|c}
  \rule[-2mm]{0pt}{4ex}$\mathcal{SSR}$ & Existence$\qquad$ & Stability$\qquad$ \\
  \hline
  \hline
  \rule[-2.5mm]{0pt}{5.5ex}Massless & $q<\sqrt6\qquad$ & $q<2\sqrt{1-c}\qquad$ \\

  \hline
  \rule[-2.5mm]{0pt}{5.5ex}Light & "\qquad & $q<\sqrt{2(1-c)}\qquad$  \\

  \hline
  \rule[-2.5mm]{0pt}{5.5ex}Heavy & "\qquad & $q<\sqrt{3(1-c)}\qquad$  \\
\end{tabular*}
\end{center}
\caption{Constant dimensionless model parameters}\label{Table-SSR-Ex1}
\end{table}

Clearly we require that $c<1$ for the stability conditions to be well defined. We also find the following generic results for the standard slow-roll solutions, from Eq. (\ref{SSR-gen-slwParam})
\begin{equation}
\epsilon_{\textrm{H}} =\epsilon\,,\qquad
\eta_{\textrm{H}}=\frac12\eta\qquad\text{and}\qquad
\gamma_{\phi}=\frac{2}{3}\epsilon.
\end{equation}

At the $\mathcal{SSR}$ critical point the vector field energy density and backreaction vanishes $z_c=s_c=0$. The energy density of the Universe is dominated by the scalar potential $y_c\approx1$, if $q\ll\sqrt{6}$, corresponding then to a period of inflation.\footnote{Inflation with an exponential potential is
power-law and it takes place for
\mbox{$q<\sqrt 2\Leftrightarrow\gamma_\phi<2/3$}. One can consider that
inflation is approximately (quasi)de Sitter only if
\mbox{$q\ll 1\Leftrightarrow\gamma_\phi\rightarrow 0$}. Power-law anisotropic inflation has been recently considered in Ref.~\cite{sodanew}.}
In this case the slow-roll conditions are satisfied $\epsilon_{\textrm{H}}=\eta_{\textrm{H}}\ll1$. Therefore slow-roll inflation is the late-time attractor. The slow-roll conditions guarantee the stability of the critical point as long as $c\not\approx1$. In this model however, once the solution reaches the attractor, inflation would never end. This is because the slow-roll
parameters are constant and the slow-roll conditions would never be violated. One would then have to introduce some kind of
hybrid mechanism or modification of the potential to end the phase
of inflation, i.e. to destabilise the critical point.

The results for the vector scaling solution in this model are given in Table \ref{Table-VSS-Ex1}.
\begin{table}[h]
\begin{center}
\begin{tabular*}{0.62\textwidth}{@{\extracolsep{\fill}} c||c|c|c|c}
  \rule[-2mm]{0pt}{4ex}$\mathcal{VSS}$ & Existence & Stability & $\mathcal{R}$ & $\Sigma_c$\\
  \hline
  \hline
  \rule[-2.5mm]{0pt}{5.5ex}Massless & $c>1$ & $q\ll2\sqrt{c},\,q\ll2\sqrt{6}c$ & $\frac{c-1}{4c^{2}}q^{2}$ & $\frac23\mathcal{R}$   \\

  \hline
  \rule[-2.5mm]{0pt}{5.5ex}Light: $\mathcal{VSS}_{\it{1}}$ & " & " & " & "  \\

  \hline
  \rule[-2.5mm]{0pt}{5.5ex}Light: $\mathcal{VSS}_{\it{2}}$ & " & " & $\frac{c-1}{2c^{2}}q^{2}$ & $-\frac23\mathcal{R}$  \\

  \hline
  \rule[-2.5mm]{0pt}{5.5ex}Light: $\mathcal{VSS}_{\it{3}}$ & $c>\frac43$ & " & $\frac{9c-8}{24c^{2}}q^{2}$ & \large$-\frac{1}{6c}\epsilon$  \\

  \hline
  \rule[-2.5mm]{0pt}{5.5ex}Heavy & $c>1$ & " & $\frac{c-1}{3c^{2}}q^{2}$ & $0$ \\

\end{tabular*}
\end{center}
\caption{Constant dimensionless model parameters}\label{Table-VSS-Ex1}
\end{table}

We also find the following generic results for the vector scaling solutions, from Eq. (\ref{VSS-gen-slwrollparam})
\begin{eqnarray}
\epsilon_{\textrm{H}} =\frac1c\epsilon\,,\qquad
\eta_{\textrm{H}}=\frac{1}{2c}\eta\qquad\text{and}\qquad
\gamma_{\phi}=\frac{2}{3c^2}\epsilon.
\end{eqnarray}

When $c>1$ the energy density of the vector field and the backreaction grows. Therefore solutions will be driven towards the vector scaling attractors. At the critical point the energy density of the vector field tracks that of the scalar field at a constant ratio $\mathcal{R}$.
However inflation is not spoilt as the vector field energy
density remains sub-dominant. The sufficient but not necessary stability conditions in Table \ref{Table-VSS} guarantee that the energy density of the Universe is dominated by the scalar potential, $y_c\approx1$.

The backreaction, due to the vector field, becomes proportional to the scalar slope $V'(\phi)$ and a negative contribution. The backreaction and the effective slope seen by the scalar field, given by Eq. (\ref{RatiosBR-gen}) is then
\begin{equation}\label{VBratioForm01}
  \mathcal{R_{B}}=-\frac{c-1}{c} \qquad\text{and}\qquad
  V'_{\textrm{eff}}= V'+\mathcal{B}_A=\frac{1}{c}V',
\end{equation}
hence the effective slope becomes flatter and the field slows down when the backreaction takes effect.
This effect can also be seen from the slow-roll conditions $\epsilon_{\textrm{H}}=\frac{1}{c}\epsilon=\eta_{\textrm{H}}=\frac{1}{2c}\eta\ll1$,
which we observe are smaller than that of the $\mathcal{SSR}$. The $\mathcal{VSS}$ therefore corresponds to a new stage of inflation with non-vanishing anisotropy $\Sigma$. The anisotropy becomes constant at the critical point but the stability conditions guarantee that it remains small $\Sigma_{c}\ll1$. This is key to not spoiling observational requirements. The anisotropy however vanishes if the vector field becomes heavy.

We also notice that for $c>1$ the $\mathcal{SSR}$ is an unstable saddle point. It is unstable because the backreaction grows and will eventually effect the dynamics of the system. Therefore the solution can move from the $\mathcal{SSR}$ to the $\mathcal{VSS}$ and in the process the scalar field would slow down as it rolls down its potential. The dynamical evolution of the model for a variety of initial conditions and choices of the dimensionless model parameters is shown in Fig.~\ref{fig-PowerLaw}.

\begin{figure}[h]
\includegraphics[width=85mm,angle=0]{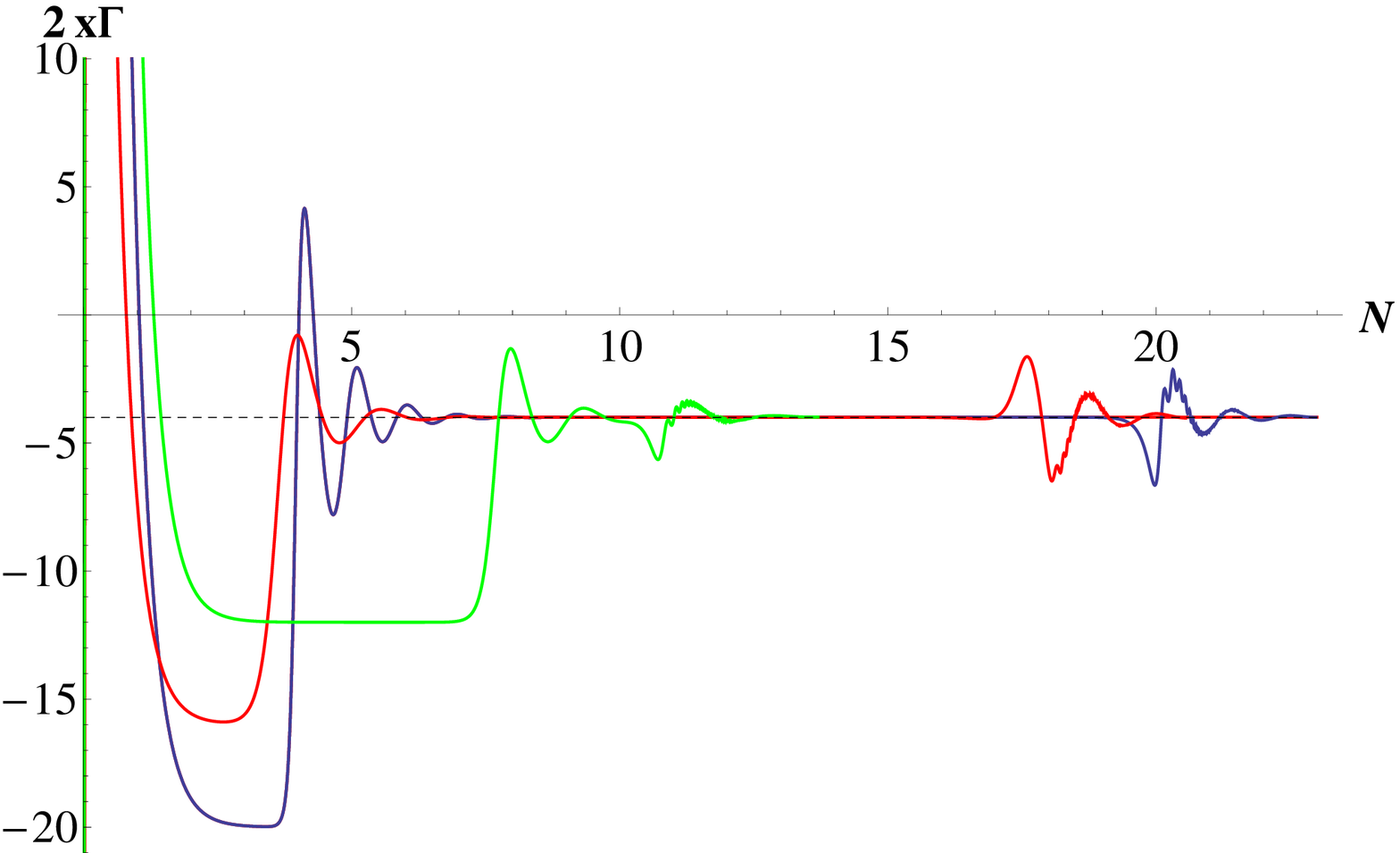}
\includegraphics[width=85mm,angle=0]{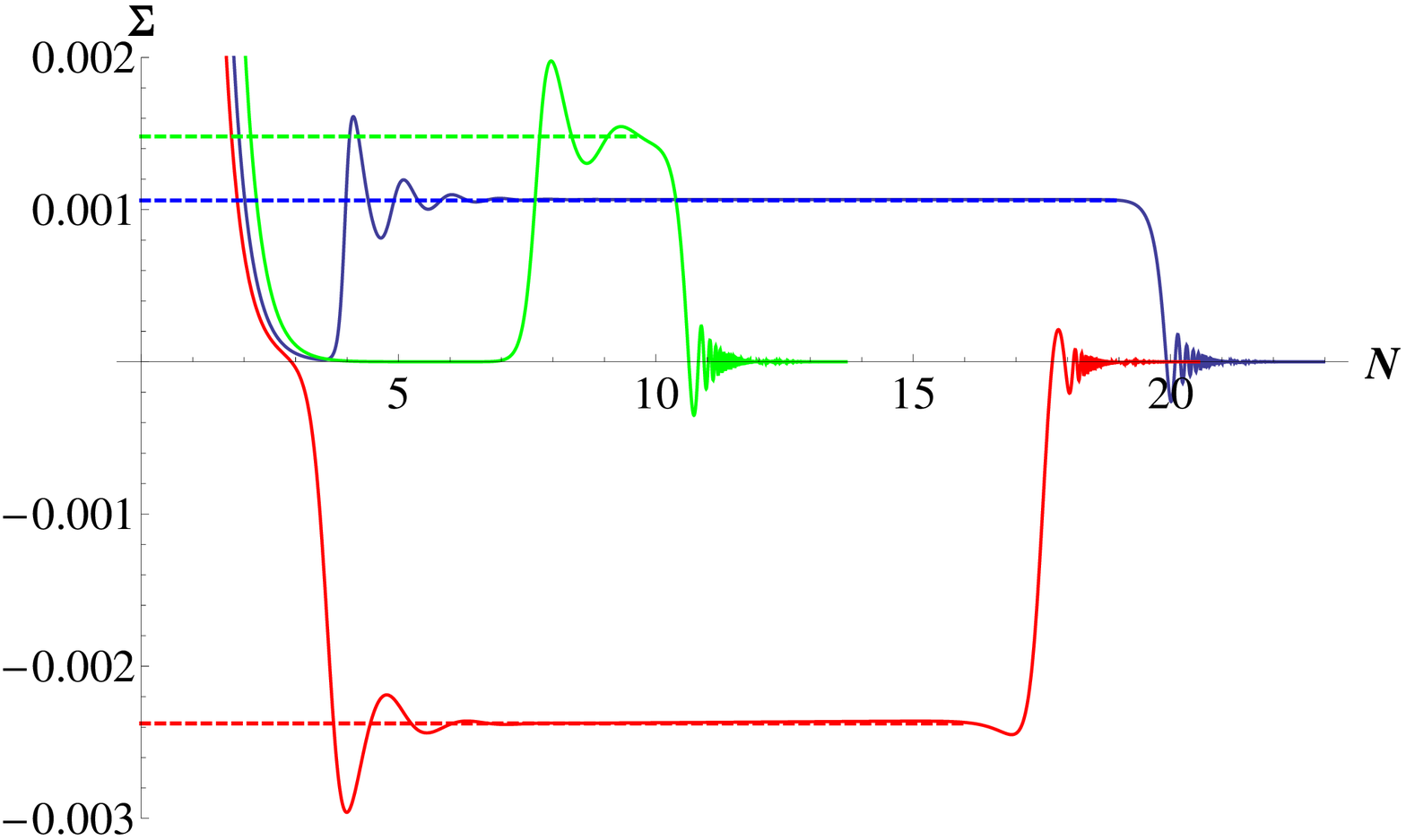}
\caption{These numerical plots show the evolution of the kinetic function scaling $\frac{1}{f}\frac{\ud f}{\ud \alpha}=2x\Gamma$, given by Eq. (\ref{scalingSol1}), and anisotropy $\Sigma$, with respect to the number of elapsing e-folds $N$, for a massive vector field, for the constant dimensionless model parameters example. The plots show numerical solutions for three different values of the dimensionless model parameters and initial conditions. We observe that the $\mathcal{SSR}$ is obtained prior to the $\mathcal{VSS}$ for certain initial conditions. At the $\mathcal{SSR}$ the kinetic function takes on a variety of scalings given by different values of $2x\Gamma$. All solutions move then to the $\mathcal{VSS}$ where the same attractor $2x\Gamma=-4$ is obtained, giving $f_{\textrm{att}}\propto a^{-4}$ and $m_{\textrm{att}}\propto a$, when considering the connection in Eq.~(\ref{connetion}). The plots clearly show that the same attractor $2x\Gamma=-4$ is also obtained when the vector field becomes heavy. As expected the anisotropy $\Sigma$ vanishes at the $\mathcal{SSR}$ but not at the $\mathcal{VSS}$ if the vector field remains light. The anisotropy does however vanish once again when the vector field becomes heavy. We also observe that $\Sigma$ can be positive or negative depending on which $\mathcal{VSS}$ the solution approaches nearest, see Table \ref{Table-VSS}. The plots of $\Sigma$ also show (in dashed lines) the analytic solutions obtained for the $\mathcal{VSS}$ attractors, see Table \ref{Table-VSS}. We can see how well they agree with the full numerical solutions.}\label{fig-PowerLaw}
\end{figure}

%---------------------------------------------------------------------------------------------------------------------
%---------------------------------------------------------------------------------------------------------------------
\subsection{The Chaotic Potential}

The archetypal example for an inflationary potential is the chaotic potential. In Ref.~\cite{anisinf} the authors considered this potential together with the following
kinetic function,
\begin{equation}
V(\phi)=\frac{1}{2}m_{\phi}^{2}\phi^{2}\qquad\text{and}\qquad f(\phi)=f_{i}e^{\frac{c}{m_{P}^{2}}\left(\phi^{2}-\phi_{i}^{2}\right)}.
\end{equation}
The dimensionless model parameters Eq. (\ref{model_param}) are given by $\lambda(\phi)=2\sqrt{\frac{3}{2}}\left(\frac{m_{P}}{\phi}\right)$
and $\Gamma(\phi)=2c\sqrt{\frac{3}{2}}\left(\frac{\phi}{m_{P}}\right)$
where $m_{\phi}$ is the mass of the scalar field and $c$ is a real positive constant. The slow-roll parameters defined in Eq. (\ref{slow-roll-param}) are $\epsilon=\eta=2\left(\frac{m_{P}}{\phi}\right)^{2}$.

When we come to consider a massive vector field we need to give a functional form for the mass $m(\phi)$. Considering the connection Eq. (\ref{connetion}), the model parameter now becomes $\Xi=-c\sqrt{\frac{3}{2}}\left(\frac{\phi}{m_{P}}\right)$. Equivalently we consider the function
\begin{equation}
m(\phi)=m_{i}e^{-\frac{c}{4m_{P}^{2}}\left(\phi^{2}-\phi_{i}^{2}\right)}.
\end{equation}

The results for the standard slow-roll solution in this model are given in Table \ref{Table-SSR-Ex2}.
\begin{table}[h]
\begin{center}
\begin{tabular*}{0.48\textwidth}{@{\extracolsep{\fill}} c||c|c}
  \rule[-2mm]{0pt}{4ex}$\mathcal{SSR}$ & Existence$\qquad$ & Stability$\qquad$ \\
  \hline
  \hline
  \rule[-2.5mm]{0pt}{5.5ex}Massless & $\left(\frac{\phi}{m_{P}}\right)^{2}>\frac{2}{3}\qquad$ &
$\left(\frac{\phi}{m_{P}}\right)^{2}>\frac{1}{1-c}\qquad$ \\

  \hline
  \rule[-2.5mm]{0pt}{5.5ex}Light & "$\qquad$ & $\left(\frac{\phi}{m_{P}}\right)^{2}>\frac{2}{1-c}\qquad$ \\

  \hline
  \rule[-2.5mm]{0pt}{5.5ex}Heavy & "$\qquad$ & $\left(\frac{\phi}{m_{P}}\right)^{2}>\frac{4}{3(1-c)}\qquad$ \\
\end{tabular*}
\end{center}
\caption{Chaotic Potential}\label{Table-SSR-Ex2}
\end{table}

Clearly we require that $c<1$ for the stability conditions to be well defined. We also find the following generic results for the standard slow-roll solutions, from Eq. (\ref{SSR-gen-slwParam})
\begin{equation}
\epsilon_{\textrm{H}} =\epsilon\,,\qquad
\eta_{\textrm{H}}=0 \qquad\text{and}\qquad
\gamma_{\phi}=\frac{2}{3}\epsilon.
\end{equation}
The energy density of the Universe is dominated by the scalar potential $y_c\approx1$,
if $\left(\frac{\phi}{m_{P}}\right)^{2}\gg\frac{2}{3}$, corresponding
then to a period of inflation. In this case, the slow-roll conditions $\epsilon_{\textrm{H}}\ll1$ and
$\eta_{\textrm{H}}=0$ are satisfied. Therefore slow-roll inflation is the late-time attractor.
It is well known that the above potential satisfies the slow-roll
conditions in the large field regime. The slow-roll conditions guarantee the stability of the critical point as long as $c\not\approx1$. The period of inflation ends when either the existence or stability conditions are violated, i.e. when the slow-roll condition is violated.

As the dimensionless model parameters are varying in this example, the critical point
is therefore moving in phase-space, and so we have the additional
constraints coming from Eqs. (\ref{SSR-mov-cond}), (\ref{SSR-mh-mov-cond}) and (\ref{SSR-heavy-mov-cond}) for the massless, light and heavy vector field cases respectively. Considering the slow-roll limit $\epsilon\ll1$, the conditions that must be satisfied are respectively
\begin{equation}
\left(\frac{\phi}{m_{P}}\right)^{2}>\frac{1}{2(1-c)}\,,\qquad
\left(\frac{\phi}{m_{P}}\right)^{2}>\frac{2}{1-c}\qquad\text{and}\qquad
\left(\frac{\phi}{m_{P}}\right)^{2}>\frac{2}{3(1-c)}.
\end{equation}
These constraints are clearly satisfied if the critical point is stable. Therefore the critical point evolves at a slower rate than
the approach of the solution to the point. The attractor solution
can then be obtained and solutions are dragged along with the motion of the critical point.

The results for the vector scaling solution in this model are given in Table \ref{Table-VSS-Ex2}.
\begin{table}[h]
\begin{center}
\begin{tabular*}{0.7\textwidth}{@{\extracolsep{\fill}} c||c|c|c|c}
  \rule[-2mm]{0pt}{4ex}$\mathcal{VSS}$ & Existence & Stability & $\mathcal{R}$ & $\Sigma_c$ \\
  \hline
  \hline
  \rule[-2.5mm]{0pt}{5.5ex}Massless & $c>1$ & $\left(\frac{\phi}{m_{P}}\right)^{2}\gg\frac{1}{c}
,\,\left(\frac{\phi}{m_{P}}\right)^{2}\gg\frac{1}{6c^2}$ & $\frac{c-1}{c^{2}}\left(\frac{m_{P}}{\phi}\right)^{2}$ & $\frac23\mathcal{R}$ \\

  \hline
  \rule[-2.5mm]{0pt}{5.5ex}Light: $\mathcal{VSS}_{\it{1}}$ & " & " & " & " \\

  \hline
  \rule[-2.5mm]{0pt}{5.5ex}Light: $\mathcal{VSS}_{\it{2}}$ & " & " & $2\frac{c-1}{c^{2}}\left(\frac{m_{P}}{\phi}\right)^{2}$ & $-\frac23\mathcal{R}$ \\

  \hline
  \rule[-2.5mm]{0pt}{5.5ex}Light: $\mathcal{VSS}_{\it{3}}$ & $c>\frac43$ & " & $\frac{9c-8}{6c^{2}}\left(\frac{m_{P}}{\phi}\right)^{2}$ & $-\frac{1}{6c}\epsilon$ \\

  \hline
  \rule[-2.5mm]{0pt}{5.5ex}Heavy & $c>1$ & " & $\frac43\frac{c-1}{c^{2}}\left(\frac{m_{P}}{\phi}\right)^{2}$ & $0$  \\

\end{tabular*}
\end{center}
\caption{Chaotic Potential}\label{Table-VSS-Ex2}
\end{table}

We also find the following generic results for the vector scaling solutions, from Eq. (\ref{VSS-gen-slwrollparam})
\begin{eqnarray}
\epsilon_{\textrm{H}} =\frac1c\epsilon\,,\qquad
\eta_{\textrm{H}}=0 \qquad\text{and}\qquad
\gamma_{\phi}=\frac{2}{3c^2}\epsilon.
\end{eqnarray}

The stability conditions guarantee that the scalar potential dominates
the energy density $y_c\approx1$. Inflation is not spoilt
as the vector field energy density is sub-dominant, $\mathcal{R}\ll1$ for all cases. However in this model $\mathcal{R}$ and $\Sigma_c$ are not exactly constant but are slowly growing as the scalar field rolls down its potential. The backreaction and the effective slope seen by the scalar field, given by Eq. (\ref{RatiosBR-gen}) is then
\begin{equation}
  \mathcal{R_{B}}=-\frac{c-1}{c} \qquad\text{and}\qquad
  V'_{\textrm{eff}}= V'+\mathcal{B}_A=\frac{1}{c}V'.
\end{equation}
The backreaction becomes
proportional to the scalar slope $V'(\phi)$ and a negative contribution. Therefore the effective slope
as seen by the scalar field becomes flatter and the field slows down. This effect can be seen from the slow-roll
conditions $\epsilon_{\textrm{H}}=\frac1c\epsilon=\frac1c\eta\ll1$,
which are smaller than that of the $\mathcal{SSR}$. The $\mathcal{VSS}$ corresponds again to a new stage of slow-roll inflation with non-vanishing anisotropy. The anisotropy remains small in the large field regime as to not spoil observational
requirements. The anisotropy will however vanish if the vector field becomes heavy. We find that all the results in the massless vector field case are in agreement with the findings of Ref. \cite{anisinf}.

We also notice that in the case where $c>1$ the $\mathcal{SSR}$ is an unstable saddle point. As the previous model the vector field energy density and backreaction are growing and will eventually effect the scalar dynamics.
If the solution moves from the $\mathcal{SSR}$ to the $\mathcal{VSS}$
the scalar field would slow down as it rolls down its potential.

As the dimensionless model parameters are varying in this example, the critical point
is therefore moving in phase-space, and so we have the additional
constraints coming from Eqs. (\ref{cp-motion-m0-VSS}), (\ref{cp-motion-mh-VSS2}), (\ref{cp-motion-mh-VSS3}) and (\ref{cp-motion-heavy-VSS}) for the massless, light ($\mathcal{VSS}_{\it{1,2,3}}$) and heavy vector field cases respectively. All vector scaling solutions have the same eigenvalue of smallest magnitude, namely $|\textrm{Re}[m_{2,3}]|\simeq\frac32$, where we have neglected the small eigenvalues $m_4$ because these correspond to flows between the scaling solutions as described at the end of Sec. \ref{Sec-mh-VSS3}.

\begin{figure}[h]
\includegraphics[width=85mm,angle=0]{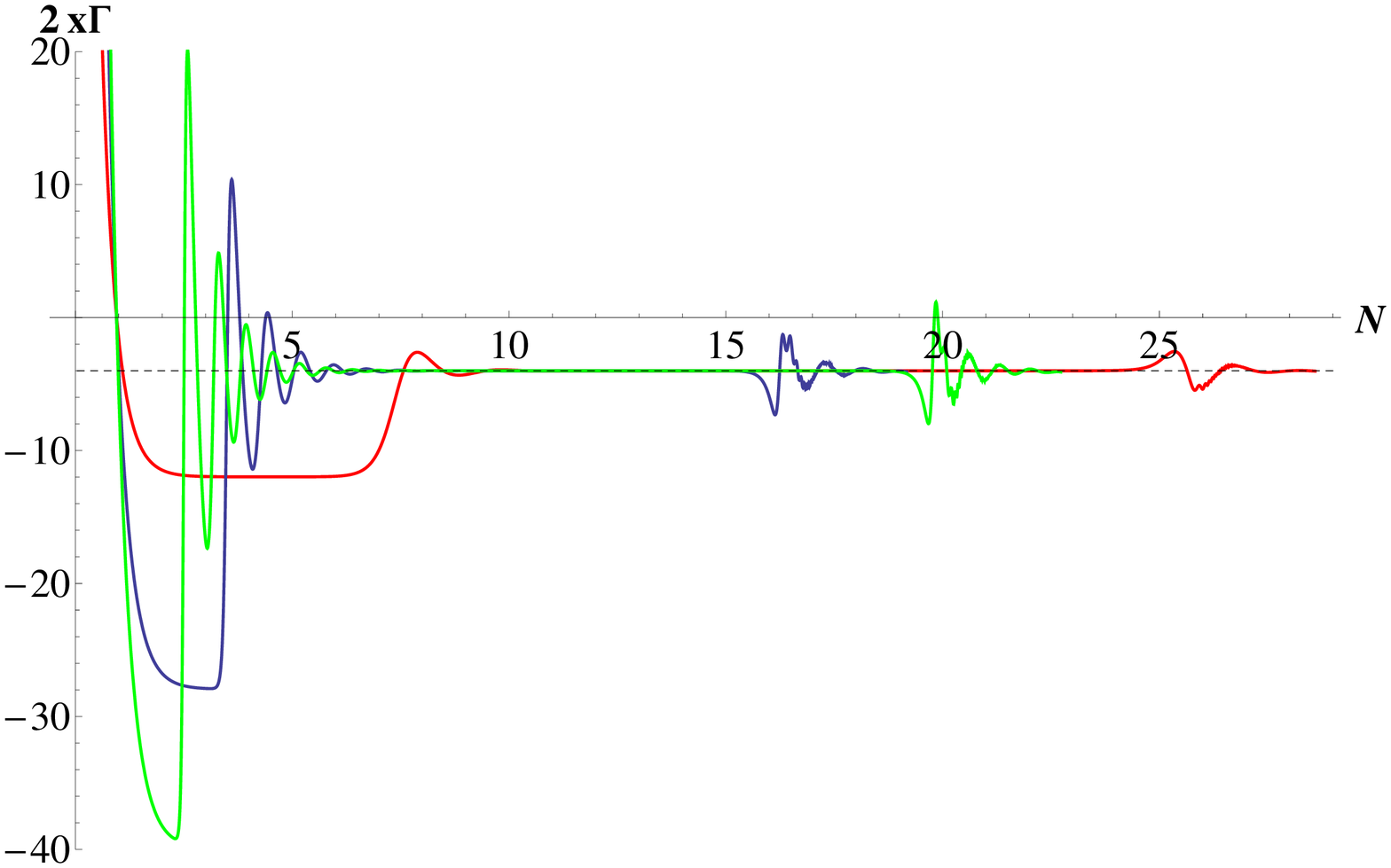}
\includegraphics[width=85mm,angle=0]{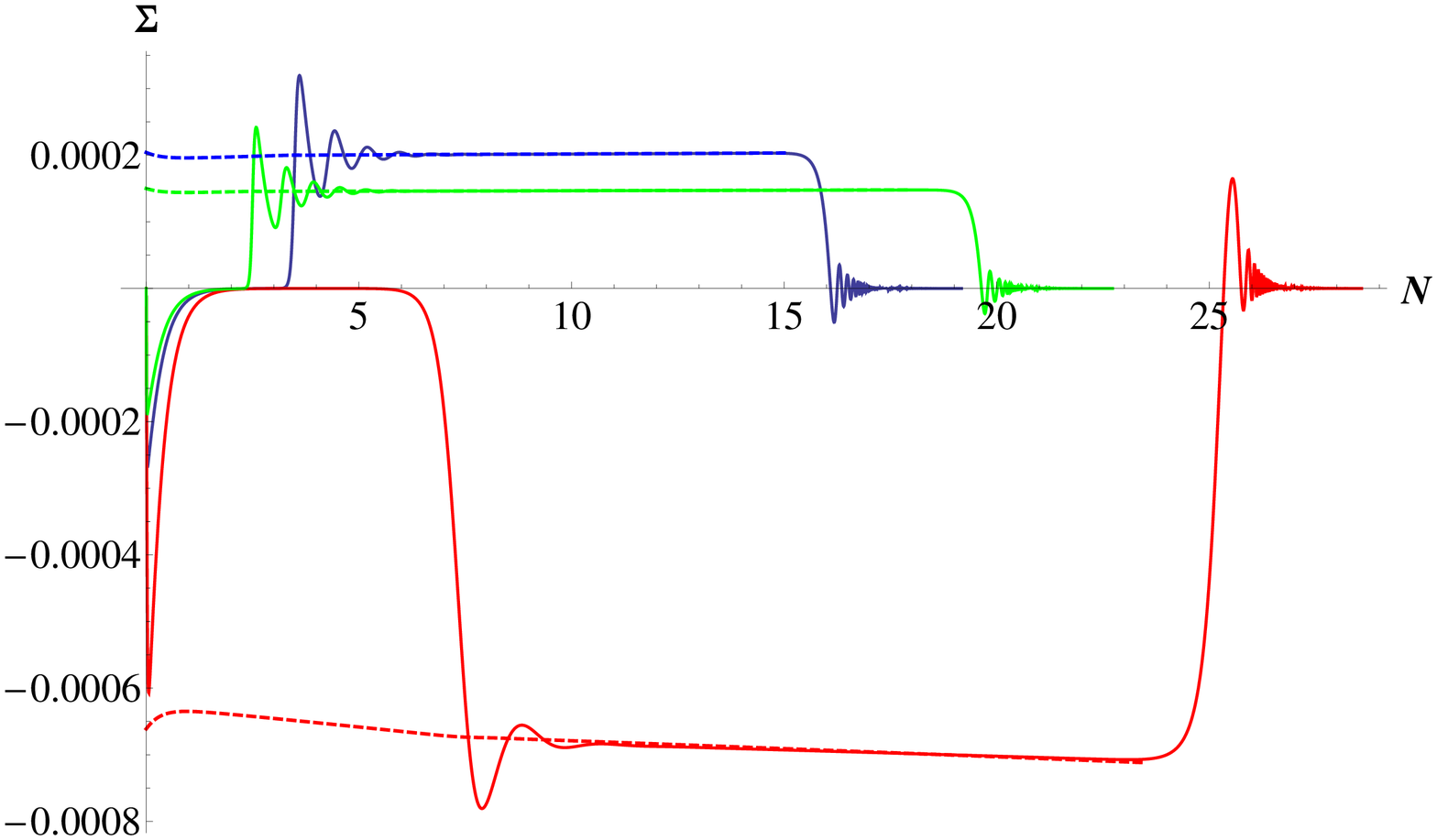}
\caption{These numerical plots show the evolution of the kinetic function scaling $\frac{1}{f}\frac{\ud f}{\ud \alpha}=2x\Gamma$, given by Eq. (\ref{scalingSol1}), and anisotropy $\Sigma$, with respect to the number of elapsing e-folds $N$, for a massive vector field, for the chaotic potential model example. The plots show numerical solutions for three different values of the dimensionless model parameters and initial conditions. We observe that the $\mathcal{SSR}$ is obtained prior to the $\mathcal{VSS}$ for certain initial conditions. At the $\mathcal{SSR}$ the kinetic function takes on a variety of scalings given by different values of $2x\Gamma$. All solutions move then to the $\mathcal{VSS}$ where the same attractor $2x\Gamma=-4$ is obtained, giving $f_{\textrm{att}}\propto a^{-4}$ and $m_{\textrm{att}}\propto a$, when considering the connection in Eq.~(\ref{connetion}). The plots clearly show that the same attractor $2x\Gamma=-4$ is also obtained when the vector field becomes heavy. As expected the anisotropy $\Sigma$ vanishes at the $\mathcal{SSR}$ but not at the $\mathcal{VSS}$ if the vector field remains light. We observe that $\Sigma$ is increasing for this example because the dimensionless model parameters are time-dependent. The anisotropy does however vanish once again when the vector field becomes heavy. We also observe that $\Sigma$ can be positive or negative depending on which $\mathcal{VSS}$ the solution approaches nearest, see Table \ref{Table-VSS}. The plots of $\Sigma$ also show (in dashed lines) the analytic solutions obtained for the $\mathcal{VSS}$ attractors, see Table \ref{Table-VSS}. We can see how well they agree with the full numerical solutions.}\label{fig-Chaotic}
\end{figure}

For this model we find $\lambda'\Gamma+\lambda\Gamma'=0$ and from the existence and stability condition described above, we find that the dimensionless model parameters satisfy $\Gamma'\gg\lambda'$,
hence the constraints are approximated by
\begin{equation}
\left(\frac{\phi}{m_{P}}\right)^{2}>\frac{4}{3c}\qquad\text{and}\qquad
\left(\frac{\phi}{m_{P}}\right)^{2}>\frac{8}{3c}.
\end{equation}
The first constraint above applies to the massless, light ($\mathcal{VSS}_{\it{1,3}}$) and heavy vector field cases. The second condition applies to the light ($\mathcal{VSS}_{\it{2}}$) critical point. These constraints are easily satisfied through the stability condition $\Gamma\gg\lambda$.
Therefore the critical point evolves at a slower rate than the approach
of the solution to the point. The attractor solution can then be obtained and solutions are dragged along with the motion of the critical points. The dynamical evolution of the model for a variety of initial conditions and choices of the dimensionless model parameters is shown in Fig.~\ref{fig-Chaotic}.

%---------------------------------------------------------------------------------------------------------------------
%---------------------------------------------------------------------------------------------------------------------
\subsection{The Supergravity Inspired Potential}

A natural framework where our models can be realised is in the context of
supergravity theories. This is because, in such theories the bosonic part of
the action is determined by three fundamental functions of the fields
of the theory: the superpotential $W$, the K\"ahler potential $K$ and the gauge
kinetic function $f$. The latter is a holomorphic function of the scalar fields
of the theory. Hence, if our vector field is a gauge field of a supergravity
theory, then it is natural to expect that $f=f(\phi)$, i.e. the gauge kinetic
function is modulated by the inflaton field.

Now, the F-term scalar potential in supergravity is given by
\begin{equation}
V_F=e^{K/m_P^2}\left[
\left(W_m+\frac{W}{m_P^2}K_m\right)^*K^{m^*n}
\left(W_n+\frac{W}{m_P^2}K_n\right)-3\frac{|W|^2}{m_P^2}\right],
\end{equation}
where $K^{m^*n}$ is the inverse of the so-called K\"ahler metric %matrix
\mbox{$K_{m^*n}\equiv\frac{\partial^2K}{\partial\Phi_n\partial\Phi_m^*}$},
subscripts denote derivatives with respect to the corresponding fields
(i.e. \mbox{$W_n\equiv\frac{\partial W}{\partial\Phi_n}$} and
\mbox{$K_n\equiv\frac{\partial K}{\partial\Phi_n}$}) and summation over
repeated indexes is assumed. The kinetic terms of the scalar fields are given
by
\begin{equation}
{\cal L}_{\rm kin}=K_{m^*n}\partial\Phi_m^*\partial\Phi_n\;.
\end{equation}
The scalar fields are canonically normalised if the K\"ahler potential is
minimal, i.e. if it is of the form \mbox{$K=\Phi_n\Phi_n^*$}. Assuming a
minimal K\"ahler potential then the K\"ahler metric becomes Euclidean
(\mbox{$K_{m^*n}=\delta_{mn}$}, where $\delta_{mn}$ is Kronecker's delta)
and the above scalar potential can be written as
\begin{equation}
V_F=e^{K/m_P^2}\left[
\sum_n\left|\frac{\partial W}{\partial\Phi_n}\right|^2+
m_P^{-2}\left(W^*\frac{\partial W}{\partial \Phi_n}\Phi_n+{\rm c.c.}\right)+
\frac{|W|^2}{m_P^2}\left(\frac{K}{m_P^2}-3\right)
\right].
\end{equation}
Suppose now that the inflaton field corresponds to a supersymmetric flat
direction so that \mbox{$W\neq W(\Phi)$} (where $\Phi$ here denotes the
corresponding superfield). In this case the only dependence of $V_F$ on the
inflaton is through the K\"ahler potential. Considering also that, for
a renormalisable superpotential we expect
\mbox{$V_F^{\rm susy}>\frac{|W|^2}{m_P^2}$}, (where \mbox{$V_F^{\rm susy}\equiv
\sum_n\left|\frac{\partial W}{\partial\Phi_n}\right|^2 $} is the F-term scalar
potential in global supersymmetry) we can readily observe that the dependence
of $V_F$ on the inflaton is only through the exponential prefactor
$e^{K/m_P^2}$. Since for the inflaton field we have
\mbox{$K=\Phi\Phi^*=|\Phi|^2$} we can write
\begin{equation}
V(\phi)=V_{0}e^{\frac{1}{2}\left(\frac{\phi}{m_{P}}\right)^{2}},
\label{sugrapot}
\end{equation}
where we defined the canonically normalised real inflaton field as
\mbox{$\phi\equiv|\Phi|/\sqrt 2$}. Below we employ the above scalar potential
with two different choices for the functional form of the gauge kinetic
function. Since we are considering a gauge field, we assume that its mass
$m$ is obtained through the Higgs mechanism. To consider that the mass is being
modulated by the inflaton as well, we assume that there is some connection
between the inflaton field $\phi$ and the Higgs field $\psi$ responsible for
$m$. An example of such a possibility is to consider a model of smooth hybrid
inflation, where the inflaton and the Higgs fields are related as
\mbox{$\psi\propto\phi^{-1}$} during inflation \cite{smooth}.

\subsubsection{Model 1}

In our first example we choose a power law kinetic function
\begin{equation}
f(\phi)=f_{i}\left(\frac{\phi}{\phi_{i}}\right)^{4c},
\end{equation}
where $c$ is a real positive constant.
The dimensionless model parameters in Eq. (\ref{model_param}) are given by $\lambda(\phi)=\sqrt{\frac{3}{2}}\left(\frac{\phi}{m_{P}}\right)$
and $\Gamma(\phi)=4c\sqrt{\frac{3}{2}}\left(\frac{m_{P}}{\phi}\right)$.
The slow-roll parameters defined in Eq. (\ref{slow-roll-param}) are $\epsilon=\frac{1}{2}\left(\frac{\phi}{m_{P}}\right)^{2}$ and
$\eta=1+\left(\frac{\phi}{m_{P}}\right)^{2}$.

When we come to consider a massive vector field we need to give a functional form for the mass $m(\phi)$. Considering the connection Eq. (\ref{connetion}), the dimensionless model parameter now becomes
$\Xi=-2c\sqrt{\frac{3}{2}}\left(\frac{m_{P}}{\phi}\right)$. Equivalently we consider the function
\begin{equation}
m(\phi)=m_{i}\left(\frac{\phi}{\phi_{i}}\right)^{-c},
\end{equation}
which means that the Higgs field responsible for the mass of the vector boson
is related with the inflaton as \mbox{$\psi\propto\phi^{-c}$}.

The results for the standard slow-roll solution in this model are given in Table \ref{Table-SSR-Ex3}.
\begin{table}[h]
\begin{center}
\begin{tabular*}{0.48\textwidth}{@{\extracolsep{\fill}} c||c|c}
  \rule[-2mm]{0pt}{4ex}$\mathcal{SSR}$ & Existence$\qquad$ & Stability$\qquad$ \\
  \hline
  \hline
  \rule[-2.5mm]{0pt}{5.5ex}Massless & $\left(\frac{\phi}{m_{P}}\right)^{2}<6\qquad$ &
$\left(\frac{\phi}{m_{P}}\right)^{2}<4(1-c)\qquad$ \\

  \hline
  \rule[-2.5mm]{0pt}{5.5ex}Light & "$\qquad$ & $\left(\frac{\phi}{m_{P}}\right)^{2}<2(1-c)\qquad$ \\

  \hline
  \rule[-2.5mm]{0pt}{5.5ex}Heavy & "$\qquad$ & $\left(\frac{\phi}{m_{P}}\right)^{2}<3(1-c)\qquad$ \\
\end{tabular*}
\end{center}
\caption{SUGRA Potential, Model 1}\label{Table-SSR-Ex3}
\end{table}

Clearly we require that $c<1$ for the stability conditions to be well defined. We also find the following generic results for the standard slow-roll solutions, from Eq. (\ref{SSR-gen-slwParam})
\begin{equation}
\epsilon_{\textrm{H}} =\epsilon\,,\qquad
\eta_{\textrm{H}}=1+\epsilon \qquad\text{and}\qquad
\gamma_{\phi}=\frac{2}{3}\epsilon.
\end{equation}

The energy density of the Universe is dominated by the scalar potential $y_c\approx1$, if $\left(\frac{\phi}{m_{P}}\right)^{2}\ll6$,
corresponding then to a period of inflation. In the small field regime, the slow-roll condition $\epsilon_{\textrm{H}}\ll1$ is satisfied, but $\eta_{\textrm{H}}\sim1$ violating the second slow-roll condition. This is expected from the
form of the scalar potential and corresponds to the well-known $\eta$-problem
of inflation model-building in the context of supergravity.
Therefore we are considering a period of fast-roll inflation. The slow-roll condition guarantees the stability of the critical point as long as $c\not\approx1$.

As the dimensionless model parameters are also varying in this example, the critical point
is therefore moving in phase-space, and so we have the additional
constraints coming from Eqs. (\ref{SSR-mov-cond}), (\ref{SSR-mh-mov-cond}) and (\ref{SSR-heavy-mov-cond}) for the massless, light and heavy vector field cases respectively.
Considering the slow-roll limit $\epsilon\ll1$, even with $|\eta|\sim1$, the conditions that must be satisfied are
\begin{equation}
c<\frac{3}{4}\,,\qquad
c<0\quad\text{and}\quad
c<\frac23.
\end{equation}

Clearly the constraint on the light vector field case cannot be met as we have set $c>0$.
In the limit $c\rightarrow 0$ the motion of the critical point becomes the same
as the approach of solutions to it. Therefore, in the light vector field case it is
unclear whether solutions can reach the $\mathcal{SSR}$ critical point,
numerical simulations are needed to verify this (see Fig.~\ref{fig-SUGRA-mod1}).
%The other two conditions are however readily satisfied, therefore the critical point evolves at a slower rate than the approach of solutions to the
%point. The attractor solution may then be obtained and solutions will be dragged along with them.

The results for the vector scaling solution in this model are given in Table \ref{Table-VSS-Ex3}.
\begin{table}[h]
\begin{center}
\begin{tabular*}{0.7\textwidth}{@{\extracolsep{\fill}} c||c|c|c|c}
  \rule[-2mm]{0pt}{4ex}$\mathcal{VSS}$ & Existence & Stability & $\mathcal{R}$ & $\Sigma_c$  \\
  \hline
  \hline
  \rule[-2.5mm]{0pt}{5.5ex}Massless & $c>1$ & $\left(\frac{\phi}{m_{P}}\right)^{2}\ll4c
,\,\left(\frac{\phi}{m_{P}}\right)^{2}\ll24c^2$ & $\frac{c-1}{4c^{2}}\left(\frac{\phi}{m_{P}}\right)^{2}$ & $\frac23\mathcal{R}$ \\

  \hline
  \rule[-2.5mm]{0pt}{5.5ex}Light: $\mathcal{VSS}_{\it{1}}$ & " & " & " & "  \\

  \hline
  \rule[-2.5mm]{0pt}{5.5ex}Light: $\mathcal{VSS}_{\it{2}}$ & " & " & $\frac{c-1}{2c^{2}}\left(\frac{\phi}{m_{P}}\right)^{2}$ & $-\frac23\mathcal{R}$ \\

  \hline
  \rule[-2.5mm]{0pt}{5.5ex}Light: $\mathcal{VSS}_{\it{3}}$ & $c>\frac43$ & " & $\frac{9c-8}{24c^{2}}\left(\frac{\phi}{m_{P}}\right)^{2}$  & $-\frac{1}{6c}\epsilon$ \\

  \hline
  \rule[-2.5mm]{0pt}{5.5ex}Heavy & $c>1$ & " & $\frac{c-1}{3c^{2}}\left(\frac{\phi}{m_{P}}\right)^{2}$ & $0$  \\

\end{tabular*}
\end{center}
\caption{SUGRA Potential, Model 1}\label{Table-VSS-Ex3}
\end{table}

We also find the following generic results for the vector scaling solutions, from Eq. (\ref{VSS-gen-slwrollparam})
\begin{eqnarray}
\epsilon_{\textrm{H}} =\frac1c\epsilon\,,\qquad
\eta_{\textrm{H}}=\frac1c \eta \qquad\text{and}\qquad
\gamma_{\phi}=\frac{2}{3c^2}\epsilon.
\end{eqnarray}

As in the previous model we can see that inflation is not spoilt as
the vector field energy density remains sub-dominant in the small field limit. The stability conditions guarantee that the scalar potential dominates the
energy density. The backreaction due to the vector field becomes proportional
to the scalar slope $V'(\phi)$ and a negative contribution, therefore
the effective slope as seen by the scalar field becomes flatter and
the field slows down, \mbox{$V_{\rm eff}'=\frac{1}{c}V'$}.
% see Eq. (\ref{VBratioForm01}).
This effect can be seen from the slow-roll conditions
$\epsilon_{\textrm{H}}=\frac1c\epsilon\ll1$
and $\eta_{\textrm{H}}=\frac{1}{c}\eta$.
In the $\mathcal{VSS}$, inflation is now possible for $\eta_{\textrm{H}}\ll1$ if $c$ is large enough. Therefore even if the $\mathcal{SSR}$ corresponds to fast-roll inflation, it is possible to obtain a period of slow-roll inflation if the solution reaches the $\mathcal{VSS}$
attractor. In that way one can overcome the $\eta$-problem which plagues
supergravity models of inflation.

The anisotropy remains small as long as the slow-roll condition is met, and
vanishes if the vector field becomes heavy. In this model $\mathcal{R}$ and $|\Sigma_c|$ are not exactly constant but are slowly decreasing as the scalar field rolls down its potential. As in previous models, when $c>1$ the $\mathcal{SSR}$
is an unstable saddle point due to the growing vector backreaction.

As the dimensionless model parameters are also varying in this example, the critical point
is therefore moving in phase-space, and so we have the additional
constraints coming from Eqs. (\ref{cp-motion-m0-VSS}), (\ref{cp-motion-mh-VSS2}), (\ref{cp-motion-mh-VSS3}) and (\ref{cp-motion-heavy-VSS}) for the massless, light ($\mathcal{VSS}_{\it{1,2,3}}$) and heavy vector field cases respectively. All vector scaling solutions have the same eigenvalue of smallest magnitude, namely $|\textrm{Re}[m_{2,3}]|\simeq\frac32$, where we have neglected the small eigenvalues $m_4$ because these correspond to flows between the scaling solutions as described at the end of Sec. \ref{Sec-mh-VSS3}.
\begin{figure}[h]
\includegraphics[width=85mm,angle=0]{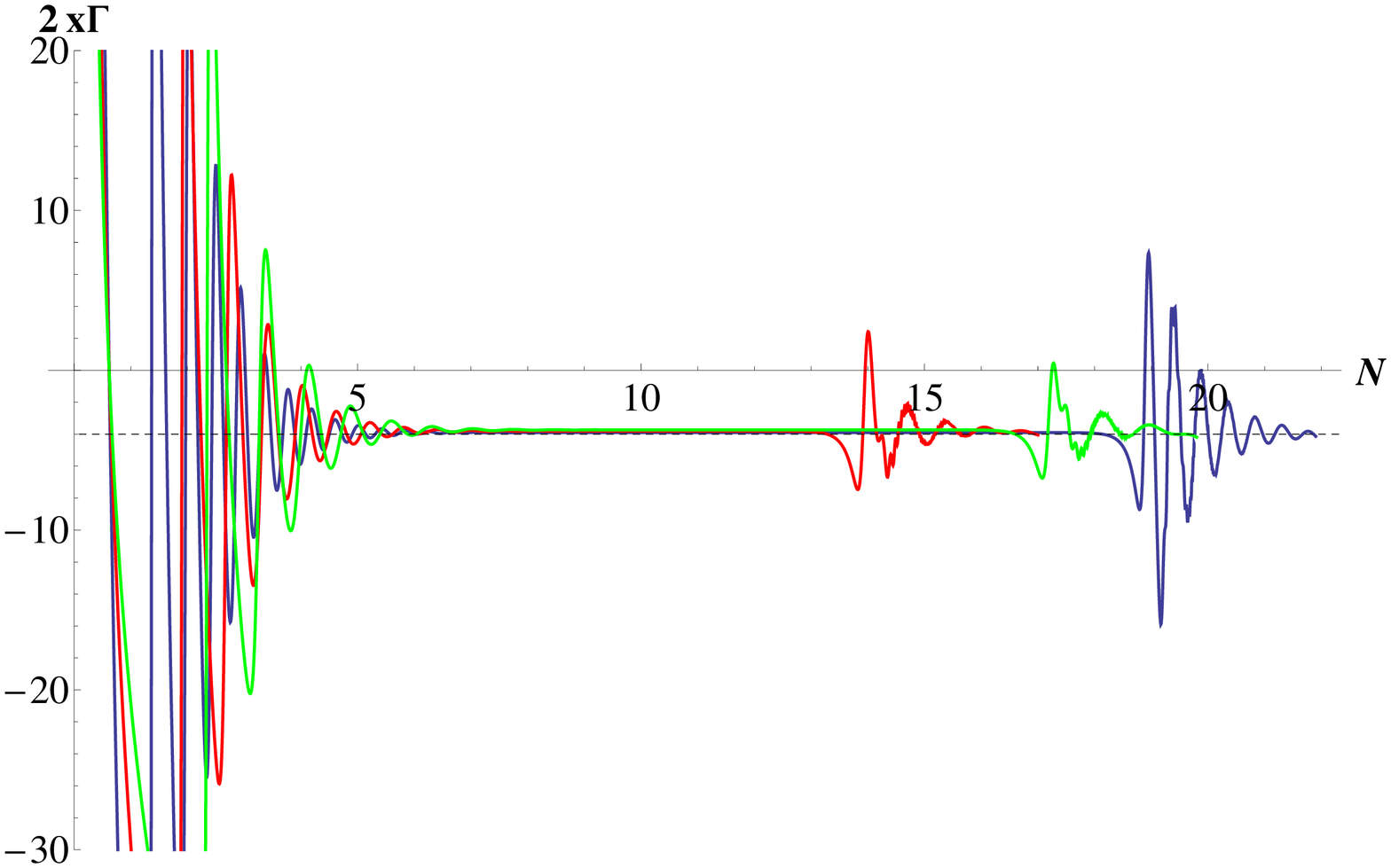}
\includegraphics[width=85mm,angle=0]{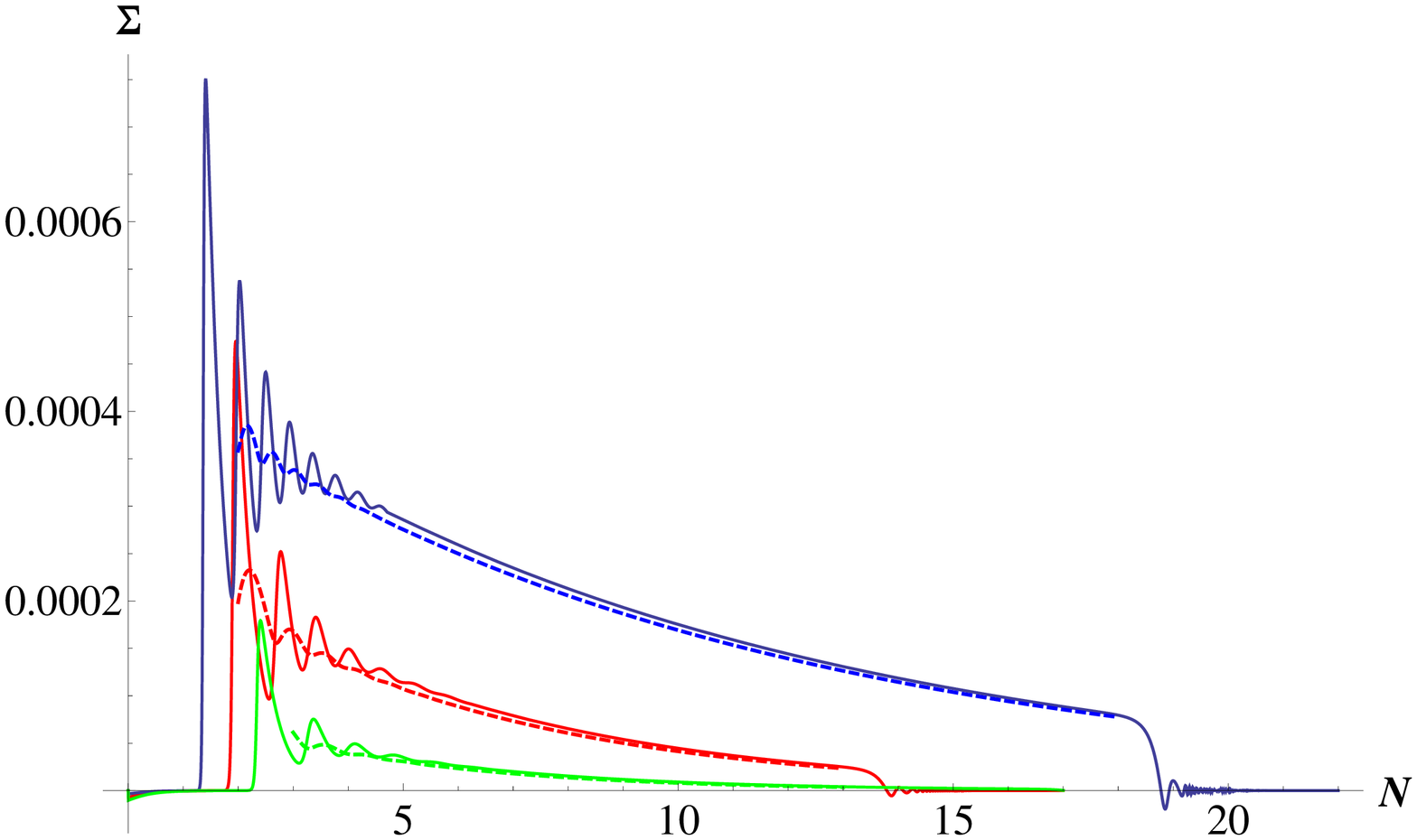}
\caption{These numerical plots show the evolution of the kinetic function scaling $\frac{1}{f}\frac{\ud f}{\ud \alpha}=2x\Gamma$, given by Eq. (\ref{scalingSol1}), and anisotropy $\Sigma$, with respect to the number of elapsing e-folds $N$, for a massive vector field, for the SUGRA inspired model 1 example. The plots show numerical solutions for three different values of the dimensionless model parameters and initial conditions. We observe that there is no $\mathcal{SSR}$ solution as the scalar potential is too steep and therefore would correspond to a period of fast-roll inflation. Despite the wide range of initial conditions, all solutions move to the $\mathcal{VSS}$ where the same attractor $2x\Gamma=-4$ is obtained, giving $f_{\textrm{att}}\propto a^{-4}$ and $m_{\textrm{att}}\propto a$, when considering the connection in Eq.~(\ref{connetion}). The plots clearly show that the same attractor $2x\Gamma=-4$ is also obtained when the vector field becomes heavy. As expected the anisotropy $\Sigma\neq0$ at the $\mathcal{VSS}$ if the vector field remains light. We observe that $\Sigma$ is decreasing for this example because the dimensionless model parameters are time-dependent. The plots of $\Sigma$ also show (in dashed lines) the analytic solutions obtained for the $\mathcal{VSS}$ attractors, see Table \ref{Table-VSS}. We can see how well they agree with the full numerical solutions.}\label{fig-SUGRA-mod1}
\end{figure}

For this model we also find that $\lambda'\Gamma+\lambda\Gamma'=0$ and from the existence and stability condition described above, we find that the dimensionless model parameters satisfy $\Gamma'\gg\lambda'$,
hence the constraints are approximated by
\begin{equation}
c>\frac23\qquad\text{and}\qquad
c>\frac43.
\end{equation}
The first constraint above applies to the massless, light ($\mathcal{VSS}_{\it{1,3}}$) and heavy vector field cases. The second condition applies to the light ($\mathcal{VSS}_{\it{2}}$) critical point. These constraints are readily satisfied through the existence condition.
Therefore the critical point evolves at a slower rate than the approach
of the solution to the point. The attractor solution can then be obtained and
the system is dragged along with the motion of the critical points. The dynamical evolution of the model for a variety of initial conditions and choices of the dimensionless model parameters is shown in Fig.~\ref{fig-SUGRA-mod1}.

%---------------------------------------------------------------------------------------------------------------------
%---------------------------------------------------------------------------------------------------------------------
\subsubsection{Model 2}

We now consider an exponential kinetic function of the form
\begin{equation}
%V(\phi)=V_{0}e^{\frac{1}{2}\left(\frac{\phi}{m_{P}}\right)^{2}}\qquad\text{and}\qquad
f(\phi)=f_{i}e^{4c\left(\frac{\phi}{\phi_{i}}-1\right)},
\end{equation}
where $c$ is a real positive constant.
The dimensionless model parameters in Eq. (\ref{model_param}) are given by $\lambda(\phi)=\sqrt{\frac{3}{2}}\left(\frac{\phi}{m_{P}}\right)$
and $\Gamma(\phi)=4c\sqrt{\frac{3}{2}}\left(\frac{m_{P}}{\phi_{i}}\right)$.
The slow-roll parameters defined in Eq. (\ref{slow-roll-param}) are
$\epsilon=\frac12\left(\frac{\phi}{m_P}\right)^2$ and $\eta=1+\left(\frac{\phi}{m_P}\right)^2$.

When we come to consider a massive vector field we need to give a functional form for the mass $m(\phi)$. Considering the connection Eq. (\ref{connetion}), the model parameter now becomes
$\Xi=-2c\sqrt{\frac{3}{2}}\left(\frac{m_{P}}{\phi_i}\right)$. Equivalently we consider the function
\begin{equation}
m(\phi)=m_{i}e^{-c\left(\frac{\phi}{\phi_{i}}-1\right)}.
\end{equation}

The results for the standard slow-roll solution in this model are given in Table \ref{Table-SSR-Ex4}.
\begin{table}[h]
\begin{center}
\begin{tabular*}{0.48\textwidth}{@{\extracolsep{\fill}} c||c|c}
  \rule[-2mm]{0pt}{4ex}$\mathcal{SSR}$ & Existence$\qquad$ & Stability$\quad$ \\
  \hline
  \hline
  \rule[-2.5mm]{0pt}{5.5ex}Massless & $\left(\frac{\phi}{m_{P}}\right)^{2}<6\qquad$ &
$\left(\frac{\phi}{m_{P}}\right)^{2}<4\left(1-c\frac{\phi}{\phi_i}\right)\quad$ \\

  \hline
  \rule[-2.5mm]{0pt}{5.5ex}Light & "$\qquad$ & $\left(\frac{\phi}{m_{P}}\right)^{2}<2\left(1-c\frac{\phi}{\phi_i}\right)\quad$  \\

  \hline
  \rule[-2.5mm]{0pt}{5.5ex}Heavy & "$\qquad$ & $\left(\frac{\phi}{m_{P}}\right)^{2}<3\left(1-c\frac{\phi}{\phi_i}\right)\quad$  \\
\end{tabular*}
\end{center}
\caption{SUGRA Potential, Model 2}\label{Table-SSR-Ex4}
\end{table}

Clearly we require that $c\frac{\phi}{\phi_i}<1$ for the stability conditions to be well defined. We also find the following generic results for the standard slow-roll solutions, from Eq. (\ref{SSR-gen-slwParam})
\begin{equation}
\epsilon_{\textrm{H}} =\epsilon\,,\qquad
\eta_{\textrm{H}}=1+\epsilon \qquad\text{and}\qquad
\gamma_{\phi}=\frac{2}{3}\epsilon.
\end{equation}

As in the previous model, the energy density of the Universe is dominated
by the scalar potential if $\left(\frac{\phi}{m_{P}}\right)^{2}\ll6$,
corresponding then to a period of inflation. This potential
satisfies the $\epsilon\ll1$ slow-roll condition in the small field regime,
this also guarantees the stability of the critical point, however we
see that $\eta\sim1$ violating this slow-roll condition. Again, this is the
well-known $\eta$-problem of inflation in the context of supergravity. Thus,
this turns out to be a period of fast-roll inflation.

As the dimensionless model parameters are also varying in this example, the critical point
is therefore moving in phase-space, and so we have the additional
constraints coming from Eqs. (\ref{SSR-mov-cond}), (\ref{SSR-mh-mov-cond}) and (\ref{SSR-heavy-mov-cond}) for the massless, light and heavy vector field cases respectively.
Considering the slow-roll limit $\epsilon\ll1$, even with $|\eta|\sim1$, the conditions that must be satisfied are
\begin{equation}
c\left(\frac{\phi}{\phi_i}\right)<\frac{3}{4}\,,\qquad
c\left(\frac{\phi}{\phi_i}\right)<0\quad\text{and}\quad
c\left(\frac{\phi}{\phi_i}\right)<\frac23.
\end{equation}

Clearly the constraint on the light vector field case cannot be met as we have set
$c\left(\frac{\phi}{\phi_i}\right)>0$. In the limit
$c\left(\frac{\phi}{\phi_i}\right)\rightarrow 0$ the motion of the
critical point becomes the same as the approach of solutions to it.
Therefore, in the light vector field case it is unclear whether solutions
can reach the $\mathcal{SSR}$ critical point, numerical simulations are needed to verify this (see Fig.~\ref{fig-SUGRA-mod2}).
%The other two conditions are however readily satisfied, therefore the critical point evolves at a slower rate than the approach of solutions to the
%point. The attractor solution may then be obtained and solutions will be dragged along with them.

The results for the vector scaling solution in this model are given in Table \ref{Table-VSS-Ex4}.
\begin{table}[h]
\begin{center}
\begin{tabular*}{0.75\textwidth}{@{\extracolsep{\fill}} c||c|c|c|c}
  \rule[-2mm]{0pt}{4ex}$\mathcal{VSS}$ & Existence & Stability & $\mathcal{R}$ & $\Sigma_c$   \\
  \hline
  \hline
  \rule[-2.5mm]{0pt}{5.5ex}Massless & $\frac{\phi}{\phi_i}>\frac1c$ & $\left(\frac{\phi}{m_{P}}\right)^{2}\ll4c\frac{\phi}{\phi_i}
,\,\left(\frac{\phi_i}{m_{P}}\right)^{2}\ll24c^2$ & $\frac{1}{4c^{2}}\left[c\frac{\phi}{\phi_i}-1\right]\left(\frac{\phi_i}{m_{P}}\right)^{2}$ & $\frac23\mathcal{R}$ \\

  \hline
  \rule[-2.5mm]{0pt}{5.5ex}Light: $\mathcal{VSS}_{\it{1}}$ & " & " & " & "  \\

  \hline
  \rule[-2.5mm]{0pt}{5.5ex}Light: $\mathcal{VSS}_{\it{2}}$ & " & " & $\frac{1}{2c^{2}}\left[c\frac{\phi}{\phi_i}-1\right]\left(\frac{\phi_i}{m_{P}}\right)^{2}$ & $-\frac23\mathcal{R}$  \\

  \hline
  \rule[-2.5mm]{0pt}{5.5ex}Light: $\mathcal{VSS}_{\it{3}}$ & $\frac{\phi}{\phi_i}>\frac{4}{3c}$ & " & $\frac{1}{24c^{2}}\left[9c\frac{\phi}{\phi_i}-8\right]\left(\frac{\phi_i}{m_{P}}\right)^{2}$ & $-\frac{\epsilon}{6c}\frac{\phi_i}{\phi}\propto\phi$  \\

  \hline
  \rule[-2.5mm]{0pt}{5.5ex}Heavy & $\frac{\phi}{\phi_i}>\frac{1}{c}$ & " & $\frac{1}{3c^{2}}\left[c\frac{\phi}{\phi_i}-1\right]\left(\frac{\phi_i}{m_{P}}\right)^{2}$ & $0$  \\

\end{tabular*}
\end{center}
\caption{SUGRA Potential, Model 2}\label{Table-VSS-Ex4}
\end{table}

We also find the following generic results for the vector scaling solutions, from Eq. (\ref{VSS-gen-slwrollparam})
\begin{eqnarray}
\epsilon_{\textrm{H}} =\frac1c\left(\frac{\phi_i}{\phi}\right)\epsilon\,,\qquad
\eta_{\textrm{H}}=\frac{1}{2c} \left(\frac{\phi_i}{\phi}\right) \eta \qquad\text{and}\qquad
\gamma_{\phi}=\frac{2}{3c^2}\left(\frac{\phi_i}{\phi}\right)^2\epsilon.
\end{eqnarray}

Once again we can see that inflation is not spoilt as the stability conditions
guarantee that the scalar potential dominates
the energy density. However for this model we also notice that $\mathcal{R}$
and $|\mathcal{R_{B}}|$ are deceasing as the scalar field evolves down
its potential. The backreaction due to the vector field is a
negative contribution to the scalar slope $V'(\phi)$,
\begin{equation}\label{VBratioForm02}
    V'_{\textrm{eff}}=V'+\mathcal{B}_A=\frac{1}{c}\left(\frac{\phi_i}{\phi}\right)V'.
\end{equation}
Therefore the effective slope as seen by the scalar field becomes flatter and the
field slows down. The slow-roll condition $\epsilon_{\textrm{H}}\ll1$ is satisfied as long as the existence condition is met.
In the $\mathcal{VSS}$ it is now possible for $\eta_{\textrm{H}}\ll1$ if $\frac{\phi}{\phi_i}\gg\frac{1}{2c}$.
As seen previously the slow-roll parameters are smaller than that
of the $\mathcal{SSR}$. In that way, therefore, one can overcome the
$\eta$-problem.

The anisotropy remains small under the stability conditions, but is
however decreasing as the scalar field evolves down
its potential. The anisotropy vanishes as the vector field becomes heavy.

\begin{figure}[h]
\includegraphics[width=85mm,angle=0]{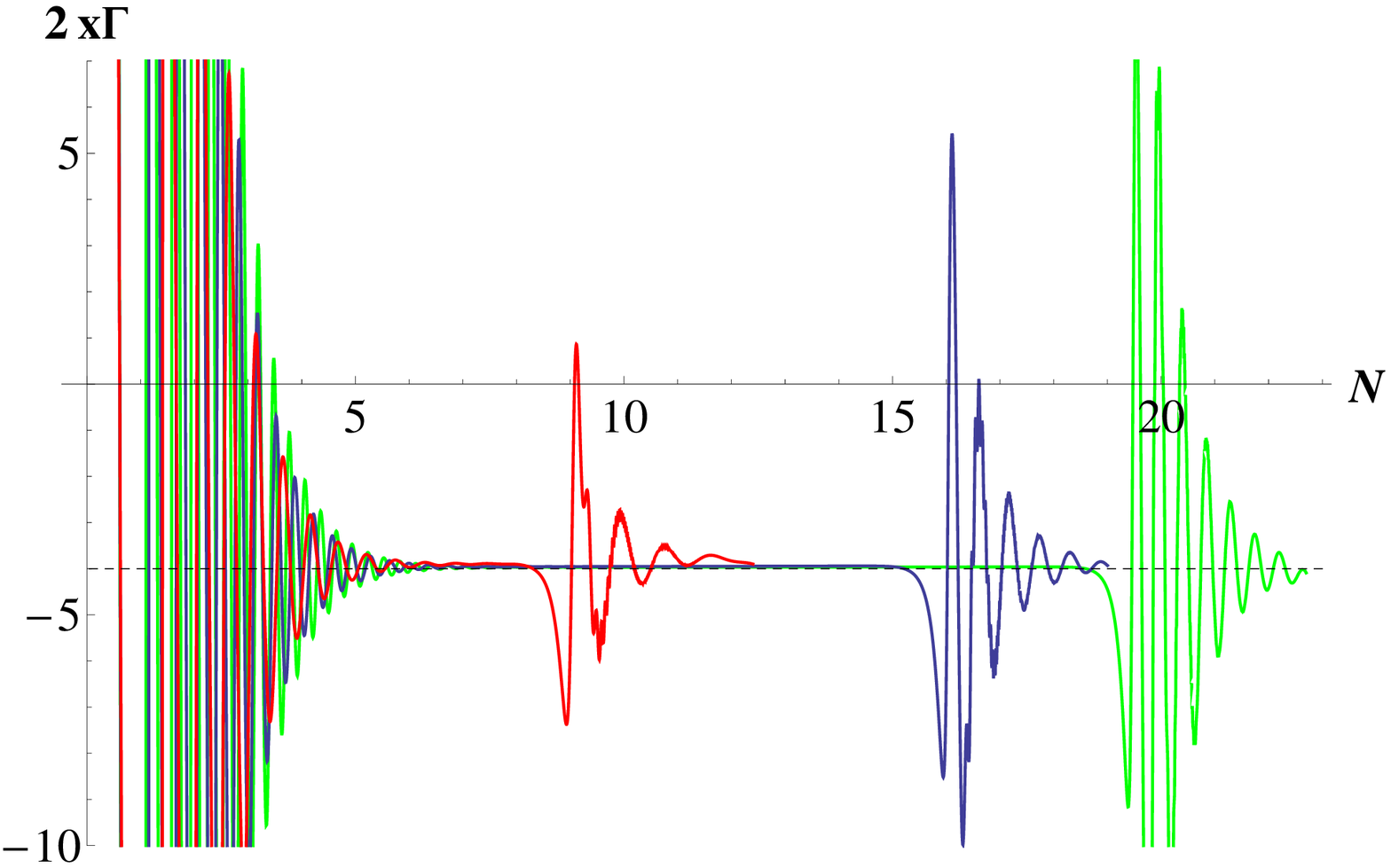}
\includegraphics[width=85mm,angle=0]{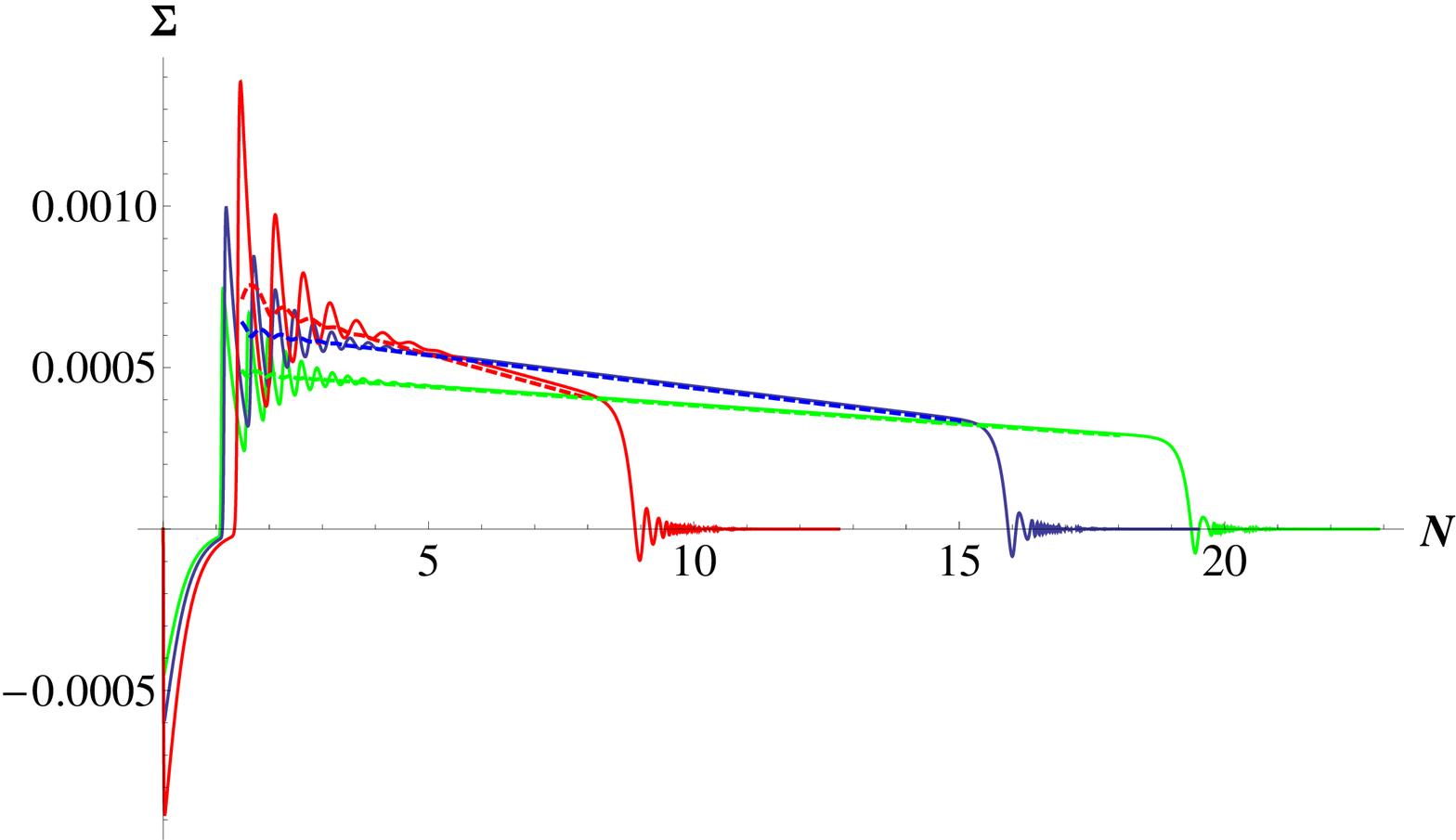}
\caption{These numerical plots show the evolution of the kinetic function scaling $\frac{1}{f}\frac{\ud f}{\ud \alpha}=2x\Gamma$, given by Eq. (\ref{scalingSol1}), and anisotropy $\Sigma$, with respect to the number of elapsing e-folds $N$, for a massive vector field, for the SUGRA inspired model 2 example. The plots show numerical solutions for three different values of the dimensionless model parameters and initial conditions. We observe that there is no $\mathcal{SSR}$ solution as the scalar potential is too steep and therefore would correspond to a period of fast-roll inflation. Despite the wide range of initial conditions, all solutions move to the $\mathcal{VSS}$ where the same attractor $2x\Gamma=-4$ is obtained, giving $f_{\textrm{att}}\propto a^{-4}$ and $m_{\textrm{att}}\propto a$, when considering the connection in Eq.~(\ref{connetion}). The plots clearly show that the same attractor $2x\Gamma=-4$ is also obtained when the vector field becomes heavy. As expected the anisotropy $\Sigma\neq0$ at the $\mathcal{VSS}$ if the vector field remains light. We observe that $\Sigma$ is decreasing for this example because the dimensionless model parameters are time-dependent. The plots of $\Sigma$ also show (in dashed lines) the analytic solutions obtained for the $\mathcal{VSS}$ attractors, see Table \ref{Table-VSS}. We can see how well they agree with the full numerical solutions.}\label{fig-SUGRA-mod2}
\end{figure}

This critical point is moving in phase space due to the
parameter $\lambda=\lambda(\phi)$. Therefore we have the additional
constraints coming from Eqs. (\ref{cp-motion-m0-VSS}), (\ref{cp-motion-mh-VSS2}), (\ref{cp-motion-mh-VSS3}) and (\ref{cp-motion-heavy-VSS}) for the massless, light ($\mathcal{VSS}_{\it{1,2,3}}$) and heavy vector field cases respectively. All vector scaling solutions have the same eigenvalue of smallest magnitude, namely $|\textrm{Re}[m_{2,3}]|\simeq\frac32$, where we have neglected the small eigenvalues $m_4$ because these correspond to flows between the scaling solutions as described at the end of Sec. \ref{Sec-mh-VSS3}. For this model, the parameter $\Gamma$ is constant. Then these constraints are reduced to
\begin{equation}
\frac{\phi}{\phi_i}>\frac{5}{3c}\qquad\text{and}\qquad
\frac{\phi}{\phi_i}>\frac{4}{3c}.
\end{equation}
The first constraint above applies to the massless and the light ($\mathcal{VSS}_{\it{3}}$) vector field cases. The second condition applies to the light ($\mathcal{VSS}_{\it{1,2}}$) and the heavy vector field cases. If these constraints are satisfied, the critical points evolve at a slower rate than the approach
of the solution to the point. The attractor solution can then be obtained and system is dragged along with the motion of the critical points. The dynamical evolution of the model for a variety of initial conditions and choices of the dimensionless model parameters is shown in Fig.~\ref{fig-SUGRA-mod2}.

%---------------------------------------------------------------------------------------------------------------------
%---------------------------------------------------------------------------------------------------------------------

%\begin{figure}[t]
%\includegraphics[width=100mm,angle=0]{flatP.ps}
%\vspace{-4cm}
%\caption{
%}\label{flatP}
%\end{figure}

%---------------------------------------------------------------------------------------------------------------------
%---------------------------------------------------------------------------------------------------------------------
\section{Conclusions}\label{conclusions}

In this paper we studied the dynamics of a model consisting of a massive Abelian vector field, with a Maxwell-type kinetic term and with a kinetic function $f(\phi)$ and mass $m(\phi)$ modulated by a scalar field $\phi$. This scalar field is driving a period of inflation.

Except for very special cases, analytical solutions to the system of field equations together with Einstein's equations are either extremely difficult if not impossible to obtain due to their non-linear nature. However we may understand the qualitative behaviour of solutions by phase-space analysis. As we are considering systems of 3 or higher dimension we cannot do much more than analyse the stationary points and their stability. We have studied the qualitative behaviour of solutions for general scalar potentials $V(\phi)$ and functional forms of the vector field kinetic function $f(\phi)$ and mass $m(\phi)$. We parametrise the model dependence through the so-called dimensionless model parameters defined as $\lambda\equiv\sqrt{\frac{3}{2}}m_{P}\left(\frac{V'}{V}\right)$,
$\Gamma\equiv\sqrt{\frac{3}{2}}m_{P}\left(\frac{f'}{f}\right)$ and
$\Xi\equiv\sqrt{6}m_{P}\left(\frac{m'}{m}\right)$. We have obtained the critical points and studied their stability for three cases; massless, light and heavy vector field.

Consider firstly the massless vector field case. This case has already been studied by Ref. \cite{anisinf,anisinf+} for a particular model. However our method is more general and produces their results as a special case. We have found 2 critical points of interest in this system. The first critical point corresponds to the standard slow-roll inflation attractor. For the solution to be inflationary i.e. $3m_PH^2\simeq V(\phi)$, the slow-roll condition $\epsilon\ll1$, must be satisfied. At the critical point the energy density of the vector field and the backreaction vanishes. The condition for this to occur is $\lambda\Gamma<6$. We also notice how the anisotropic stress, given by $\Sigma$, vanishes at the critical point $\Sigma=0$, providing an example of the cosmic no-hair theorem. Under the conditions described above, the standard slow-roll critical point is stable i.e. a late-time attractor solution.

Only exponential potentials and kinetic functions give a closed system of equations and therefore exact stationary points. However, with more general forms for the potential and the kinetic function, critical points become time-dependent. The motion of the critical points in phase-space is given by the time-dependence of the dimensionless model parameters. If solutions approach the critical points faster than their evolution then we can expect the attractors to be reached and solutions to be dragged along with them. The (strong) condition for this to happen leads to the second slow-roll condition $|\eta|\ll1$, as long as the bifurcation value is not realised $\lambda\Gamma\not\approx6$. The condition may even be satisfied if the $\eta$ slow-roll condition is marginally violated $|\eta|\sim1$. In other words, the standard slow-roll attractor can also apply to fast-roll inflation.

The second critical point of interest is the so-called vector scaling solution. This critical point arises due to the backreaction effects of the vector field on the dynamics of the scalar field. It therefore requires $\lambda\Gamma>6$. This condition guarantees that the backreaction, and therefore also the vector field energy density, grows. However, before the vector field energy density grows so much that it spoils inflation, the backreaction on the dynamics of the scalar field intervenes and inhibits further growth. The critical point corresponds to an inflationary solution, if $\Gamma\gg\lambda$ and $\Gamma\gg1$. These conditions also guarantee that the critical point is stable i.e. a late-time attractor. The vector field backreaction has the effect of flattening the effective potential slope experienced by the inflaton $V'_{\textrm{eff}}=\frac{6}{\lambda\Gamma}V'$. The evolution of the inflaton consequently slows down with the slow-roll parameter reduced by $2\frac{\lambda}{\Gamma}\epsilon_{\textrm{H}}$, compared to the standard slow-roll case. The effect of the backreaction generates a new stage of slow-roll inflation where  the energy density of the vector field is kept subdominant $\mathcal{R}\ll1$, where $\mathcal{R}$ is the vector to scalar energy density ratio. The energy density of the vector field tracks that of the scalar field with constant $\mathcal{R}$ if the dimensionless model parameters are constant, otherwise $\mathcal{R}$ varies slowly depending on the time-dependence of the dimensionless model parameters.

The anisotropic stress $\Sigma$, however does not vanish at the critical point.
Indeed, there is a residual anisotropy given by $\Sigma\simeq\frac23\mathcal{R}$. This provides a counter-example to the cosmic no-hair theorem and has attracted some interest. Such a prolonged anisotropy can be used to generate statistical anisotropy in the curvature perturbation and in gravitational waves, see
Ref.~\cite{anisinf,anisinf+}.

In analogy to the standard slow-roll solution, the motion of the vector scaling critical point leads to a second `slow-roll' condition given by $m_P\Big|\frac{\Gamma'}{\Gamma^2}\Big|\ll1$. This result is derived when considering a flat potential, i.e. $|\eta-2\epsilon|\lesssim1$ and as long as the bifurcation value is not approximately realised $\lambda\Gamma\not\approx6$.

The vector scaling attractor gives the solution $f_{\rm att}\propto e^{-4\alpha}$ in the limits described above. This particular scaling for the kinetic function enables the generation of a scale-invariant transverse component of the vector field perturbation spectrum as seen in Ref. \cite{sugravec,varkin}.

Now, suppose that the vector field has non-zero mass $m$. To allow particle production to occur the mass of the physical vector field \mbox{$M\equiv m/\sqrt f$} has to be small ($M\ll H$) when the cosmological scales exit the horizon. If the mass is originally negligible, in order for it to have any effect we need to consider that its magnitude is increasing during inflation. Thus, it is possible that the field becomes heavy ($M\gtrsim H$) by the end of inflation. In this case, when considering a massive vector field, we may break the problem down into two regimes; light and heavy field. When considering a light vector field, as in the massless case, we find two types of critical points which are of interest. The standard slow-roll solution is of course still a critical point of the system. The conditions for the vector field energy density and backreaction to vanish are now $\lambda\Gamma<6$ and $|\lambda\Xi|<3$. When considering a massive vector field there are two sources of backreaction to the scalar field dynamics. And therefore, depending on which dominates first, there are different vector scaling solutions. In fact we find three distinct vector scaling solutions. However considering the limits $\Gamma,|\Xi|\gg\lambda$ and $\Gamma,|\Xi|\gg1$, all vector scaling solutions are inflationary with attractor solutions either $f_{\rm att} \propto a^{-4}$ or $m_{\rm att} \propto a$.

To simultaneously attain those attractor solutions we impose the connection
$f'/f=-4m'/m$, i.e. $\Xi=-\frac12\Gamma$.
By imposing this connection, we obtain the scaling of the kinetic function and mass which generates a scale invariant vector field spectrum.
If the vector field perturbations contribute to the curvature perturbation
$\zeta$
then it can generate statistical anisotropy in the spectrum and bispectrum
of $\zeta$ throughout the entire range of the cosmological scales.
The conditions to obtain an inflationary universe with the above scaling for the kinetic function and mass are minimal, namely $\Gamma\gg1$ and $\Gamma\gg\lambda$. Under these limits we have found that there is a small flow of solutions to one particular vector scaling solution, but the attractors for the kinetic function and mass are not affected. In the vector scaling solutions the vector field contribution to the density generates anisotropic stress and, if it is non-negligible, it may lead to anisotropic inflation, which can produce statistically anisotropic perturbations for the inflaton field itself. They can also be used as a source of statistical anisotropy in $\zeta$ as in the massless vector field case.

Once the vector field becomes heavy it starts to oscillate rapidly \cite{vecurv}. As expected from Ref. \cite{vecurv}, the anisotropy vanishes once the field becomes heavy $\Sigma=0$. As in the previous vector scaling solutions the conditions to obtain an inflationary attractor, are $\Gamma\gg\lambda$, $\Gamma\gg1$ and $\lambda\Gamma>6$, when considering the connection $\Xi=-\frac12\Gamma$ described above.

So, in general, we find that if $\lambda\Gamma>6$ then the backreaction grows and solutions will flow to a vector scaling solution where $f_{\rm att} \propto a^{-4}$. If we then impose the connection  $f'/f=-4m'/m$, we also obtain the solution $m_{\rm att} \propto a$. The general conditions for these to be attractor solutions are simply $\Gamma\gg\lambda$ and $\Gamma\gg1$. The above conditions correspond to the requirement that the backreaction of the vector field onto the roll of the inflaton is not decreasing in time, so that it can eventually affect the inflationary dynamics and drive the system to the vector scaling solution(s). The bounds on the dimensionless model parameters $\Gamma$ and $\lambda$ allow half the parameter space to lead to the attractor solution which generates scale invariant spectra for the perturbations of the vector field components. As demonstrated in our examples in Sec.~\ref{examples}, the parameter space includes natural values for our dimensionless model parameters.

The next step is to realise this model in the context of realistic theories
beyond the standard model. A promising possibility is within supergravity
theories. Indeed, our vector field could be a gauge boson of some grand
unified theory (GUT) in the framework of supergravity. In this case, the gauge
kinetic function is a holomorphic function of the scalar fields of the theory.
Hence it is natural to expect that the rolling of the inflaton would modulate
the kinetic function. Similarly, the mass will be modulated by a Higgs field,
which could well be connected to the inflaton through the supersymmetric GUT
group. In supergravity, K\"{a}hler corrections to the scalar potential result
in masses of order $H$ during inflation \cite{randall}, which implies that the
inflaton and the Higgs fields are fast-rolling down the potential slopes,
causing
significant variation in the gauge kinetic function, allowing for our vector
scaling solution to be attained. As we have shown, after the vector scaling
attractor is reached, the backreaction of the vector field slows down the
roll of the inflaton. This gives rise to a slow-roll phase of inflation,
overcoming thereby the infamous $\eta$-problem, which plagues inflationary
models in supergravity. We have clearly demonstrated this in the last two of
our examples in Sec.~\ref{examples}. This is an added bonus to our setup.

%All in all, our paper demonstrates that, under fairly general conditions,
%a scale invariant spectrum of perturbations for the components of a vector
%field, massive or not, whose kinetic function (and mass) is modulated
%by the inflaton field is an attractor solution. If the field is massless, or
%if it remains light until the end of inflation, this attractor solution
%also generates anisotropic stress, which can render inflation weakly
%anisotropic. The above two characteristics of the attractor solution can
%source (independently or combined together) significant statistical anisotropy
%in the curvature perturbation, which may well be observable in the near future.

%---------------------------------------------------------------------------------------------------------------------
%---------------------------------------------------------------------------------------------------------------------
%\begin{acknowledgements}
\acknowledgements

\noindent
This work was supported (in part) by the European Union through the
Marie Curie Research and Training Network
\textquotedbl{}UniverseNet\textquotedbl{}(MRTN-CT-2006-035863).
J.M.W. is also supported by the Lancaster University Physics
Department.

%I would like to thank D.~H.~Lyth and J.~McDonald for discussions and the
%referee for insightful comments.
%\end{acknowledgements}

\end{widetext}
%---------------------------------------------------------------------------------------------------------------------
%---------------------------------------------------------------------------------------------------------------------

%\newpage

\appendix
\section{Eigenvalues}
The following plots show how the real parts of the eigenvalues
of the matrix $\mathcal{M}$ behave as a function of the dimensionless model parameters $\lambda$, $\Gamma$ and $\Xi$, for the different cases considered. In the massive vector field cases, light and heavy, we assume the connection of Eq. (\ref{connetion}).

\subsection{Massless Vector Field - Vector Scaling Solution, $\mathcal{VSS}$.}\label{App-VSS-m0}
Here we show numerical solutions to the real parts of the eigenvalues of matrix $\mathcal{M}$. These correspond to the $\mathcal{VSS}$ critical point of the massless vector field case, Eq. (\ref{VSS-m0}). We show numerical solutions because the analytical solutions are far too complicated to reproduce here. However we can clearly see from the plots, that in the limits $\Gamma\gg\lambda$ and $\Gamma\gg1$, the real parts
of the eigenvalues are approximated by [c.f. Eq. (\ref{VSS-m0-Remi})]
\begin{equation}
\textrm{Re}[m_{1}]\simeq-3\qquad\text{and}\qquad\textrm{Re}[m_{2,3}]\simeq-\frac{3}{2}.
\end{equation}
In the limits considered the eigenvalues depend weakly on the dimensionless model parameters. This can be seen as plateaus in the plots. This therefore ensures the validity of our method in establishing the stability of the critical point.
\bigskip
\begin{figure}[h]
\includegraphics[width=50mm,angle=0]{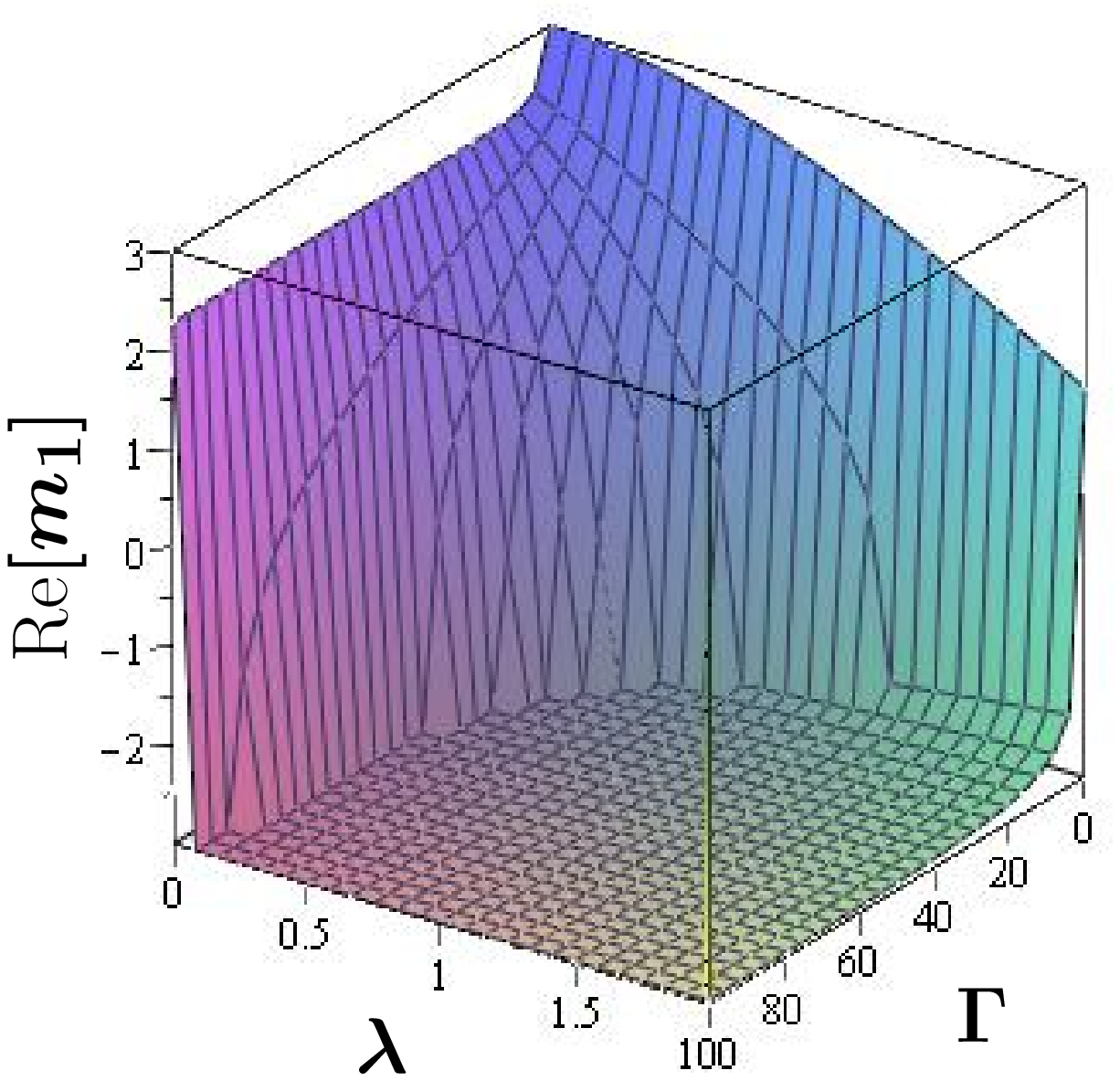}
\includegraphics[width=50mm,angle=0]{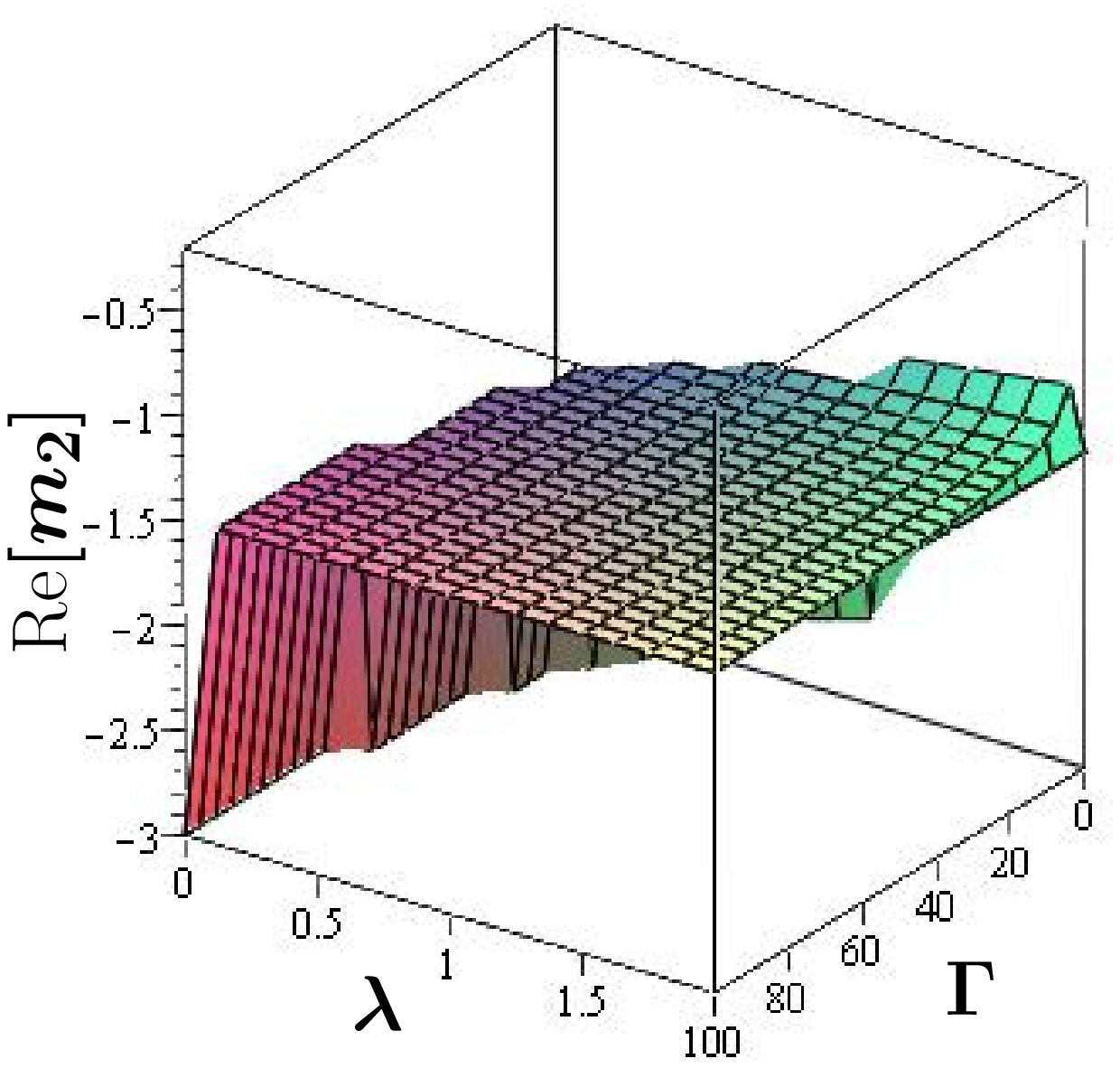}
\includegraphics[width=50mm,angle=0]{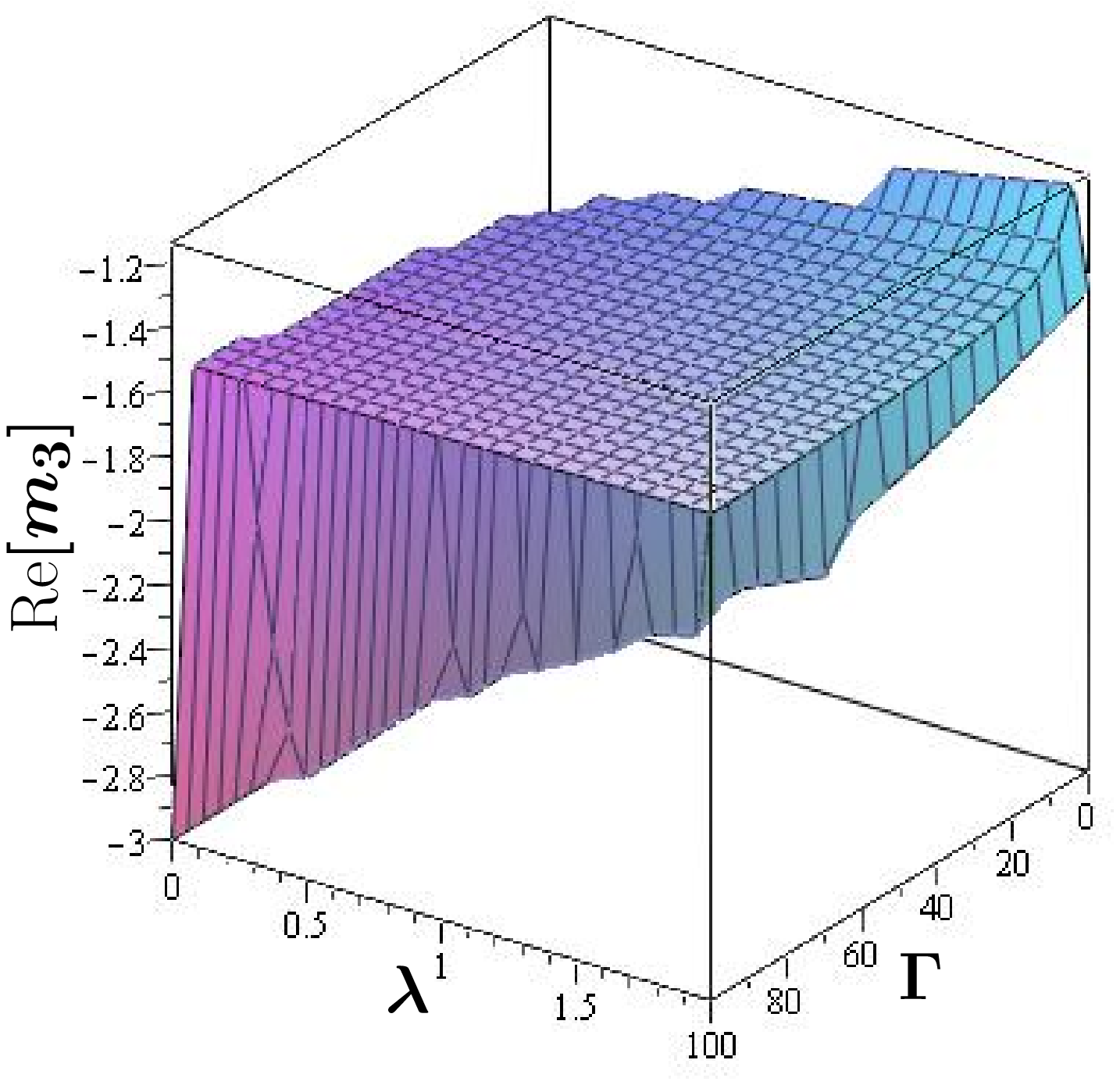}
\caption{These plots show the eigenvalues of matrix $\mathcal{M}$ for the $\mathcal{VSS}$ critical point in the massless vector field case.}
\end{figure}
%---------------------------------------------------------------------------------------------------------------------

\subsection{Light Vector Field - Vector Scaling Solution, $\mathcal{VSS}_{\it1}$.}\label{App-VSS1-mh-light}
The eigenvalues of the matrix $\mathcal{M}$ for the $\mathcal{VSS}_{\it1}$ critical point of the light vector field case, Eq. (\ref{VSS1}), are given by Eq. (\ref{VSS1-fulleigen}). It is clear from the analytical solutions that in the limits $\Gamma\gg\lambda$ and $\Gamma\gg1$ the real parts of the eigenvalues are approximated by [c.f. Eq. (\ref{VSS1-eigen})]
\begin{equation}
\textrm{Re}[m_{1}]\simeq-3\:,\qquad \textrm{Re}[m_{2,3}]\simeq-\frac{3}{2}\:,\quad\text{and}\quad \textrm{Re}[m_{4}]\simeq\frac{3\lambda\Gamma-12}{\Gamma^{2}}.
\end{equation}
In the limits considered the eigenvalues $m_{1,2,3}$ clearly depend weakly on the dimensionless model parameters. Here we are showing a numerical solution to the real part of the eigenvalue $m_4$ to demonstrate that it also depends weakly on the dimensionless model parameters. This can be seen as a plateau in the plot, and therefore ensures the validity of our method in establishing the stability of the critical point.

\begin{figure}[h]
\includegraphics[width=50mm,angle=0]{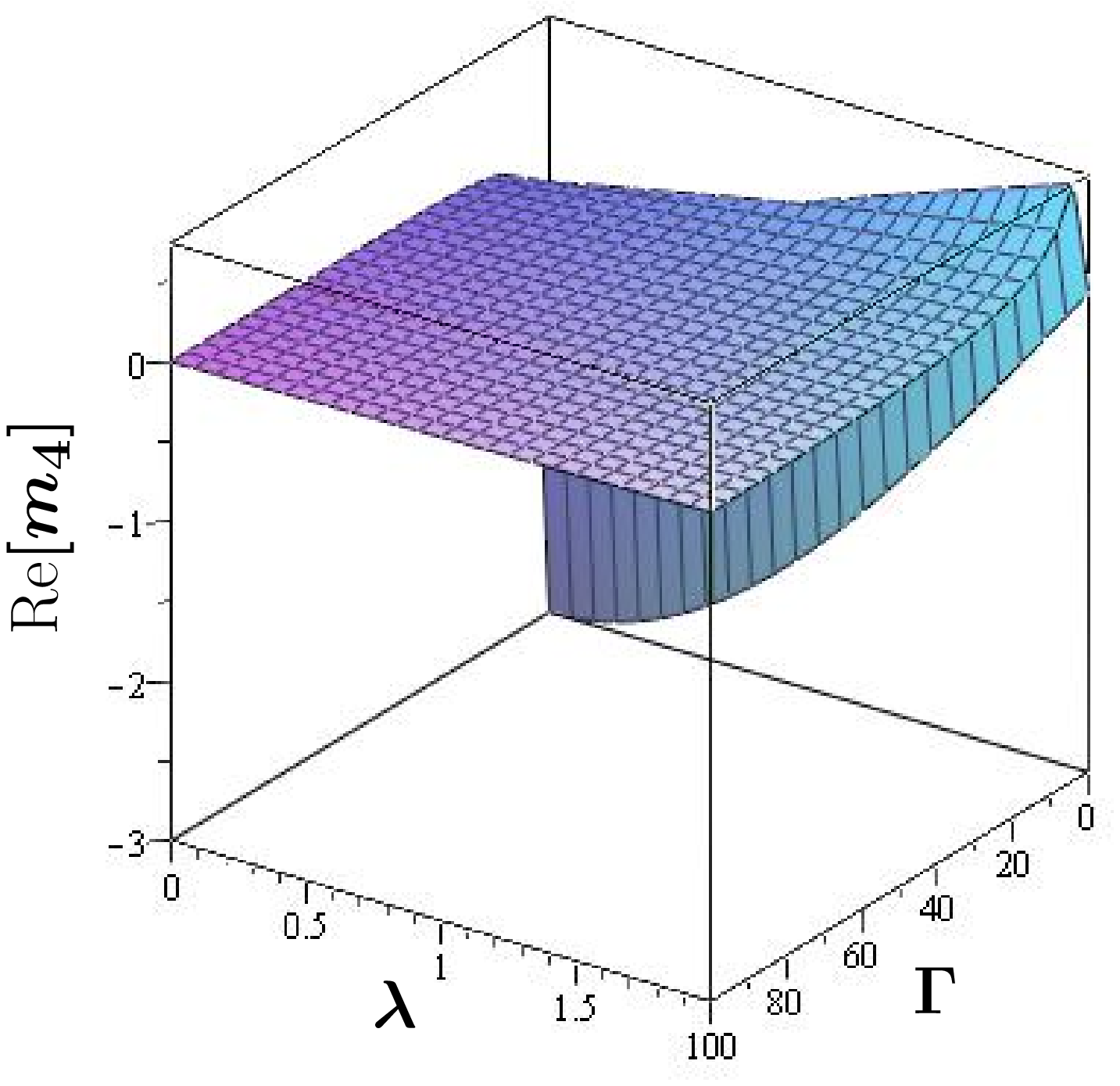}
\caption{This plot shows the eigenvalue $m_4$ of matrix $\mathcal{M}$ for the $\mathcal{VSS}_{\it1}$ critical point in the light vector field case.}
\end{figure}
%---------------------------------------------------------------------------------------------------------------------
\newpage
\subsection{Light Vector Field - Vector Scaling Solution, $\mathcal{VSS}_{\it2}$.}\label{App-VSS2-mh-light}
Here we show numerical solutions to the real parts of the eigenvalues of matrix $\mathcal{M}$ which correspond to the $\mathcal{VSS}_{\it2}$ critical point of the light vector field case, Eq. (\ref{VSS2}). We show numerical solutions because the analytical solutions are far too complicated to reproduce here. However we can clearly see from the plots, that in the limits $|\Xi|\gg\lambda$ and $|\Xi|\gg1$, the real parts of the eigenvalues are approximated by [c.f. Eq. (\ref{VSS2-eigen})] ($m_4$ is obtained by the analytic expression)
\begin{equation}
\textrm{Re}[m_{1}]\simeq-3,\quad
\textrm{Re}[m_{2,3}]\simeq-\frac{3}{2},\quad
\textrm{Re}[m_{4}]\simeq-\frac{3\left(\lambda\Xi+4\right)}{\Xi^{2}}
=\frac{6\lambda\Gamma-48}{\Gamma^{2}}.
\end{equation}
In the limits considered the eigenvalues depend weakly on the dimensionless model parameters. This can be seen as plateaus in the plots. This therefore ensures the validity of our method in establishing the stability of the critical point.

\bigskip
\begin{figure}[h]
\includegraphics[width=50mm,angle=0]{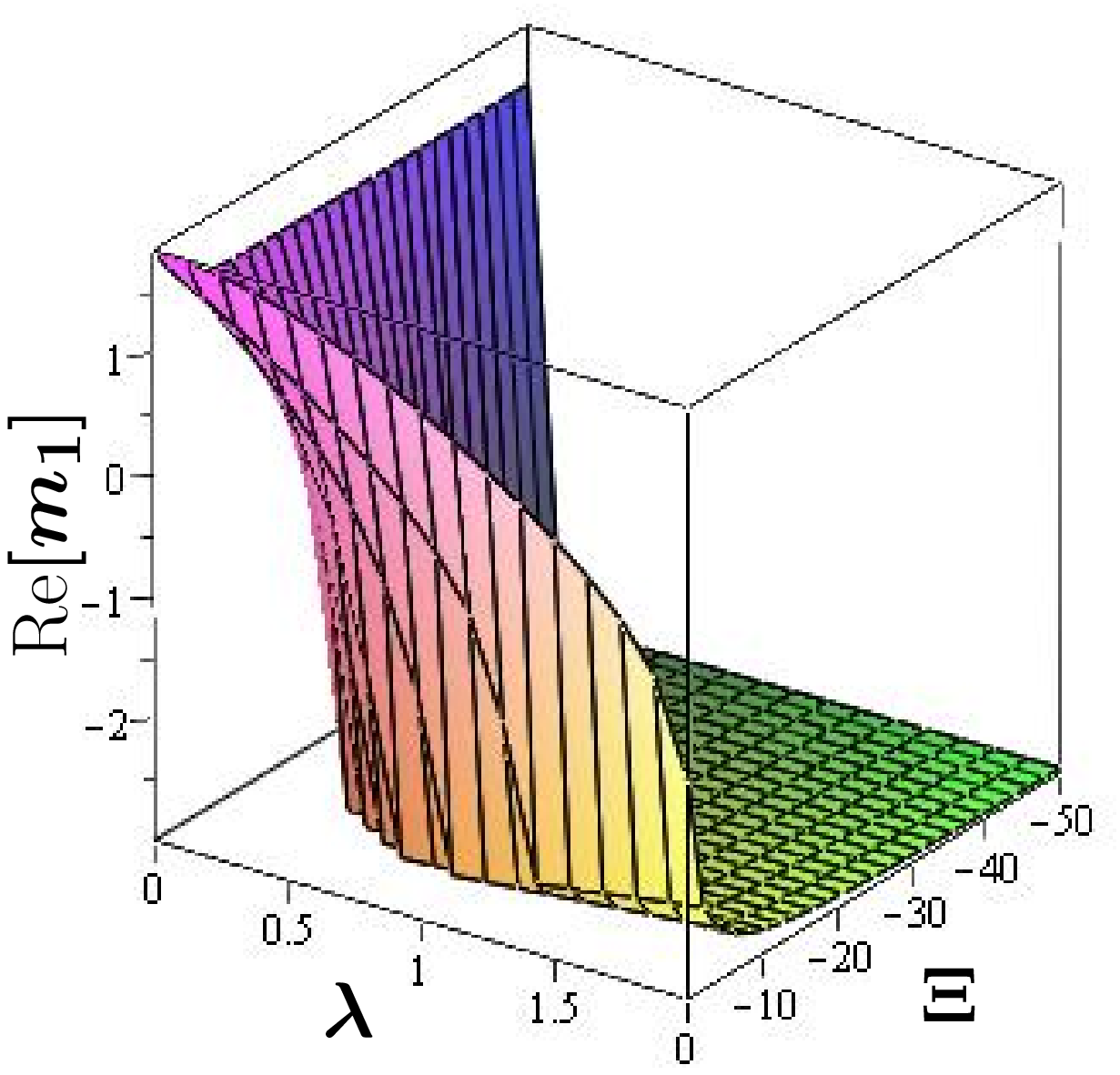}
\includegraphics[width=50mm,angle=0]{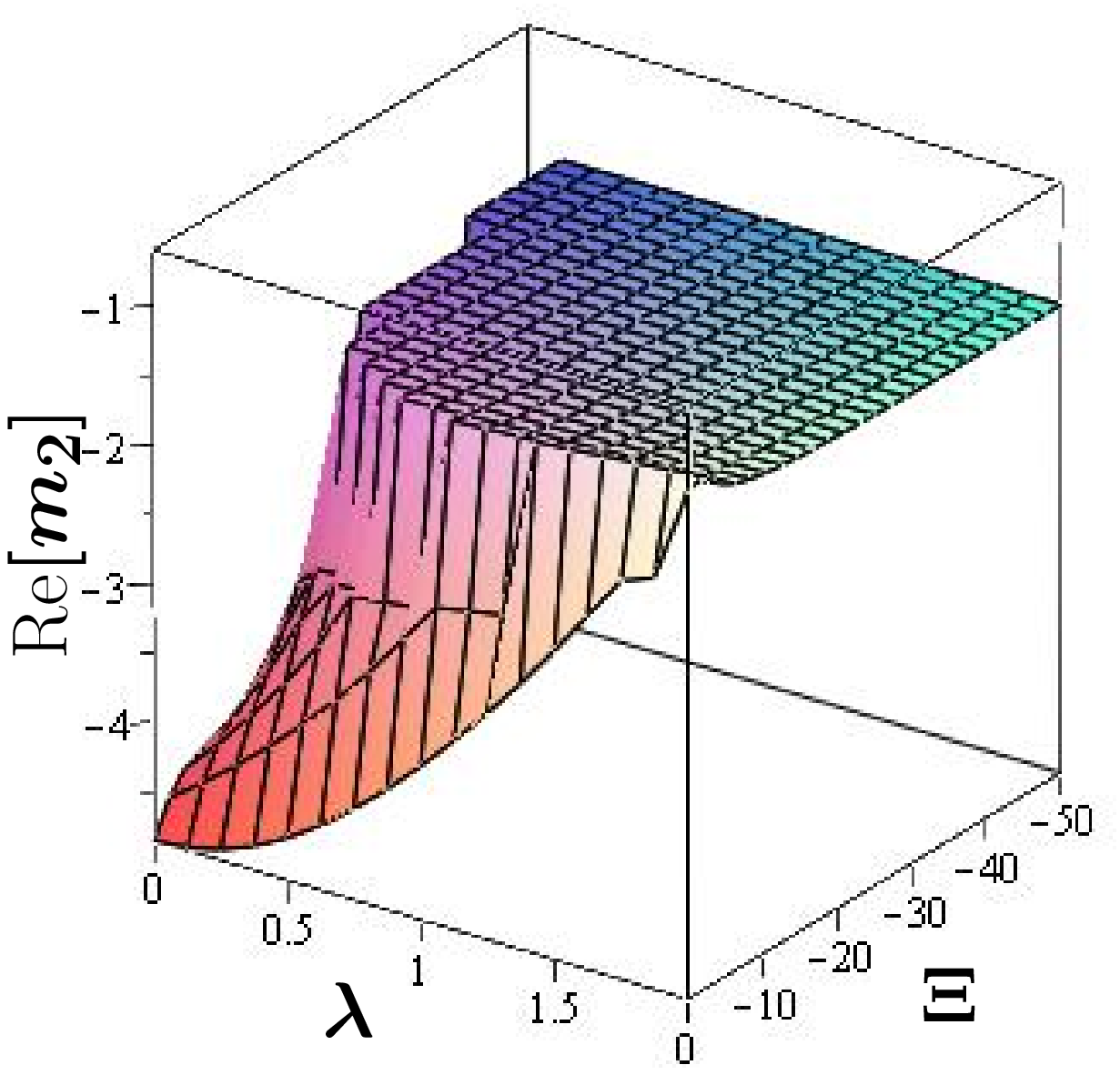}
\includegraphics[width=50mm,angle=0]{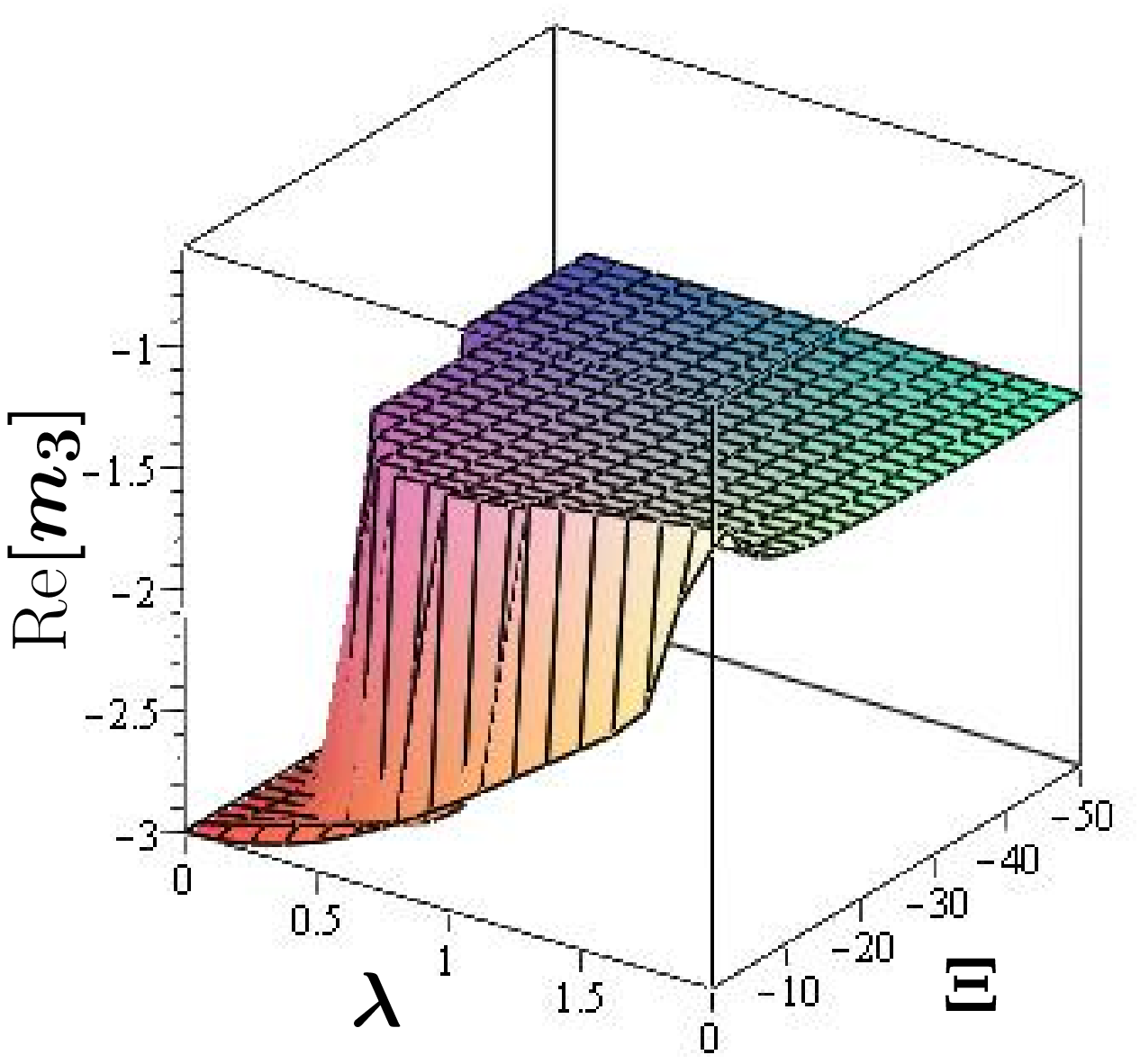}
\includegraphics[width=50mm,angle=0]{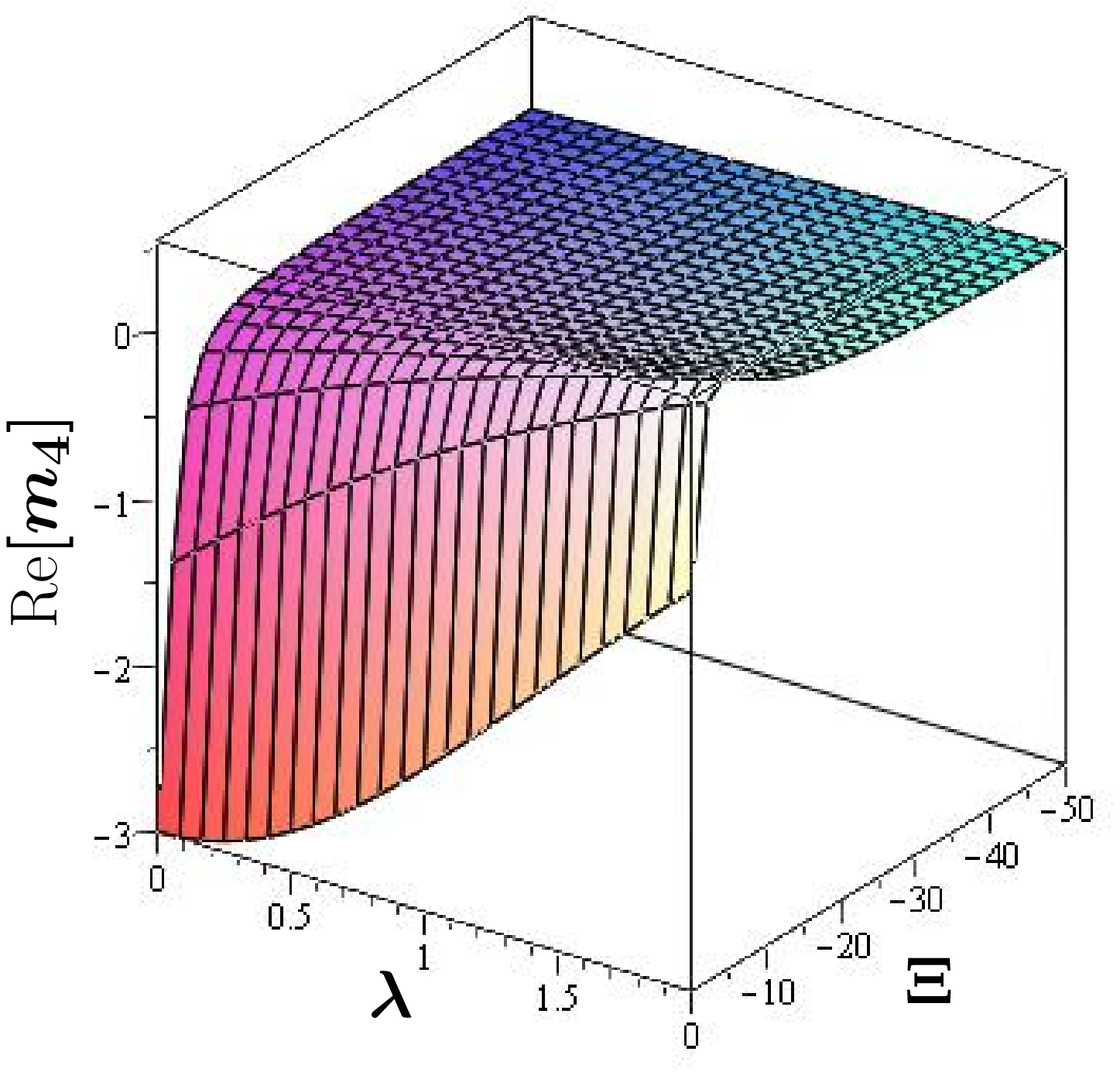}
\caption{These plots show the eigenvalues of matrix $\mathcal{M}$ for the $\mathcal{VSS}_{\it2}$ critical point in the light vector field case.}
\end{figure}

%---------------------------------------------------------------------------------------------------------------------

\subsection{Light Vector Field - Vector Scaling Solution, $\mathcal{VSS}_{\it3}$.}\label{App-VSS3-mh-light}
Here we show numerical solutions to the real parts of the eigenvalues of matrix $\mathcal{M}$ which correspond to the $\mathcal{VSS}_{\it3}$ critical point of the light vector field case, Eq. (\ref{VSS3}). We show numerical solutions because the analytical solutions are far too complicated to reproduce here. However we can clearly see from the plots, that in the limits $\Gamma\gg\lambda$ and $\Gamma\gg1$, the real parts
of the eigenvalues are approximated by [c.f. Eq. (\ref{VSS3-eigen})] ($m_4$ is obtained by the analytic expression)
\begin{equation}
 \textrm{Re}[m_{1}]\simeq-3
 \,,\qquad
 \textrm{Re}[m_{2,3}]\simeq-\frac{3}{2}\qquad\text{and}\qquad
 \textrm{Re}[m_4]\simeq -12\,{\frac { \left(\Gamma\lambda-8 \right)
 \left(\Gamma\lambda-4 \right) }{ \left( 3 \Gamma \lambda-20 \right)
 {\Gamma }^{2}}}.
\end{equation}
In the limits considered the eigenvalues depend weakly on the dimensionless model parameters. This can be seen as plateaus in the plots. This therefore ensures the validity of our method in establishing the stability of the critical point.

\begin{figure}[h]
\includegraphics[width=50mm,angle=0]{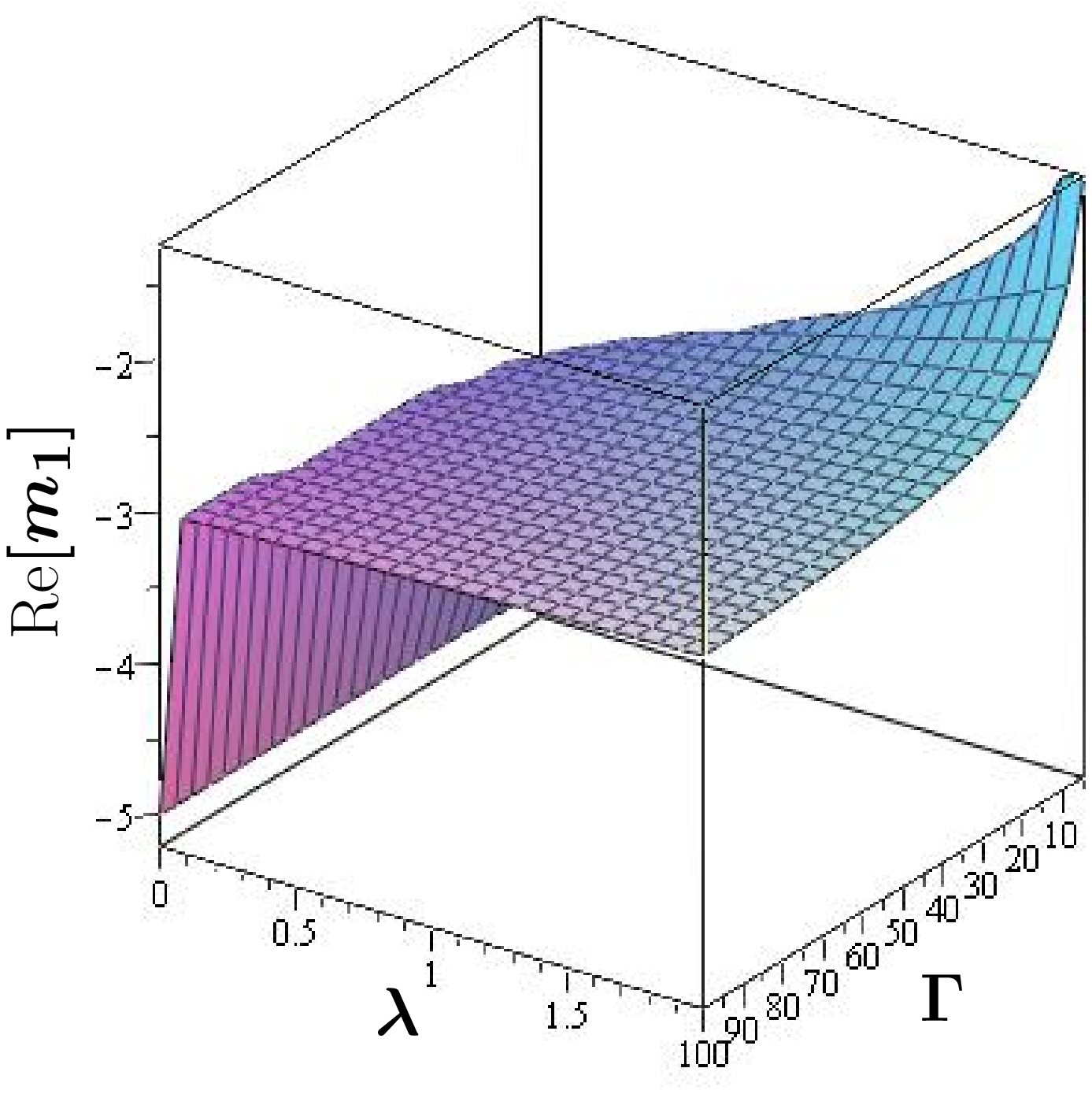}
\includegraphics[width=50mm,angle=0]{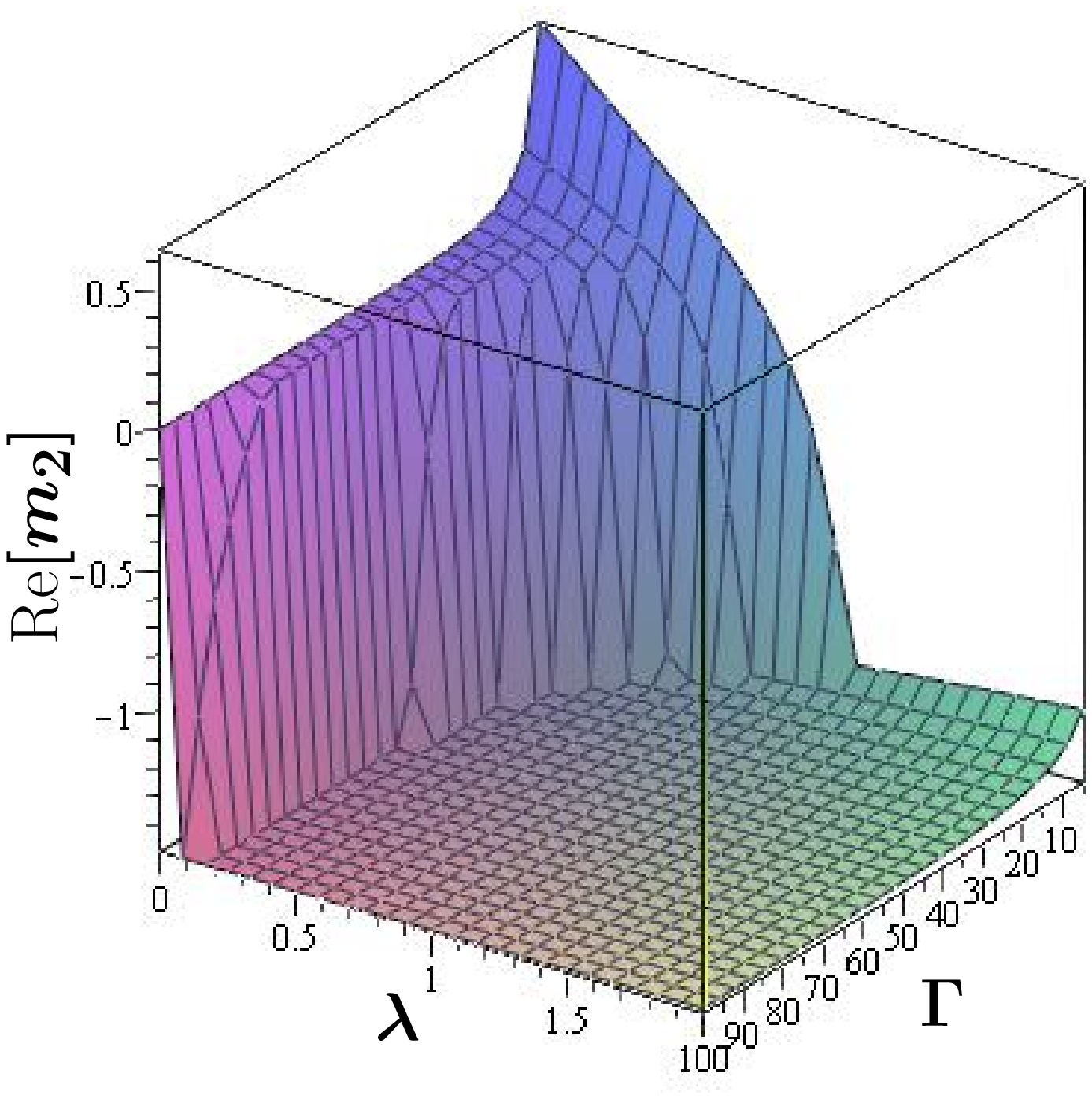}
\includegraphics[width=50mm,angle=0]{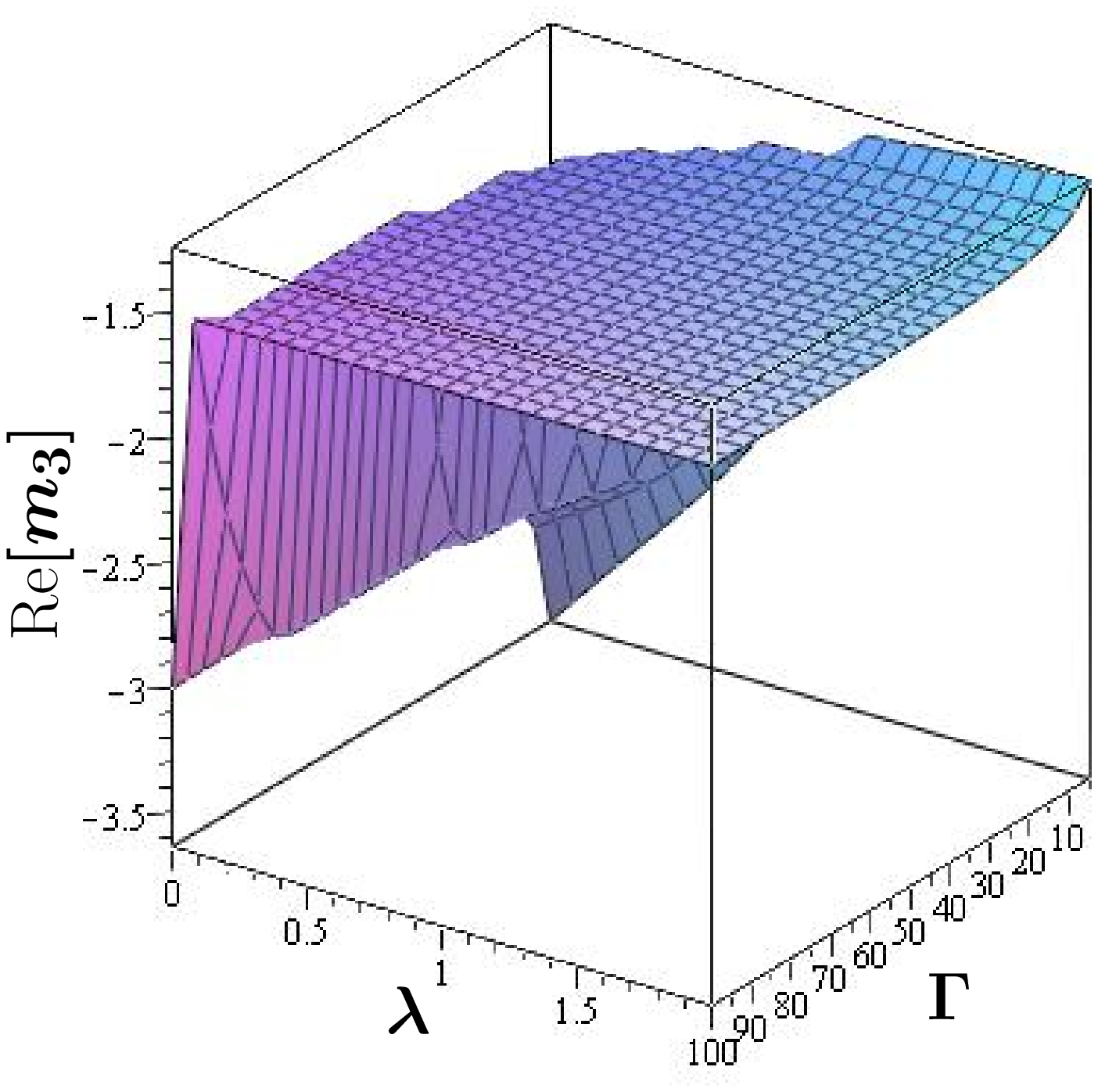}
\includegraphics[width=50mm,angle=0]{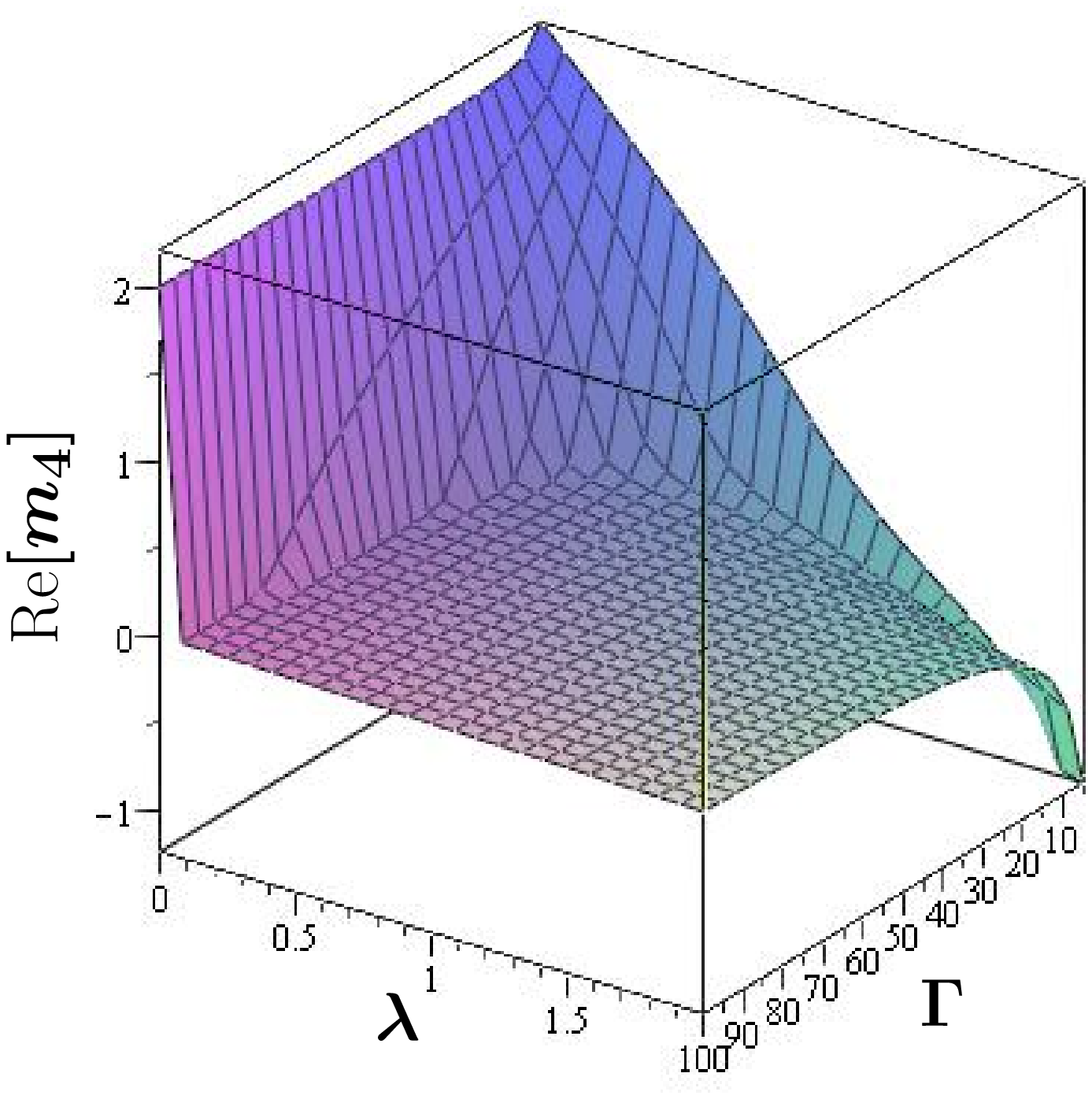}
\caption{These plots show the eigenvalues of matrix $\mathcal{M}$ for the $\mathcal{VSS}_{\it3}$ critical point in the light vector field case.}
\end{figure}

\newpage
%---------------------------------------------------------------------------------------------------------------------
%---------------------------------------------------------------------------------------------------------------------

\begin{thebiblio}{03}

%---------------------------------------------------------------------------------------------------------
%------My References-------------------------------------------------------------------------------------
\bibitem{inf}
A.~A.~Starobinsky,
%A new type of isotropic cosmological models without singularity,
Phys.\ Lett.\   B {\bf 91} (1980) 99;
  %%CITATION = PHLTA,B91,99;%%
A.~H.~Guth,
%The Inflationary Universe: A Possible Solution To The Horizon And Flatness Problems
Phys.\ Rev.\  D {\bf 23} (1981) 347.
%%CITATION = PHRVA,D23,347;%%

\bibitem{book}
A.~R.~Liddle and D.~H.~Lyth,
{\it The Primordial Density Perturbation: Cosmology, Inflation and the origin of Structure}
(Cambridge Univ. Press, Cambridge U.K., 2009).

\bibitem{leandros}
L.~Perivolaropoulos,
  %``The rise and fall of the cosmic string theory for cosmological
  %perturbations,''
  Nucl.\ Phys.\ Proc.\ Suppl.\  {\bf 148} (2005) 128
  %[arXiv:astro-ph/0501590].

\bibitem{curv}
D.~H.~Lyth and D.~Wands,
%``Generating the curvature perturbation without an inflaton,''
Phys.\ Lett.\ B {\bf 524} (2002) 5;
K.~Enqvist and M.~S.~Sloth,
%``Adiabatic CMB perturbations in pre big bang string cosmology,''
Nucl.\ Phys.\ B {\bf 626} (2002) 395;
%[arXiv:hep-ph/0109214].
%%CITATION = HEP-PH 0109214;%%
T.~Moroi and T.~Takahashi,
%``Effects of cosmological moduli fields on cosmic microwave background,''
Phys.\ Lett.\ B {\bf 522} (2001) 215
[Erratum-ibid.\ B {\bf 539} (2002) 303];
%[arXiv:hep-ph/0110096].
%%CITATION = HEP-PH 0110096;%%

\bibitem{wmap}
Komatsu~E {\it et al.},
%Seven-Year Wilkinson Microwave Anisotropy Probe (WMAP) Observations:
%Cosmological Interpretation
%arXiv:
1001.4538 [astro-ph.CO].
%%CITATION = ARXIV:1001.4538;%%

\bibitem{inhom}
%\bibitem{eriksen}
  H.~K.~Eriksen, F.~K.~Hansen, A.~J.~Banday, K.~M.~Gorski and P.~B.~Lilje,
  %``Asymmetries in the CMB anisotropy field,''
  Astrophys.\ J.\  {\bf 605} (2004) 14
  [Erratum-ibid.\  {\bf 609} (2004) 1198];
%  [arXiv:astro-ph/0307507].
  %%CITATION = ASJOA,605,14;%%

\bibitem{AoE}
K.~Land and J.~Magueijo,
%``The axis of evil,''
  Phys.\ Rev.\ Lett.\  {\bf 95} (2005) 071301;
% [arXiv:astro-ph/0502237].
  %%CITATION = PRLTA,95,071301;%%
Mon.\ Not.\ Roy.\ Astron.\ Soc.\ (2007) {\bf 378} 153.
%({\it Preprint} astro-ph/0611518)
%%CITATION = MNRAA,378,153;%%

\bibitem{acker}
L.~Ackerman, S.~M.~Carroll and M.~B.~Wise,
  %``Imprints of a Primordial Preferred Direction on the Microwave Background,''
  Phys.\ Rev.\  D {\bf 75} (2007) 083502.
%  [arXiv:astro-ph/0701357].
  %%CITATION = PHRVA,D75,083502;%%

\bibitem{forg}
  A.~de Oliveira-Costa and M.~Tegmark,
  %``CMB multipole measurements in the presence of foregrounds,''
  Phys.\ Rev.\  D {\bf 74} (2006) 023005.
%  [arXiv:astro-ph/0603369].
  %%CITATION = PHRVA,D74,023005;%%

\bibitem{hansen}
F.~K.~Hansen, A.~J.~Banday and K.~M.~Gorski
%Testing the cosmological principle of isotropy: local power spectrum estimates of the WMAP data,''
Mon.\ Not.\ Roy.\ Astron.\ Soc.\  {\bf 354} (2004) 641.
%({\it Preprint} astro-ph/0404206)
%%CITATION = MNRAA,354,641;%%

\bibitem{GE}
%N.~E.~Groeneboom and H.~K.~Eriksen,
% %``Bayesian analysis of sparse anisotropic universe models and application to
% %the 5-yr WMAP data,''
%  Astrophys.\ J.\  {\bf 690} (2009) 1807.
%%  [arXiv:0807.2242 [astro-ph]].
%  %%CITATION = ASJOA,690,1807;%%
N.~E.~Groeneboom, L.~Ackerman, I.~K.~Wehus and H.~K.~Eriksen,
  %``Bayesian analysis of an anisotropic universe model: systematics and
  %polarization,''
%  arXiv:
0911.0150 [astro-ph.CO].
  %%CITATION = ARXIV:0911.0150;%%

\bibitem{ref}
D.~Hanson and A.~Lewis,
  %``Estimators for CMB Statistical Anisotropy,''
  Phys.\ Rev.\  D {\bf 80} (2009) 063004.
%  [arXiv:0908.0963 [astro-ph.CO]].
  %%CITATION = PHRVA,D80,063004;%%

\bibitem{fnlanis}
 M.~Kar\v{c}iauskas, K.~Dimopoulos and D.~H.~Lyth,
  %``Anisotropic non-Gaussianity from vector field perturbations,''
  Phys.\ Rev.\  D {\bf 80} (2009) 023509.
%  [arXiv:0812.0264 [astro-ph]].
  %%CITATION = PHRVA,D80,023509;%%

\bibitem{cesar}
C.~A.~Valenzuela-Toledo, Y.~Rodriguez and D.~H.~Lyth,
  %``Non-gaussianity at tree- and one-loop levels from vector field
  %perturbations,''
  Phys.\ Rev.\  D {\bf 80} (2009) 103519;
%  [arXiv:0909.4064 [astro-ph.CO]].
  %%CITATION = PHRVA,D80,103519;%%
C.~A.~Valenzuela-Toledo and Y.~Rodriguez,
  %``Non-gaussianity from the trispectrum and vector field perturbations,''
  Phys.\ Lett.\  B {\bf 685} (2010) 120.
%  [arXiv:0910.4208 [astro-ph.CO]].
  %%CITATION = PHLTA,B685,120;%%

\bibitem{planck}
A.~R.~Pullen and M.~Kamionkowski
%Cosmic Microwave Background Statistics for a Direction-Dependent Primordial Power Spectrum
Phys.\ Rev.\  D {\bf 76} (2007) 103529.
%({\it Preprint} 0709.1144 [astro-ph])
%%CITATION = PHRVA,D76,103529;%%

\bibitem{vecurv}
K.~Dimopoulos,
%``Can a vector field be responsible for the curvature perturbation in the
%universe?,''
Phys.\ Rev.\  D {\bf 74} (2006) 083502.
%[arXiv:hep-ph/0607229].
%%CITATION = PHRVA,D74,083502;%%

\bibitem{stanis}
K.~Dimopoulos, M.~Kar\v{c}iauskas, D.~H.~Lyth and Y.~Rodriguez,
  %``Statistical anisotropy of the curvature perturbation from vector field
  %perturbations,''
  JCAP {\bf 0905} (2009) 013.
%  [arXiv:0809.1055 [astro-ph]].
  %%CITATION = JCAPA,0905,013;%%

\bibitem{anisinf0}
S.~Kanno, M.~Kimura, J.~Soda and S.~Yokoyama,
  %``Anisotropic Inflation from Vector Impurity,''
  JCAP {\bf 0808} (2008) 034.
%  [arXiv:0806.2422 [hep-ph]];
  %%CITATION = JCAPA,0808,034;%%

\bibitem{anisinf}
  M.~A.~Watanabe, S.~Kanno and J.~Soda,
  %``Inflationary Universe with Anisotropic Hair,''
  Phys.\ Rev.\ Lett.\  {\bf 102} (2009) 191302.
  %[arXiv:0902.2833 [hep-th]].
  %%CITATION = PRLTA,102,191302;%%
%\cite{Watanabe:2010fh}

\bibitem{anisinf+}
  M.~a.~Watanabe, S.~Kanno and J.~Soda,
  %``The Nature of Primordial Fluctuations from Anisotropic Inflation,''
  Prog.\ Theor.\ Phys.\  {\bf 123} (2010) 1041;
  %[arXiv:1003.0056 [astro-ph.CO]].
  %%CITATION = PTPKA,123,1041;%%
  T.~R.~Dulaney and M.~I.~Gresham,
  %``Primordial Power Spectra from Anisotropic Inflation,''
  Phys.\ Rev.\  D {\bf 81} (2010) 103532;
  %[arXiv:1001.2301 [astro-ph.CO]].
  %%CITATION = PHRVA,D81,103532;%%
  A.~E.~Gumrukcuoglu, B.~Himmetoglu and M.~Peloso,
  %``Scalar-Scalar, Scalar-Tensor, and Tensor-Tensor Correlators from
  %Anisotropic Inflation,''
  Phys.\ Rev.\  D {\bf 81} (2010) 063528;
  %[arXiv:1001.4088 [astro-ph.CO]].
  %%CITATION = PHRVA,D81,063528;%%
  B.~Himmetoglu,
  %``Spectrum of Perturbations in Anisotropic Inflationary Universe with Vector
  %Hair,''
  JCAP {\bf 1003} (2010) 023.
  %[arXiv:0910.3235 [astro-ph.CO]].
  %%CITATION = JCAPA,1003,023;%%

\bibitem{varkin}
K.~Dimopoulos, M.~Karciauskas and J.~M.~Wagstaff,
  %``Vector Curvaton with varying Kinetic Function,''
  Phys.\ Rev.\  D {\bf 81} (2010) 023522;
%  [arXiv:0907.1838 [hep-ph]].
  %%CITATION = PHRVA,D81,023522;%%
%\bibitem{noinsta}
%K.~Dimopoulos, M.~Karciauskas and J.~M.~Wagstaff,
  %``Vector Curvaton without Instabilities,''
  Phys.\ Lett.\  B {\bf 683} (2010) 298.
%  [arXiv:0909.0475 [hep-ph]].
  %%CITATION = PHLTA,B683,298;%%

\bibitem{pmfrev}
D.~Grasso and H.R.~Rubinstein,
  %``Magnetic fields in the early universe,''
  Phys.\ Rept.\  {\bf 348} (2001) 163;
%  [arXiv:astro-ph/0009061].
  %%CITATION = PRPLC,348,163;%%
M.~Giovannini,
  %``The magnetized universe,''
  Int.\ J.\ Mod.\ Phys.\  D {\bf 13} (2004) 391.
%  [arXiv:astro-ph/0312614].
  %%CITATION = IMPAE,D13,391;%%

\bibitem{mine}
A.~C.~Davis, K.~Dimopoulos, T.~Prokopec and O.~Tornkvist,
  %``Primordial spectrum of gauge fields from inflation,''
  Phys.\ Lett.\  B {\bf 501} (2001) 165;
%  [Phys.\ Rev.\ Focus {\bf 10} (2002) STORY9]
%  [arXiv:astro-ph/0007214].
  %%CITATION = 00627,10,STORY9;%%
K.~Dimopoulos, T.~Prokopec, O.~Tornkvist and A.~C.~Davis,
  %``Natural magnetogenesis from inflation,''
  Phys.\ Rev.\  D {\bf 65} (2002) 063505.
%  [arXiv:astro-ph/0108093].
  %%CITATION = PHRVA,D65,063505;%%

\bibitem{TW}
M.~S.~Turner and L.~M.~Widrow,
  %``Inflation Produced, Large Scale Magnetic Fields,''
  Phys.\ Rev.\  D {\bf 37} (1988) 2743.
  %%CITATION = PHRVA,D37,2743;%%

\bibitem{nonmin}
K.~Dimopoulos and M.~Kar\v{c}iauskas,
  %``Non-minimally coupled vector curvaton,''
  JHEP {\bf 0807} (2008) 119.
%  [arXiv:0803.3041 [hep-th]].
  %%CITATION = JHEPA,0807,119;%%

\bibitem{peloso}
B.~Himmetoglu, C.~R.~Contaldi and M.~Peloso,
  %``Instability of the ACW model, and problems with massive vectors during
  %inflation,''
  Phys.\ Rev.\  D {\bf 79} (2009) 063517;
%  [arXiv:0812.1231 [astro-ph]];
  %%CITATION = PHRVA,D79,063517;%%
%B.~Himmetoglu, C.~R.~Contaldi and M.~Peloso,
  %``Instability of anisotropic cosmological solutions supported by vector
  %fields,''
  Phys.\ Rev.\ Lett.\  {\bf 102} (2009) 111301.
%  [arXiv:0809.2779 [astro-ph]].
  %%CITATION = PRLTA,102,111301;%%

\bibitem{RA2save}
M.~Kar\v{c}iauskas and D.~H.~Lyth,
%On the health of a vector field with (R A$^2$)/6 coupling to gravity
1007.1426 [astro-ph.CO].
%%CITATION = ARXIV:1007.1426;%%

\bibitem{gaugekin}
M.~Giovannini,
  %``On the variation of the gauge couplings during inflation,''
  Phys.\ Rev.\  D {\bf 64} (2001) 061301;
%  [arXiv:astro-ph/0104290].
  %%CITATION = PHRVA,D64,061301;%%
K.~Bamba and J.~Yokoyama,
  %``Large-scale magnetic fields from inflation in dilaton electromagnetism,''
  Phys.\ Rev.\  D {\bf 69} (2004) 043507;
%  [arXiv:astro-ph/0310824].
  %%CITATION = PHRVA,D69,043507;%%
%K.~Bamba and J.~Yokoyama,
  %``Large-scale magnetic fields from dilaton inflation in noncommutative
  %spacetime,''
  Phys.\ Rev.\  D {\bf 70} (2004) 083508;
%  [arXiv:hep-ph/0409237].
  %%CITATION = PHRVA,D70,083508;%%
O.~Bertolami and R.~Monteiro,
  %``Varying electromagnetic coupling and primordial magnetic fields,''
  Phys.\ Rev.\  D {\bf 71} (2005) 123525;
%  [arXiv:astro-ph/0504211].
  %%CITATION = PHRVA,D71,123525;%%
J.~M.~Salim, N.~Souza, S.~E.~Perez Bergliaffa and T.~Prokopec,
  %``Creation of cosmological magnetic fields in a bouncing cosmology,''
  JCAP {\bf 0704} (2007) 011;
%  [arXiv:astro-ph/0612281].
  %%CITATION = JCAPA,0704,011;%%
K.~Bamba and M.~Sasaki,
  %``Large-scale magnetic fields in the inflationary universe,''
  JCAP {\bf 0702} (2007) 030;
%  [arXiv:astro-ph/0611701].
  %%CITATION = JCAPA,0702,030;%%
J.~Martin and J.~Yokoyama,
  %``Generation of Large-Scale Magnetic Fields in Single-Field Inflation,''
  JCAP {\bf 0801} (2008) 025;
%  [arXiv:0711.4307 [astro-ph]].
  %%CITATION = JCAPA,0801,025;%%
K.~Bamba and S.~D.~Odintsov,
  %``Inflation and late-time cosmic acceleration in non-minimal Maxwell-$F(R)$
  %gravity and the generation of large-scale magnetic fields,''
  JCAP {\bf 0804} (2008) 024;
%  [arXiv:0801.0954 [astro-ph]].
  %%CITATION = JCAPA,0804,024;%%
K.~Bamba, C.~Q.~Geng and S.~H.~Ho,
  %``Large-scale magnetic fields from inflation due to Chern-Simons-like
  %effective interaction,''
  JCAP {\bf 0811} (2008) 013;
%  [arXiv:0806.1856 [astro-ph]].
  %%CITATION = JCAPA,0811,013;%%
V.~Demozzi, V.~Mukhanov and H.~Rubinstein,
  %``Magnetic fields from inflation?,''
  JCAP {\bf 0908} (2009) 025.
%  [arXiv:0907.1030 [astro-ph.CO]].
  %%CITATION = JCAPA,0908,025;%%

\bibitem{sugravec}
K.~Dimopoulos,
  %``Supergravity inspired Vector Curvaton,''
  Phys.\ Rev.\  D {\bf 76} (2007) 063506.
%  [arXiv:0705.3334 [hep-ph]].
  %%CITATION = PHRVA,D76,063506;%%

\bibitem{soda}
  S.~Yokoyama and J.~Soda,
  %``Primordial statistical anisotropy generated at the end of inflation,''
  JCAP {\bf 0808} (2008) 005.
%  [arXiv:0805.4265 [astro-ph]].
  %%CITATION = JCAPA,0808,005;%%

%\cite{Kanno:2009ei}
\bibitem{Kanno:2009ei}
  S.~Kanno, J.~Soda and M.~a.~Watanabe,
  %``Cosmological Magnetic Fields from Inflation and Backreaction,''
  JCAP {\bf 0912}, 009 (2009).
  %[arXiv:0908.3509 [astro-ph.CO]].
  %%CITATION = JCAPA,0912,009;%%

%\cite{Emami:2010rm}
\bibitem{Emami:2010rm}
  R.~Emami, H.~Firouzjahi, S.~M.~S.~Movahed and M.~Zarei,
  %``Anisotropic Inflation from Charged Scalar Fields,''
  %arXiv:
  1010.5495 [astro-ph.CO].
  %%CITATION = ARXIV:1010.5495;%%

\bibitem{sodanew}
S.~Kanno, J.~Soda and M.~A.~Watanabe,
  %``Anisotropic Power-law Inflation,''
%  arXiv:
1010.5307 [hep-th].
  %%CITATION = ARXIV:1010.5307;%%

\bibitem{carroll}
S.~M.~Carroll, T.~R.~Dulaney, M.~I.~Gresham and H.~Tam,
  %``Instabilities in the Aether,''
  Phys.\ Rev.\  D {\bf 79} (2009) 065011;
%  [arXiv:0812.1049 [hep-th]].
  %%CITATION = PHRVA,D79,065011;%%
T.~R.~Dulaney, M.~I.~Gresham and M.~B.~Wise,
  %``Classical stability of a homogeneous, anisotropic inflating space-time,''
  Phys.\ Rev.\  D {\bf 77} (2008) 083510
  [Erratum-ibid.\  D {\bf 79} (2009) 029903].
%  [arXiv:0801.2950 [astro-ph]];
  %%CITATION = PHRVA,D77,083510;%%

\bibitem{tikto}
  J.~M.~Cornwall, D.~N.~Levin and G.~Tiktopoulos,
  %``Derivation Of Gauge Invariance From High-Energy Unitarity Bounds On The S
  %Matrix,''
  Phys.\ Rev.\  D {\bf 10} (1974) 1145
  [Erratum-ibid.\  D {\bf 11} (1975) 972].
  %%CITATION = PHRVA,D10,1145;%%

\bibitem{WainwrightEllis} J.~Wainwright and G. F. R.~Ellis,
{\it Dynamical Systems in Cosmology},
(Cambridge University Press, Cambridge, 1997).

%\cite{Wald:1983ky}
\bibitem{wald}
  R.~M.~Wald,
  % ``Asymptotic behavior of homogeneous cosmological models in the presence of a
  %positive cosmological constant,''
  Phys.\ Rev.\  D {\bf 28}, 2118 (1983).
  %%CITATION = PHRVA,D28,2118;%%

\bibitem{FR}
A.~D.~Linde,
  %``Fast-Roll Inflation,''
  JHEP {\bf 0111} (2001) 052.
%  [arXiv:hep-th/0110195].
  %%CITATION = JHEPA,0111,052;%%

\bibitem{smooth}
G.~Lazarides and C.~Panagiotakopoulos,
  %``Smooth hybrid inflation,''
  Phys.\ Rev.\  D {\bf 52} (1995) R559
  %[arXiv:hep-ph/9506325].
  %%CITATION = PHRVA,D52,559;%%

\bibitem{randall}
%\cite{Dine:1995kz}
%\bibitem{Dine:1995kz}
M.~Dine, L.~Randall and S.~Thomas,
%``Baryogenesis from flat directions of the
%supersymmetric standard model,''
Nucl.\ Phys.\ B {\bf 458} (1996) 291;
%[arXiv:hep-ph/9507453]
%%CITATION = HEP-PH 9507453;%%
%\cite{Dine:1995uk}
%\bibitem{Dine:1995uk}
%M.~Dine, L.~Randall and S.~Thomas,
%``Supersymmetry breaking in the
%early universe,''
Phys.\ Rev.\ Lett.\  {\bf 75} (1995) 398;
%[arXiv:hep-ph/9503303].
%%CITATION = HEP-PH 9503303;%%
D.~H.~Lyth and T.~Moroi,
  %``The masses of weakly-coupled scalar fields in the early universe,''
  JHEP {\bf 0405} (2004) 004.
%  [arXiv:hep-ph/0402174].
  %%CITATION = JHEPA,0405,004;%%

\end{thebiblio}
\end{document}